\documentclass[12pt]{article}

\usepackage[
  letterpaper,
  left=1.5in,
  right=1.5in,
  top=1in,
  bottom=1in,
  footskip=0.6in
]{geometry}
\usepackage[english]{babel}
\usepackage[T1]{fontenc}
\usepackage{newtxtext}
\usepackage{microtype}
\usepackage{setspace}
\usepackage{titlesec}
\usepackage{etoolbox}

\usepackage{amsmath,amssymb,amsfonts,amsthm,mathtools}
\usepackage{newtxmath}
\usepackage{bbm}

\usepackage{graphicx}
\usepackage{caption}
\usepackage{subcaption}
\usepackage{booktabs}
\usepackage{multirow}
\usepackage{float}
\usepackage{algorithm}
\usepackage{algpseudocode}

\usepackage{comment}
\usepackage{xcolor}
\usepackage{xspace}
\usepackage{natbib}
\usepackage{authblk}
\usepackage[
  colorlinks=false,
  linkbordercolor={1 0 0},
  citebordercolor={0 1 0},
  urlbordercolor={0 1 1},
  pdfborder={0 0 1}
]{hyperref}
\usepackage{cleveref}

\AtBeginDocument{%
  \setlength{\abovedisplayskip}{6pt plus 2pt minus 1pt}%
  \setlength{\belowdisplayskip}{6pt plus 2pt minus 1pt}%
  \setlength{\abovedisplayshortskip}{3pt plus 1pt}%
  \setlength{\belowdisplayshortskip}{4pt plus 1pt minus 1pt}%
  \allowdisplaybreaks[1]%
}

\titleformat{\section}
  {\large\bfseries}{\thesection.}{0.65em}{}
\titleformat{\subsection}
  {\normalsize\bfseries}{\thesubsection.}{0.55em}{}
\titleformat{\subsubsection}
  {\normalsize\itshape}{\thesubsubsection.}{0.55em}{}
\titlespacing*{\section}{0pt}{2.4ex plus 0.6ex minus 0.2ex}
  {1.0ex plus 0.2ex}
\titlespacing*{\subsection}{0pt}{1.9ex plus 0.4ex minus 0.2ex}
  {0.7ex plus 0.2ex}
\titlespacing*{\subsubsection}{0pt}{1.5ex plus 0.3ex minus 0.2ex}
  {0.5ex plus 0.1ex}

\newtheoremstyle{statplain}
  {6pt}{6pt}{\itshape}{}
  {\bfseries}{.}{0.5em}{}
\newtheoremstyle{statdefinition}
  {6pt}{6pt}{\normalfont}{}
  {\bfseries}{.}{0.5em}{}

\theoremstyle{statplain}
\newtheorem{theorem}{Theorem}[section]
\newtheorem{lemma}[theorem]{Lemma}
\newtheorem{proposition}[theorem]{Proposition}
\newtheorem{corollary}[theorem]{Corollary}

\theoremstyle{statdefinition}

\newtheorem{comments}[theorem]{Comment}

\DeclareMathOperator*{\argmin}{arg\,min}

\DeclareMathOperator{\poly}{poly}

\DeclareMathOperator{\KL}{KL}

\newcommand{\R}{\mathbb{R}}
\newcommand{\E}{\mathbb{E}}

\AtBeginEnvironment{thebibliography}{\small\setstretch{1}}
\definecolor{myGreen}{RGB}{80,180,0}
\definecolor{myGold}{rgb}{0.75,0.6,0.12}
\definecolor{myBlue}{rgb}{0.12,0.45,0.75}

\usepackage{array}
\usepackage{bm}
\usepackage{enumitem}
\usepackage{pifont}

\let\E\relax
\let\R\relax

\let\poly\relax

\let\manuscriptusepackage\usepackage
\renewcommand{\usepackage}[2][]{}

\def\comments{1}

\ifnum\comments=1
    \newcommand{\mynote}[2]{\marginpar{\color{#1}\sf \tiny #2}}
    \newcommand{\myinline}[2]{{\color{#1}\sf [{#2}]}}
\else
    \newcommand{\mynote}[2]{}
    \newcommand{\myinline}[2]{}    
\fi

\newcommand{\E}{\mathbb{E}}

\newcommand{\R}{\mathbb{R}}

\newcommand{\obs}{^{\mathrm{obs}}}
\newcommand{\expm}{^{\mathrm{exp}}} 

\newcommand{\CV}{\mathsf{CV}}

\newcommand{\cN}{\mathcal{N}}

\newcommand{\expshort}{{\mathrm{exp}}}

\newcommand{\obsshort}{{\mathrm{obs}}}

\newcommand{\lexp}{{L^{\mathrm{exp}}}}
\newcommand{\lobs}{{L^{\mathrm{obs}}}}

\newcommand{\nexp}{{N^{\expshort}}}
\newcommand{\dexp}{d_{\expshort}}
\newcommand{\datepar}{d_{\atepar}}

\newcommand{\dobs}{d_{\obsshort}}
\newcommand{\distexp}{P^{\expshort}}

\newcommand{\ate}{\tau^{\star}}
\newcommand{\trueate}{\tau^{\star}}
\newcommand{\boundatediff}{B_{\ate,0}}
\newcommand{\boundatelin}{B_{\ate,1}}
\newcommand{\boundate}{B_{\ate}}

\newcommand{\boundatenum}{B_{\ate,\mathrm{num}}}

\newcommand{\empmean}[1]{\widehat{\E}_{#1}}

\newcommand{\atepar}{\eta}
\newcommand{\atepardiff}{\widehat{\Delta}}
\usepackage{mathabx}
\newcommand{\normaldiff}{\widebar{\Delta}}
\newcommand{\covernum}{{\mathcal N}}

\newcommand{\trueatepar}{\eta^{\star}}
\newcommand{\strongcvx}{\gamma}

\newcommand{\ateparspace}{{\sf H}}
\newcommand{\boundatepar}{B_{\ateparspace}}

\newcommand{\estatepar}{\widehat\eta}

\newcommand{\dset}{D}
\newcommand{\dsetobs}{D^{\obsshort}}

\newcommand{\regu}{\lambda}
\newcommand{\trueregu}{\lambda^{\star}}
\newcommand{\estregu}{\widehat{\lambda}}
\newcommand{\regucst}{V}

\newcommand{\estate}{\widehat\tau\expm}

\newcommand{\Par}{\theta}
\newcommand{\Pardim}{d_{\Par}}

\newcommand{\Parspace}{\Theta}
\newcommand{\boundobspar}{B_{\Parspace}}
\newcommand{\EstPar}{\widehat\theta}

\newcommand{\atefun}{\beta}
\newcommand{\fold}{B}
\newcommand{\numfold}{K}

\newcommand{\hesshort}{T}
\newcommand{\hesshorttil}{\widetilde{T}}
\newcommand{\diffpar}{\widetilde\Delta}
\newcommand{\diffpartwo}{\widebar{\Delta}}

\newcommand{\Eone}{E}
\newcommand{\onehot}{e}

\newcommand{\Htermone}{\mathbf{T}^1}
\newcommand{\Htermtwo}{\mathbf{T}^2}
\newcommand{\boundobsh}{B_{\mathrm{obs},2}}
\newcommand{\boundobsthr}{B_{\mathrm{obs},3}}
\newcommand{\remainderterm}{{\mathcal R}}

\newcommand{\remainder}{R}
\newcommand{\lboundobsh}{b_{\mathrm{obs},2}}

\newcommand{\nobs}{{N^{\obsshort}}}
\newcommand{\expsam}{X^{\expshort}}
\newcommand{\sam}{X}

\newcommand{\expsamspace}{\mathcal{X}^{\expshort}}

\newcommand{\expres}{Y^{\expshort}}
\newcommand{\res}{Y}
\newcommand{\tre}{W}

\newcommand{\covariate}{Z}

\newcommand{\exptre}{W^{\expshort}}
\newcommand{\expcov}{Z^{\expshort}}

\newcommand{\obssam}{X^{\obsshort}}
\newcommand{\obssamspace}{\mathcal{X}^{\obsshort}}
\newcommand{\obsres}{Y^{\obsshort}}
\newcommand{\obstre}{W^{\obsshort}}
\newcommand{\obscov}{Z^{\obsshort}}
\newcommand{\prop}{p} 

\newcommand{\Zfun}{h}

\newcommand{\Zfuntilone}{\widetilde{h}}
\newcommand{\boundZfuntil}{B_{\Zfuntilone}}

\newcommand{\bZfunzero}{B_{\Zfun,0}}
\newcommand{\bZfunone}{B_{\Zfun,1}}
\newcommand{\bZfuntwo}{B_{\Zfun,2}}

\newcommand{\PopZfun}{H}
\newcommand{\PopZfprocess}{\widetilde{H}}

\newcommand{\samset}{\mathcal{J}}
\newcommand{\samsetnum}{|\mathcal{J}|}
\newcommand{\metric}{\rho}

\newcommand{\ind}{\mathbin{\perp\!\!\!\perp}}

\def\argmin{\mathrm{argmin}}

\def\R{{\mathbb R}}
\def\E{{\mathbb E}}
\def\bzero{{\boldsymbol 0}}

\def\IdMat{{\mathbf I}}
\def\cN{{\mathcal N}}

\def\eps{{\varepsilon}}

\def\KL{\mathsf{KL}}

\newcommand{\bigO}{\mathcal{O}}

\newcommand{\defn}{\coloneqq}
\newcommand{\revdef}{\eqqcolon}

\newcommand{\matsnorm}[2]{|\!|\!| #1 | \! | \!|_{{#2}}}

\newcommand{\vecnorm}[2]{\| #1\|_{#2}}

\newcommand{\opnorm}[1]{\ensuremath{\matsnorm{#1}{\tiny{\mbox{op}}}}}

\newcommand{\inprod}[2]{\ensuremath{\langle #1 , \, #2 \rangle}}

\newcommand{\Prob}{\ensuremath{{\mathbb{P}}}}
\renewcommand{\P}{\ensuremath{{\mathbb{P}}}}

\newcommand{\polyshort}{{C}}
\newcommand{\polyz}{{C(d,\gamma,B)}}

\newcommand{\poly}{C(B)}
\newcommand{\event}{\mathcal{E}}

\newcommand{\polyshortprime}{{C'}}
\newcommand{\polyprime}{C'(B)}
\newcommand{\polyprimez}{{C'(d,\gamma,B)}}

\newcommand{\expout}{Y^{\expshort}}
\newcommand{\truemean}{\tau^{\star}}
\newcommand{\estmean}{\widehat{\mu}}
\newcommand{\obserr}{\varepsilon}

\newcommand{\boundmean}{1}
\newcommand{\smallgap}{\Delta}

\newcommand{\bounderr}{1}

\newcommand{\meanconsta}{c_1}
\newcommand{\meanconstb}{c_2}
\newcommand{\meanvala}{L} 

\newcommand{\minmaxclass}{{\mathcal{M}}}

\newcommand{\obsout}{Y^{\obsshort}}

\newcommand{\myunder}[1]{\noindent{\underline{#1}}}

\newcommand{\simiid}{\overset{iid}{\sim}}

\usepackage{enumitem} 

\newenvironment{myassumption}[2]{%
  \begin{enumerate}[label=\textbf{\bf{(#1)}}]
  \item \label{#2}
}{%
  \end{enumerate}
}
\let\usepackage\manuscriptusepackage

\newcommand{\blind}{1} 
\newif\ifjrssb
\jrssbfalse

\begin{document}


\if1\blind
{
\title{\bf Cross-Validated Causal Inference: a Modern Method to Combine Experimental and Observational Data}

\setlength{\affilsep}{0pt}
    \author{\textbf{Xuelin Yang}\thanks{Imbens and Athey's work is supported by the Office of Naval Research under Grant N00014-17-1-2131; Jordan's work is supported by European Union under Grant ERC-2022-SYG-OCEAN-101071601. Code is available at \url{https://github.com/xyang23/cross_validated_causal}. Email correspondence: \texttt{xuelin@berkeley.edu}. }\\
  Department of Statistics, UC Berkeley \\
    \textbf{Licong Lin}\\
    Department of Statistics, UC Berkeley  \\ 
    \textbf{Susan Athey}\\
    Graduate School of Business, Stanford University\\ 
    \textbf{Michael I. Jordan}\\
    Department of Statistics, UC Berkeley; Inria Paris 
    \\ 
    \textbf{Guido W. Imbens}\\
    Graduate School of Business, Stanford University
    }

  \maketitle
} \fi
\if0\blind
{
  \bigskip
  \bigskip
  \bigskip
  \begin{center}

    {\LARGE\bf Cross-Validated Causal Inference: \\[0.6em]
    a Modern Method to Combine Experimental and Observational Data}

  \end{center}
  \medskip
} \fi

\vspace{-5mm}
\begin{abstract}
We develop new methods to integrate experimental and observational data in causal inference. While randomized controlled trials offer strong internal validity, they are often costly and therefore limited in sample size. Observational data, though cheaper and often with larger sample sizes, are prone to biases due to unmeasured confounders. To harness their complementary strengths, we propose a systematic framework that formulates causal estimation as an empirical risk minimization (ERM) problem. A full model containing the causal parameter is obtained by minimizing a weighted combination of experimental and observational losses---capturing the causal parameter's validity and the full model's predictive fit, respectively. The weight is chosen through cross-validation on the causal parameter across experimental folds. Our experiments on real and synthetic data show the efficacy and reliability of our method. We also provide theoretical non-asymptotic error bounds.
\end{abstract}

\noindent%
\vfill

\newpage

\section{Introduction}\label{sec:intro}

We focus on the problem of estimating the average treatment effect (ATE) in causal inference. Over the past decades, a wide range of statistical methods have been developed to draw causal conclusions from either experimental or observational studies. Experimental data, collected from randomized controlled trials (RCTs), offer high internal validity, so we are able to obtain unbiased or consistent treatment effect estimates. However, such data can be costly to obtain. In contrast, observational data are often cheaper, but their internal validity is suspect.
Specifically, ATE estimates based on observational data, assuming unconfoundedness, may suffer from biases due to unobserved confounders. 

In this paper we consider the combination of them, 
with the goal of producing robust (to the presence of unobserved confounders) and precise (by including observational data) causal conclusions. We use the experimental data to tell us how much we want to incorporate each data source. This is conducted by cross-validating combined estimates against estimates computed on splits of experimental data.

To illustrate the basic ideas, consider 
a setting with no covariates. Suppose we are interested in the experimental population.
We have an experimental sample where we observe both treated and control units, and an observational sample where we observe only control units, based on the widely used LaLonde data \citep{lalonde1986evaluating, dehejia1999causal}. Because in this setting there is no question about estimating the average outcome for the treated, for which we only have the experimental data, the question is how to estimate the average control outcome for the experimental population, $\mathbb{E}[Y_i\expm(C)]$. 
The average control outcome in the experimental sample, $\overline{Y}_C\expm$, is unbiased for this expectation (but possibly imprecise due to limited data size). 
The average of the control outcome in the observational sample, $\overline{Y}_C\obs$, may be biased for the experimental population’s average control outcome.
We consider a weighted combined estimate of them:
\begin{align} 
    \widehat\theta(\lambda) 
    = (1 - \lambda) \overline{Y}_C\expm + \lambda \overline{Y}_C\obs, \qquad \lambda\in[0,1].
\end{align}
What properties would we like the tuning parameter $\lambda$ to have?
If the experimental sample is large, then even if the bias in the observational sample is very small, as long as there is some bias  we would like $\lambda$ to be close to zero. If on the other hand the bias in the observational sample is negligible, then we would like to choose $\lambda$ close to one, to put more weight on the part with a smaller variance.  In other words, we would like to shrink our experimental estimate towards the observational estimate, but do so in a data-adaptive fashion, that is, with a data-driven $\lambda$. 
We do not know the magnitude of the bias, and, unlike existing work, do not wish to impose assumptions on it.
Therefore, we select $\lambda$ through cross-validation: we split the experimental data into $K$ folds, compute a held-out estimate (the sample mean) on the held-out fold, and compare it with the combined estimate on $K-1$ folds and all observational data. The tuning parameter $\lambda$ is selected to minimize the cross-validation objective over $K$ different held-out folds. Formally, we have
\[
\widehat\lambda= \arg\min_{\lambda\in[0,1]}  \underbrace{ \frac{1}{K} \sum_{k = 1}^{K}  \Big( \overbrace{\underbrace{\overline{Y}_{C,B_k}\expm}_{\text{held-out fold}}}^{\text{held-out estimate}} -  \Big( \overbrace{(1 - \lambda) \underbrace{\overline{Y}\expm_{C,-B_k}}_{K-1 \text{ folds} }~+ ~\lambda \overline{Y}_C\obs\Big) }^{\text{combined estimate}} \Big)^2}_{\CV(\lambda), \text{ the cross-validation objective}},
\]
where the subscripts $\{B_k, -B_{k}\}_{ k\in[K]}$ denote the complementary subsets.
If the observational estimate is very biased, then any nonzero $\lambda$ would result in a very different combined estimate than the held-out estimate. So cross-validation would select a $\lambda$ that is close to 0. If the bias is small, cross-validation would favor larger $\lambda$ because the combined estimate would be more stable across different held-out folds. 
In the paper we extend this to more general models involving covariates, where we may use a covariate-adjusted model for the combined estimate and any consistent experimental estimator of our choice for the held-out estimate.

\begin{figure}[tbp]
\ifjrssb
\vspace{1mm}
\fi
    \centering
        \subfloat[No-covariate setting.]{
        \includegraphics[width=0.45\textwidth]{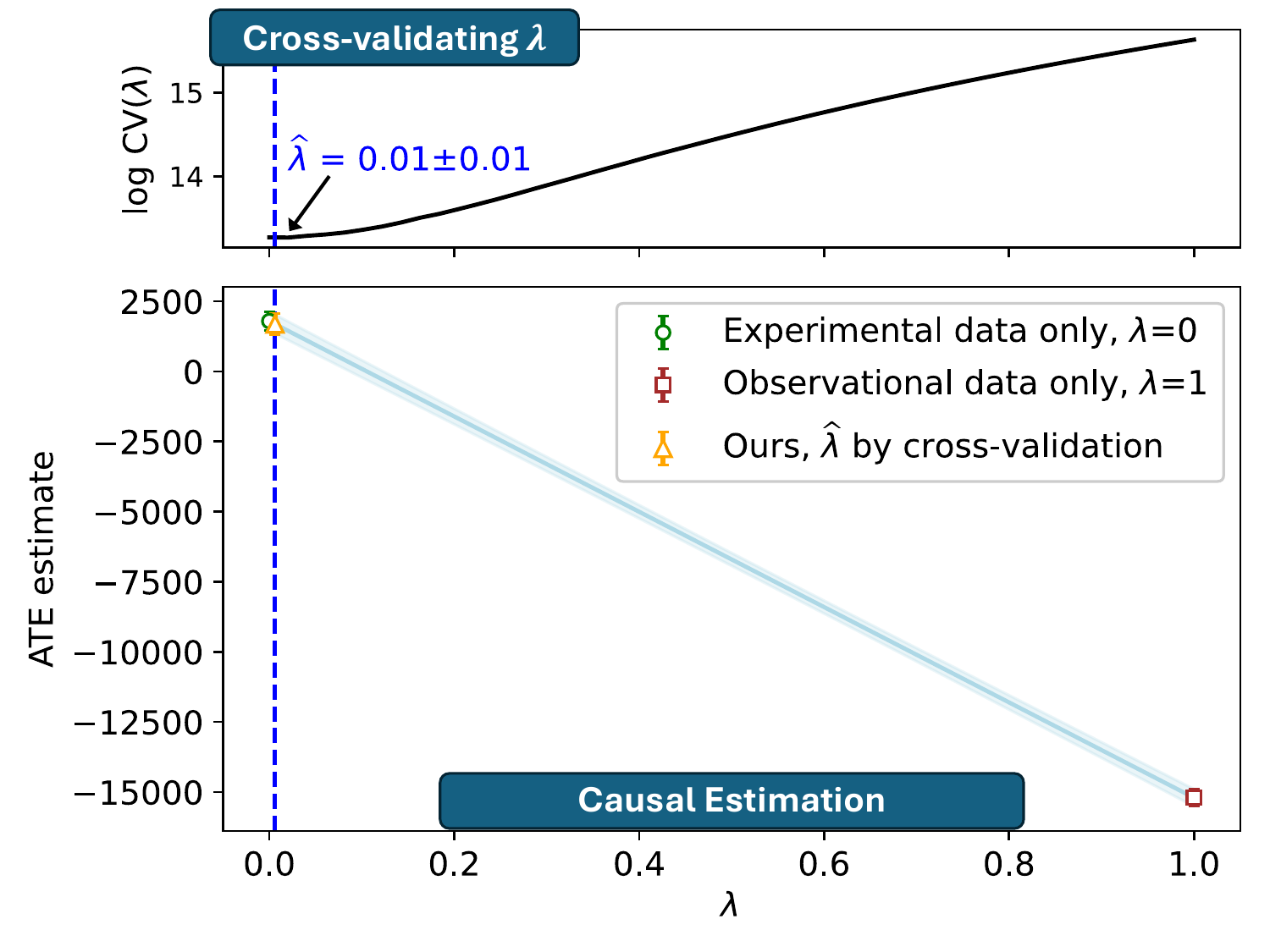}
        \label{fig:lalonde_intro_mean}
        }
    \subfloat[Covariate-adjusted linear setting.]{
    \includegraphics[width=0.45\textwidth]{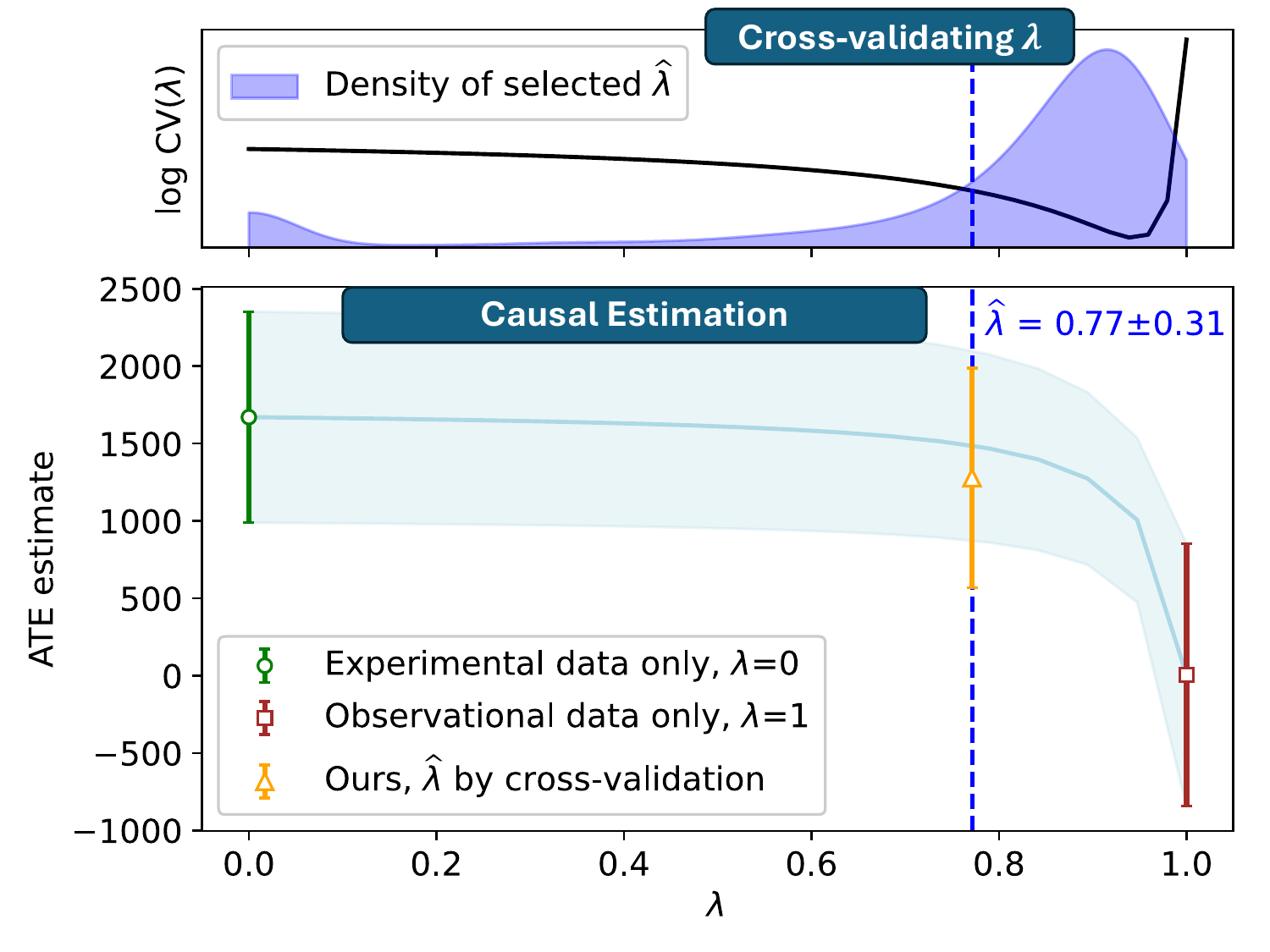}
        \label{fig:lalonde_intro_col8}
        }
    \caption{Cross-Validated Causal Inference (CVCI) using $\lambda$. Top panels: selection of $\lambda$ via the cross-validation objective $\CV(\lambda)$. The curve shows the average of $\CV(\lambda)$ over 5000 random splits of experimental data, and the blue dashed line shows the average selected $\widehat{\lambda}$. Bottom panels: treatment effect estimates for different $\lambda$. The top panel in (b) shows an estimated density for $\widehat\lambda$ across different ways of splitting the data. The top panel in (a) omits the density as $\widehat\lambda$ is almost always 0.
     We supplement the detailed setup and a discussion in Section~\ref{sec:setup_intro_figure}.
     } 
    \label{fig:lalonde_intro}
\end{figure}

In Figure~\ref{fig:lalonde_intro}, we illustrate this process based on the LaLonde data \citep{lalonde1986evaluating, dehejia1999causal}. The treatment is participation in a job training program, and the outcome is real income after the treatment. 
In the bottom two panels we present two sets of three estimates of the ATE. First, in both panels, results based on the experimental data alone (corresponding to $\lambda=0$). Second, again in both panels,
results based on the observational data alone (corresponding to $\lambda=1$). Both are intended to set the stage for our
preferred results based on the cross-validated $\widehat\lambda$. The cross-validation is based on five-fold splits, leading to a unique $\widehat\lambda$. We repeat this many times to get a distribution of selected $\widehat\lambda.$
In the case without covariates, we find that  the selected $\widehat\lambda$ is always close to or exactly equal to 0, corresponding to the experimental estimates. The cross-validation makes clear that the data can tell us that the observational data are of little value in this case. 
For a covariate-adjusted version of the observational data estimator, the cross-validated $\widehat\lambda$ is  much closer to 1, with the average value for $\widehat\lambda$ over many choices of five folds equal to 0.77. Here the data imply that  the observational data are valuable. 
The combination of the two sets of results shows that in this case our proposed method can detect when the observational sample is valuable, and when it is not, in a fully data-driven way. In the paper we will show when we can improve upon the experimental-only estimate and when we cannot (Section~\ref{sec:experiments}).

Our contributions are three-fold:
First, we introduce a novel method to systematically combine experimental and observational data. 
The methodological advantages include: $(i)$ we do not require additional model specifications or identification assumptions; $(ii)$ the method allows for the setting  where observational and experimental samples have different covariates; $(iii)$ the method allows the treatment to have a different effect in two populations.
Second, we conduct experiments on both synthetic and real data to show the effectiveness of our method. For synthetic studies, we address the most common cases (no-covariate and covariate-adjusted linear settings). We also present a case study on the LaLonde-Dehejia-Wahba dataset \citep{lalonde1986evaluating, dehejia1999causal}.
Third, we develop supporting non-asymptotic theories. Under regularity conditions, we show that our method achieves an $\bigO(1/\nexp)$ error rate regardless of the level of bias in the observational data, where $\nexp$ is the experimental sample size (Corollary~\ref{cor:robustness_atefun_estpar}). 
Moreover, in the no-covariate setting,
we characterize the improvement of CVCI  on both data sources (Proposition~\ref{prop:cv_mean_improvement}), and show that the $\bigO(1/\nexp)$-rate is minimax optimal over a class of robust estimators that combine them (Theorem~\ref{thm:mean_estimation_minimax}).

\section{Identification}\label{sec:problem_formulation}

Suppose we have access to two datasets $X\expm$ and $X\obs$.  The former is comprised of $\nexp$ experimental samples, $\expsam_i=(\expres_i,\exptre_i,\expcov_i)\in\expsamspace$, with the observed outcome $\expres_i\in\R$, binary treatment $\exptre_i\in\{0,1\}$, and covariate vector $\expcov_i\in\R^{\dexp}$. The latter consists of $\nobs$ observational samples, $\obssam_i=(\obsres_i,\obstre_i,\obscov_i)\in\obssamspace$, defined analogously. 
\ifjrssb
\else

\fi
We adopt the classical potential outcome framework~\citep{rubin1974estimating},
where the potential outcomes are denoted by $(\res^{\sf s}_i(1),\res^{\sf s}_i(0))$, and $\res^{\sf s}_i  = \res^{\sf s}_i(\tre^{\sf s}_i)$ for $\sf s\in\{\expshort,\obsshort\}$.
For the experimental data,
we make standard assumptions for the sake of identifying and consistently estimating the treatment effect: (1) $(\expres_i(1),\expres_i(0),\exptre_i,\expcov_i)\simiid\distexp$ for some distribution $\distexp$; (2) there is no unobserved confounder,  {\it i.e.}, $(\expres_i(1),\expres_i(0))\ind \exptre_i|\expcov_i$; (3) the overlap condition is satisfied, {\it i.e.}, the propensity score $\P(\exptre_i=1|\expcov_i)$ lies in the open interval $(0,1)$. These are typically enforced directly by the researcher in an experimental setting, since researchers control the assignment to treatment. 
We want to estimate the ATE on the population associated with the experimental data:
 \begin{align*}
\trueate \defn \E[ \expres(1) -\expres(0)].
 \end{align*}

For the observational data, we impose no distributional assumptions. In particular, we permit unmeasured confounders and selection bias. We do not assume that the two data sources share the same covariate distributions, thus allowing for covariate shift. We do not require their outcome models to be the same, permitting label shift.
Additionally, we allow the observational data to be non-independent and non-identically distributed (non-i.i.d.), as we take them as a whole and only split experimental data. 
We note that these are robustness guarantees. If the observational data are highly biased, the final estimator will effectively ignore them. To benefit from incorporating observational data, we need them to meet a certain quality standard, though the quality threshold here is more flexible than existing methods. 
We will elaborate on this aspect in the experiments (Section~\ref{sec:experiments}).


\section{Cross-Validated Causal Inference} \label{sec:method}

 In this section we present CVCI in the no-covariate, covariate-adjusted linear, and general parametric settings. Each of them follows the same procedure. Suppose we have a full model $\theta$ predicting some outcome using covariates (if any) and the treatment from the observational data. Denote $\beta \defn \beta(\theta)$ 
as its causal parameter used directly to estimate the treatment effect. 


\subsection{Case I: No-covariate setting}\label{sec:no-covariate}

We start with the standard no-covariate setting where only response and treatment are observed in both sources. The full model and the causal parameter are both scalars $\beta(\theta)=\beta\in\R$. For a random experimental sample $\sam=(\res, \tre) \sim P\expm$, the ATE
can be estimated by the difference in means
\[
\frac{1}{\sum_i \mathbbm{1}\{\exptre_i=1\}} \sum_{\exptre_i=1} \expres_i - \frac{1}{\sum_i \mathbbm{1}\{\exptre_i=0\}} \sum_{\exptre_i=0} \expres_i  .
\]
Here we need to estimate both the control and the treated means. For simplicity of presentation we illustrate with the control mean. The treated mean can be estimated separately with the same procedure. 
The procedure has two steps:
\paragraph{Step 1.}
Select a tuning parameter $\widehat\lambda$ by cross-validation. We go over possible candidates of $\lambda$ and find the one minimizing the cross-validation objective.  For each candidate $\lambda$, we split the experimental data into $K$ folds and compare the held-out estimate (on the held-out fold) with the combined estimate (on $K-1$ experimental folds and all observational data). That is,
{\small
\begin{align}
   \widehat\lambda &= \arg\min_{\lambda\in[0,1]} \CV(\lambda) 
   =\arg\min_{\lambda\in[0,1]}   \frac{1}{K} \sum_{k = 1}^{K}  \Big( \overbrace{\underbrace{\overline{Y}_{C,B_k}\expm}_{\text{held-out fold}}}^{\text{held-out estimate}} -  \Big( \overbrace{(1 - \lambda) \underbrace{\overline{Y}\expm_{C,-B_k}}_{K-1 \text{ folds} }+ \lambda \overline{Y}_C\obs\Big) }^{\text{combined estimate}} \Big)^2.
\end{align}}
The subscripts $\{B_k, -B_{k}\}_{ k\in[K]}$ denote the complementary subsets, and the overline denotes the sample mean.
\paragraph{Step 2. }
Plug in $\widehat\lambda$ to obtain a final estimate $\beta(\widehat\theta(\widehat\lambda)) = \widehat\theta(\widehat\lambda)$ on all data
\begin{align}
 \widehat\theta(\lambda) &= \arg\min_\theta   (1 - \lambda) (\overline{Y}\expm - \theta)^2 + \lambda  (\overline{Y}\obs - \theta)^2 
    = (1 - \lambda) \overline{Y}\expm + \lambda \overline{Y}\obs. \label{eq:mean_close_form}
\end{align}
See Section~\ref{proof:mean_close_form} for its derivation. When we estimate the control (or treated) mean separately, we may select different $\widehat \lambda$ values for different groups. An alternative way is to estimate the treatment effect (difference-in-mean) as a whole. The choice depends on the practical use case. When the observational treated and control groups have different levels of selection bias, separate estimation may be preferable as it allows each group to receive a different weight. When the biases in the two means may largely cancel, the alternative may be more effective.

\subsection{Case II: Linear setting}
\label{sec:linear_setting}

Suppose we have $\theta$ as a linear model for observational data that predicts the outcome given treatment and covariates, where its first entry $\beta:=\beta(\theta)\in\R$ corresponds to the estimated treatment effect. That is, the full model is $\theta\in\R^{\dobs+2}$ for $\dobs\ge0$ covariates, and 
$\E(Y_i^\obsshort\mid W^\obsshort_i,Z^\obsshort_i)=\beta  W^\obsshort_i+\theta_Z ^\top Z^\obsshort_i , ~  \theta^\top=(\beta \quad \theta_Z^\top).$ 
Compared to the no-covariate setting, we make two changes. First, for the held-out estimate, we replace the sample mean with a consistent experimental estimator. Second, for the combined estimate, rather than computing a weighted average, we fit the outcome model on the combined data using a combined loss, and extract its causal parameter corresponding to the treatment effect.
The procedure is formalized into two steps:
\paragraph{Step 1. } 
Select a tuning parameter $\widehat\lambda$ by cross-validation. To ensure the consistency of the combined estimate (the causal parameter), we select $\lambda$ to minimize its discrepancy from a held-out experimental estimate $\estate$:{\small
\begin{align} 
   \widehat\lambda &= \arg\min_{\lambda\in[0,1]} \CV(\lambda)  
   =\arg\min_{\lambda\in[0,1]}   \frac{1}{K} \sum_{k = 1}^{K}  \Big( \overbrace{\estate(\underbrace{X\expm_{B_k}}_{\text{held-out fold}})}^{\text{held-out estimate}} -  \overbrace{\beta(\widehat\theta(\lambda; \underbrace{X\expm_{-B_k}}_{K-1 \text{ folds}} , X\obs))}^{\text{combined estimate}} \Big)^2.
\end{align}}
This favors values of $\lambda$ that yield consistent and more stable combined estimates across held-out splits.
We use the standard squared loss as it is strongly convex in $\beta$ (which facilitates the theoretical analysis) and admits a desirable additive structure (formalized as Lemma~\ref{lem:additive_mse}) for common estimators
defined later in Eq.~\eqref{eq:m-estimate_all_avg}.

To compute the held-out estimate, we use a consistent ATE estimator of our choice, fitted on the held-out experimental sample. This could be the simple difference in means,
$
    \estate(X\expm_{B_k})= \overline{Y}_{B_k, T}^\expshort-\overline{Y}^\expshort_{B_k,C},
$
or a more complex estimator that involves covariate adjustment,
for example, a plug-in estimator with a linear outcome model $\theta\expm$ with its first entry being the treatment coefficient, 
\begin{align*}
    \estate(X\expm_{B_k})= e_1^\top \Big(\arg\min_{\theta\expm} \frac{1}{|B_k|}\sum_{i\in B_k} \Big(\res_i\expm -  \Big( \tre_i\expm ~ {\covariate_i\expm}^\top \Big) \theta\expm \Big)^2 \Big),
\end{align*}
where $e_1^\top = (1 ~ 0 \cdots 0 )$.
Other examples of experimental estimators include:
{\small
\begin{align}
    &\estate(X\expm_{B_k}) = \frac{1}{|{B_k}|} \sum_{i \in {B_k}} \phi (X_i\expm),  \label{eq:m-estimate_all_avg}\\
    \phi (X\expm_i)
    &= \begin{cases}
        \overline{Y}^\expshort_{B_k, T}-\overline{Y}^\expshort_{B_k, C}=\frac{|B_k|}{| \{\tre\expm_j = \tre\expm_i\}_{j \in B_k}| }{(2 \tre\expm_i - 1)} \res\expm_i,  &\text{(difference-in-means)} \\
        \widehat\mu(1, \covariate\expm_i) - \widehat\mu(0, \covariate\expm_i),  &\text{(plug-in estimator)}\\
        \frac{\tre\expm_i}{\widehat\pi(\covariate\expm_i)} (\res\expm_i - \widehat\mu(1, \covariate\expm_i)) + \widehat\mu(1, \covariate\expm_i) -\\ \quad\quad \left ( \frac{1-\tre\expm_i}{1-\widehat\pi(\covariate\expm_i)} (\res\expm_i - \widehat\mu(0, \covariate\expm_i) ) + \widehat\mu(0, \covariate\expm_i) \right ),  &\text{(AIPW estimator)} \label{eq:m-estimate_all}
    \end{cases}
\end{align}}
with an outcome model $\widehat\mu:\{0, 1\} ~\times~ \R^{\dexp} \rightarrow \R $ and propensity score $\widehat\pi: \R^{\dexp} \rightarrow (0, 1) $.
When propensity scores are known, one may set $\widehat\pi=\pi$. 

For the combined estimate, we want to ensure the validity of the causal parameter on the experimental data, but meanwhile leveraging the additional sample size when fitting on the potentially confounded observational data.
To encapsulate this dynamic, we design it to be a minimization problem of a combination of losses
\begin{align}\label{eq:linear_loss_combination}
    \widehat{\theta} (\lambda; X_{-B_k}\expm, X\obs) = \arg \min_{\theta} \Big\{ (1 - \lambda) \underbrace{L\expm (\beta (\theta); X_{-B_k}\expm)}_{\text{causal parameter}} + \lambda \underbrace{L\obs (\theta; X\obs)}_{\text{full model}}  \Big\}.
 \end{align}  
The experimental loss evaluates the validity of the causal parameter $\beta(\theta)\in\R$. We again use a squared loss to compare $\beta$ with the experimental estimate $\estate(X_{-B_k}\expm)$:
\[
L\expm(\beta(\theta); X\expm_{-B_k}) \defn \Big( \beta - \estate( X\expm_{-B_k})  \Big)^2.
\]
The observational loss aims at the predictive fit of the full model on observational data. We minimize a squared predictive loss where we compare the observed outcome with the predicted outcome over observational samples:
\[
L^\obsshort(\theta; X^\obsshort) \defn 
\frac{1}{N\obs}\sum_{i=1}^{N\obs} \Big(\res_i\obs -   \Big( \tre_i\obs\quad {\covariate_i\obs}^\top \Big) \theta \Big)^2
.
\]
We now put both losses together for Eq.~\eqref{eq:linear_loss_combination}. 
Denote $\tre\obs, \covariate\obs, \res\obs$ as the respective matrices containing all observational samples, where each column corresponds to one sample. Append a 1 to each $\covariate_i\obs$ to include an intercept term. 
Solving the minimization problem in Eq.~\eqref{eq:linear_loss_combination} is equivalent to solving the following linear system 
\begin{align}\label{eq:linear_close_form}
\left( (1-\lambda) e_1 e_1^\top + \frac{\lambda}{N\obs} \begin{bmatrix} \tre\obs\\ \covariate\obs \end{bmatrix} \begin{bmatrix} \tre\obs\\ \covariate\obs \end{bmatrix}^\top \right)  \theta  = (1-\lambda) \estate e_1 + \frac{\lambda}{N\obs} \begin{bmatrix} \tre\obs\\ \covariate\obs \end{bmatrix} \res\obs,
\end{align}
where $e_1 = (1 ~ 0 \cdots 0 )^\top, $ and $\estate=\estate(X_{-B_k})$ is an experimental estimate.
Intuitively, the first term on both sides regularizes the treatment coefficient toward the experimental estimate, while the second term on both sides fits the full model to the observational data. The derivation is provided in Section~\ref{proof:linear_close_form}.

\paragraph{Step 2. }
Plug in the selected $\widehat \lambda $ to obtain the final estimate $\beta(\widehat\theta(\widehat \lambda ; X\expm, X\obs))$. We solve the same minimization problem as in Eq.~\eqref{eq:linear_loss_combination} but with all data available.

 \subsection{General parametric setting}
 Suppose we have an outcome model $\theta\in\Theta$, and its causal parameter $\beta\defn\beta(\theta)\in\R$ is a linear function of parameters in $\theta$. For example, this can be a partial linear model \citep{robinson1988root}  
 $
     \E (Y_i\obs \mid W_i\obs , Z_i\obs ) = \beta W_i\obs + f_{\theta_Z}  (Z_i\obs), ~ \theta^\top=(\beta \quad \theta_Z^\top), 
$
 where $f_{\theta_Z}$ is a nonlinear model such as a neural network. Another example is letting $\theta$ be concatenated parameters of various nuisance models.
Similar to the linear setting, we solve an empirical risk minimization (ERM) problem to estimate $\theta$:
given $\lambda$, we minimize a combination of experimental and observational losses
    \begin{align}\label{eq:general_combined_losses}
    \widehat{\theta} (\lambda) \defn \widehat{\theta} (\lambda; X\expm, X\obs) = \arg \min_{\theta} \Big\{ (1 - \lambda) \underbrace{L\expm (\beta (\theta); X\expm)}_{\text{causal parameter}} + \lambda \underbrace{L\obs (\theta; X\obs)}_{\text{full model}}  \Big\}.
  \end{align}
    We have provided closed-form solutions for the most common cases (no-covariate and linear settings, under squared losses). For other cases, we can employ gradient-based, (quasi-)Newton, or other optimization techniques suited to the objective function. For example, we can use gradient descent when $\theta$ parametrizes a neural network.

We now discuss the choices of losses. The experimental loss quantifies the validity of the causal parameter on experimental data. As it evaluates a scalar $\beta$, the squared error is a natural choice. This loss serves as a benchmark for consistent causal estimation. The observational loss evaluates how well the full model (including its causal parameter) explains the observational dataset under its data-generating process. 
To facilitate the theoretical analysis, we assume strong convexity and smoothness conditions ({\it i.e.}, three times differentiable with bounded second and third derivatives), as formalized later in Assumption~\ref{ass:obs_ate}. This class includes squared loss, L$_2$ regularization (ridge loss), and L$_p$ loss ({\it i.e.}, $|y-y'|^p/p, p\ge3$). On the other hand, this class excludes L$_1$ regularization (LASSO, due to the non-differentiability at zero), elastic net (a combination of L1 and L2 regularization), and Huber loss (because it is not twice differentiable at the threshold). 

We now present the two-step procedure for the general case:

\paragraph{Step 1. } 
Select $\widehat{\lambda}$ by cross-validating the causal parameter:
   \begin{align} \widehat{\lambda} &= {\arg\min}_{\lambda\in[0,1]} \CV(\lambda; X\expm, X\obs)\nonumber\\
   &= \arg\min_{\lambda\in[0,1]}  \frac{1}{K} \sum_{k = 1}^{K} L\expm(\beta(\widehat{\theta} (\lambda; X_{-B_k}\expm, X\obs)) ; X_{B_k}\expm).\label{eq:cv_objective_general}
    \end{align}     
      $\CV$ quantifies how well the estimated treatment effect aligns with experimental evidence. 
      Under squared experimental loss,
    \begin{align}
\widehat{\lambda} &= \arg\min_{\lambda\in[0,1]}   \frac{1}{K} \sum_{k = 1}^{K}  \Big( \overbrace{\estate(X\expm_{B_k})}^{\text{held-out estimate (Eq.~\eqref{eq:m-estimate_all})} } -  \overbrace{\beta(\widehat\theta(\lambda; X\expm_{-B_k}, X\obs))}^{\text{combined estimate  (Eq.~\eqref{eq:general_combined_losses})}} \Big)^2.
    \end{align}  
    When the causal estimate from the combined data is well aligned (consistent) with the experimental evidence, and more stable (with smaller variance because of observational data) across held-out splits, cross-validation favors those models. Otherwise, it reverts toward the experimental data to control for potential bias. 
\paragraph{Step 2.} 
Fit the model with $\widehat\lambda$ on all data. The final estimate is $\beta(\widehat{\theta} (\widehat{\lambda}; X\expm, X\obs)). $



\section{Experiments}\label{sec:experiments}



\begin{figure}[H]
    \ifjrssb
    \vspace{0.5mm}
    \fi
    \centering
    \includegraphics[width=\textwidth]{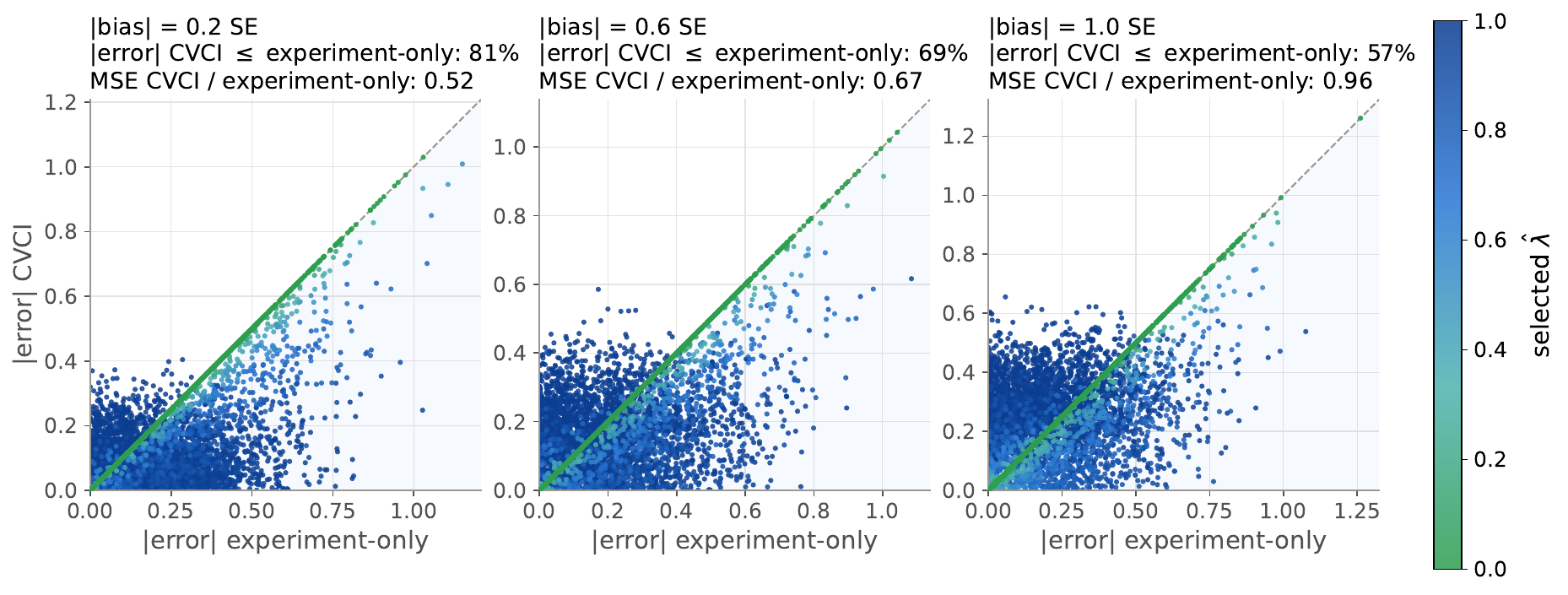}
    \caption{Illustration of comparison with the experiment-only estimate. Each point is one simulation, where points below the dashed diagonal are runs where CVCI has the lower absolute error. CVCI has a lower or equal error in $81\%$, $69\%$ and $57\%$ of the runs, when bias is $0.2$, $0.6$ and $1.0$ experimental standard errors, respectively. Linear setting design that corresponds to Column~5 of Figure~\ref{fig:bias_level_map}.} 
    \label{fig:run_level_wins}
\end{figure}

In this section, we will first compare CVCI with common methods under various scenarios of observational bias that are likely to happen in practice (Figure~\ref{fig:bias_level_map}). We will then provide a taxonomy of both the no-covariate setting (Section~\ref{sec:nocov_setting}) and the linear setting (Section~\ref{sec:experiments_linear_setting}), with a case study of the LaLonde dataset (Section~\ref{sec:lalonde}). To start with, we demonstrate in Figure~\ref{fig:run_level_wins} that CVCI improves on the experimental-only estimate when the observational bias is small.


\subsection{Comparison overview}\label{sec:experiments_overview}
\subsubsection{Settings}
We compare CVCI with the two single-sourced estimates (experimental-only and observational-only), three common test-based pooling baselines that decide whether to use the observational data based on parametric tests (the pre-test, the soft threshold, and the adaptive soft threshold of \citet{armstrong2025adapting}), and three fixed-weight combinations (the error-prone estimator of \citet{yang2020combining}, precision weighting, and pooled OLS with an indicator for data source). Each simulation run draws a new dataset and one random split of its experimental sample.

We incorporate different sources of observational bias. These include: constant bias, heavy-tailed random bias, heterogeneous bias depending on the covariate, a misspecified observational model (a linear model on a quadratic ground truth), omitted covariates, confounded assignment, and unmeasured confounding. 

For the semi-synthetic designs, we build on the Infant Health and Development Program (IHDP) study \citep{hill2011bayesian} ($747$ children, $25$ covariates) and the Tennessee Student/Teacher Achievement Ratio (STAR) class-size experiment \citep{krueger1999experimental} ($3730$ students, $7$ covariates).
We induce confounding of strength $\gamma$, either by a latent variable that shifts both the treatment probability and the outcome residual, or, similar to the setting of \citet{yang2020combining}, by assigning treatment through covariates that are then omitted from the observational data.
Mathematical formulations of the baselines and designs are provided in Section~\ref{sec:implementation_overview}.

\subsubsection{Results}
\begin{figure}[htp]
    \ifjrssb
    \vspace{0.5mm}
    \fi
    \centering
    \includegraphics[width=\textwidth, height=0.74\textheight, keepaspectratio]{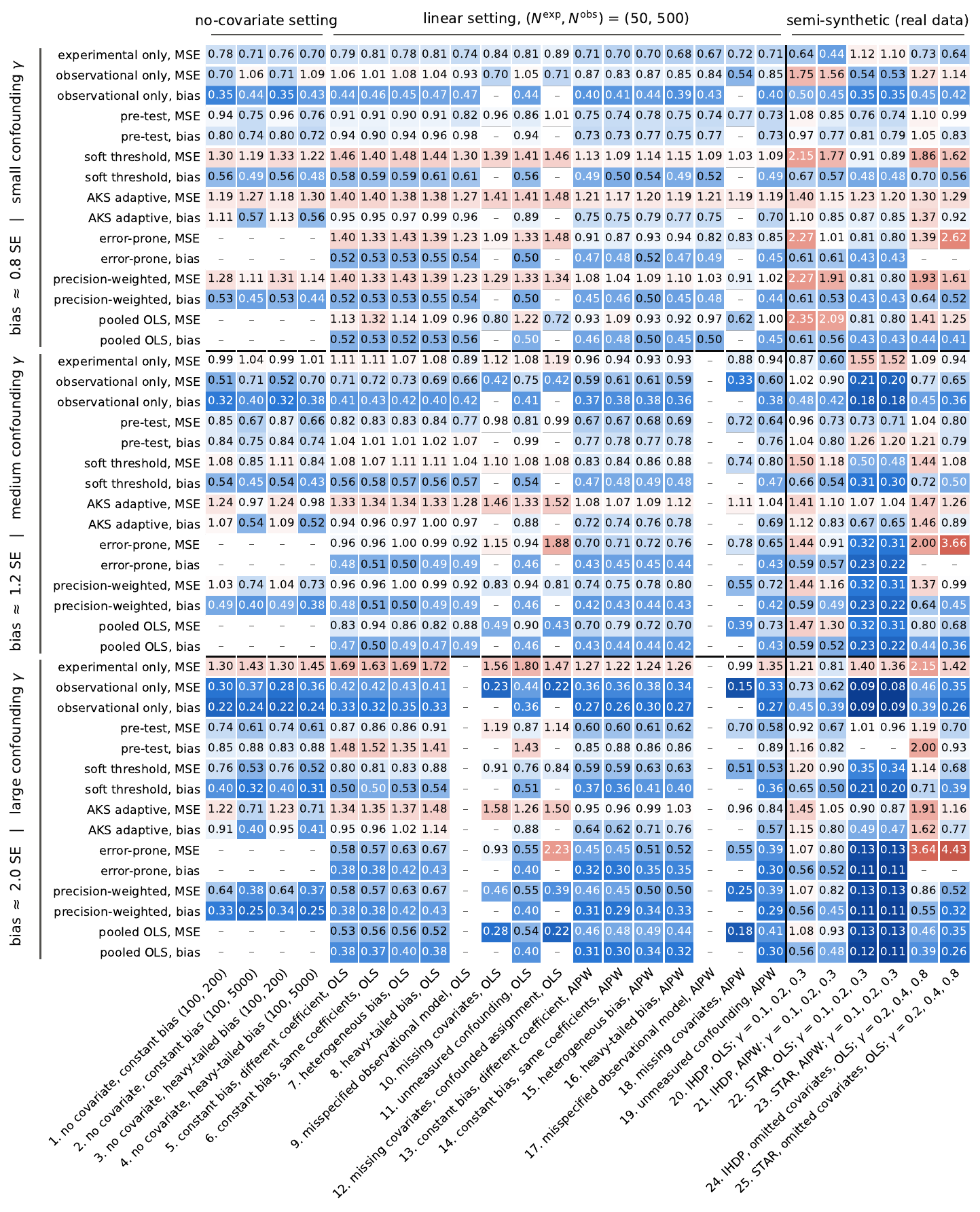}
    \caption{MSE and absolute bias ratios of CVCI to other baselines. Each row is one estimator, and each column is one design. Values below $1$ (favor CVCI) are in blue, values above $1$ are in red. 
    Columns are grouped into no-covariate setting, linear setting, and semi-synthetic linear setting (with labels at the top). Rows are grouped into three blocks of observational bias (labels on the left). 
    CVCI has the smaller bias in most settings, and the smaller MSE in many of them, in particular when the observational bias is at least about one experimental standard error. For column labels at the bottom, numbers in parentheses ``(100, 200)'' indicate sample sizes; ``OLS'' and ``AIPW'' indicate the experimental estimator. Detailed settings are provided in Section~\ref{sec:implementation_overview}.}
    \label{fig:bias_level_map}
\end{figure}

No estimator performs universally well across settings. Figure~\ref{fig:bias_level_map} includes observational bias levels that are so large that no method should be expected to improve on the experiment alone. Relative to the experimental-only estimate, CVCI lowers the mean squared error (MSE) by $10\%$ to $35\%$ when the bias level is small (about $0.8$ experimental standard errors), matches it when the bias level is intermediate (about $1.2$), and loses by a factor of $1.2$ to $1.8$ when the bias level is high (about $2$). 
We will elaborate on this aspect in Section~\ref{sec:linear_results}.

The CVCI estimate has a clear advantage in its bias: in $93\%$ of the cells, it is less biased than the baselines, by a factor of $1.5$ to $5$ against the rules that keep a fixed or capped weight on the observational data (precision weighting, pooled OLS, the soft threshold, and the error-prone estimator outside the confounded-assignment designs where its premise holds). 
We now turn to MSE: the adaptive soft threshold is designed to minimize worst-case MSE, and its MSE is smaller, as it should be.
CVCI instead keeps consistency while aiming to reduce variance---it is less biased in $51$ of $63$ cells, typically by $10\%$ to $25\%$. At the high observational bias level, CVCI has a lower MSE than every other combination rule, often by half or more, with the exception of the error-prone estimator when its premise holds.
At the intermediate bias level, CVCI matches or improves on every combination rule, where the pre-test is at its worst as it rejects only part of the time.


\subsection{No-covariate setting}\label{sec:nocov_setting}
\subsubsection{Settings}  

We now provide a taxonomy of the no-covariate setting. Without loss of generality, we take our samples to be $Y\expm_1, \ldots, Y\expm_{N\expm} \simiid \cN(\tau^\star, {\sigma}^2)$ and $Y\obs_1, \ldots, Y\obs_{N\obs} \simiid \cN(\tau^\star + \varepsilon, {\sigma}^2)$. 
We compare with single-sourced sample means and a baseline based on a Gaussian parametric t-test. 
For implementation details see Section~\ref{sec:implementation_mean}.

\subsubsection{Results}

\begin{figure}[ht]
    \centering
    \ifjrssb
    \vspace{0.5mm}
    \fi
    \captionsetup[sub]{justification=centering}
    \subfloat[$N\expm=100$, $N\obs=5000$, $\sigma^2=1$.]{%
        \includegraphics[width=0.48\textwidth]{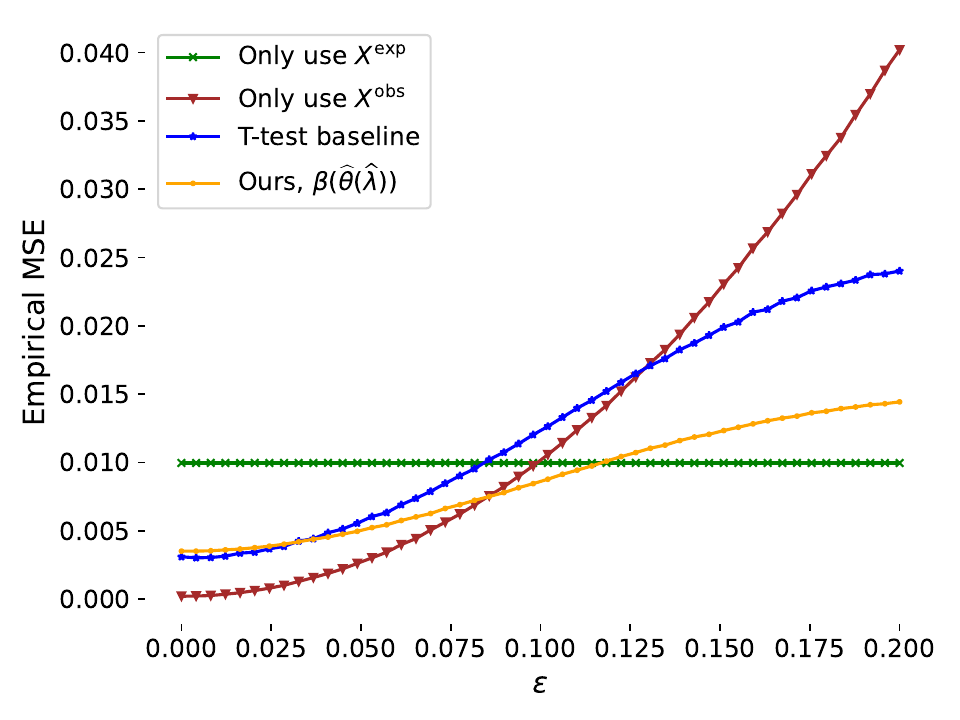}%
        \label{fig:a_mean_mse_eps}%
        }\hfill
    \subfloat[$N\expm=100$, $N\obs=200$, $\sigma^2=1$.]{%
        \includegraphics[width=0.48\textwidth]{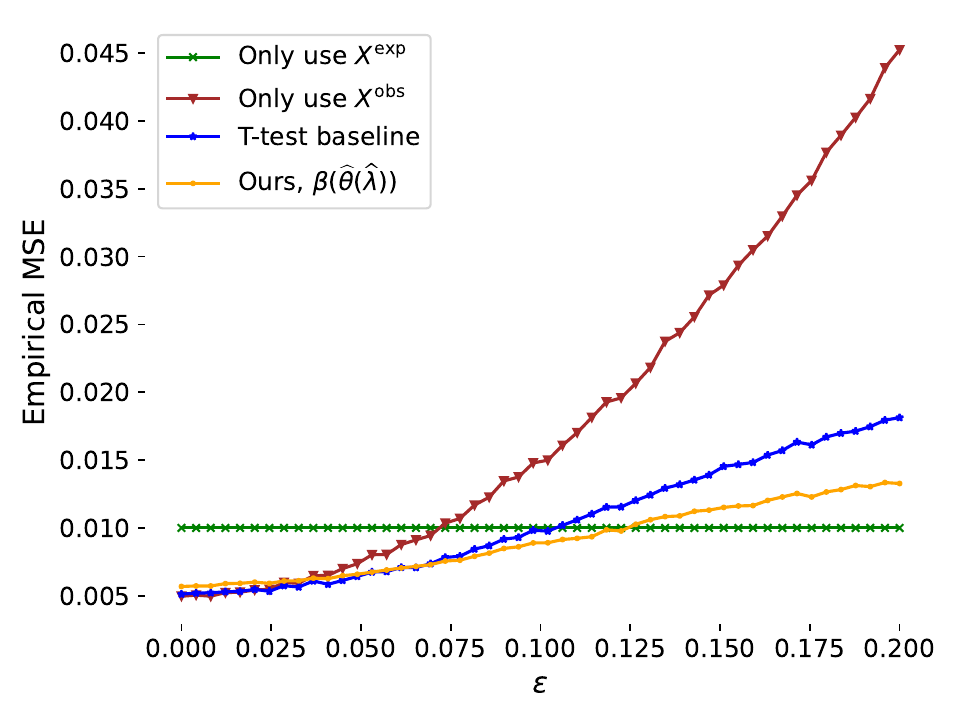}%
        \label{fig:b_mean_mse_eps}%
        }\\
    \subfloat[Same settings as (a). Inset: Zoom in.]{%
        \includegraphics[width=0.48\textwidth]{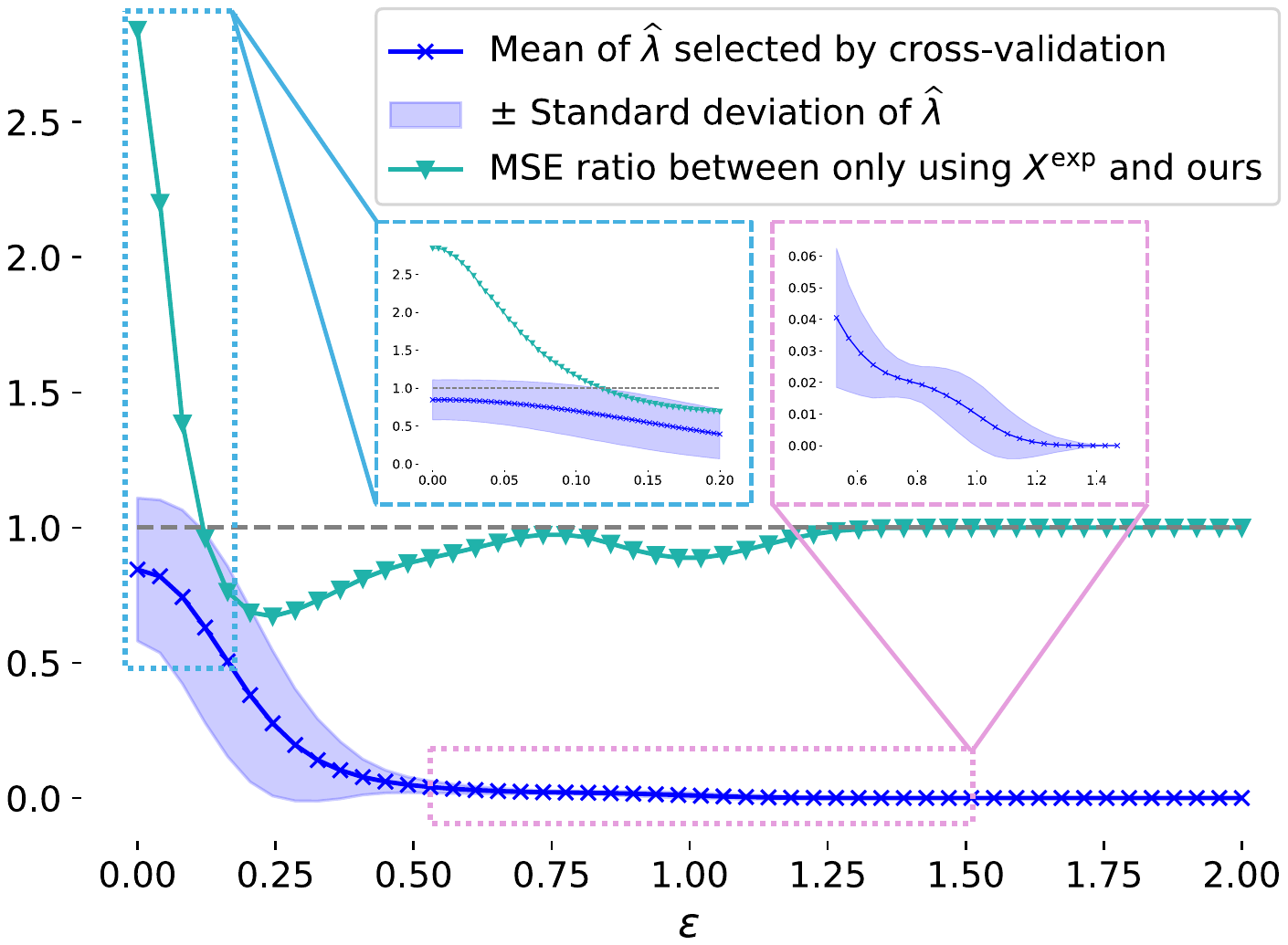}%
        \label{fig:c_mean_mse_eps}%
        }\hfill
    \subfloat[$N\expm=100$, $\varepsilon=0.1$, $\sigma^2=1$.]{%
        \includegraphics[width=0.48\textwidth]{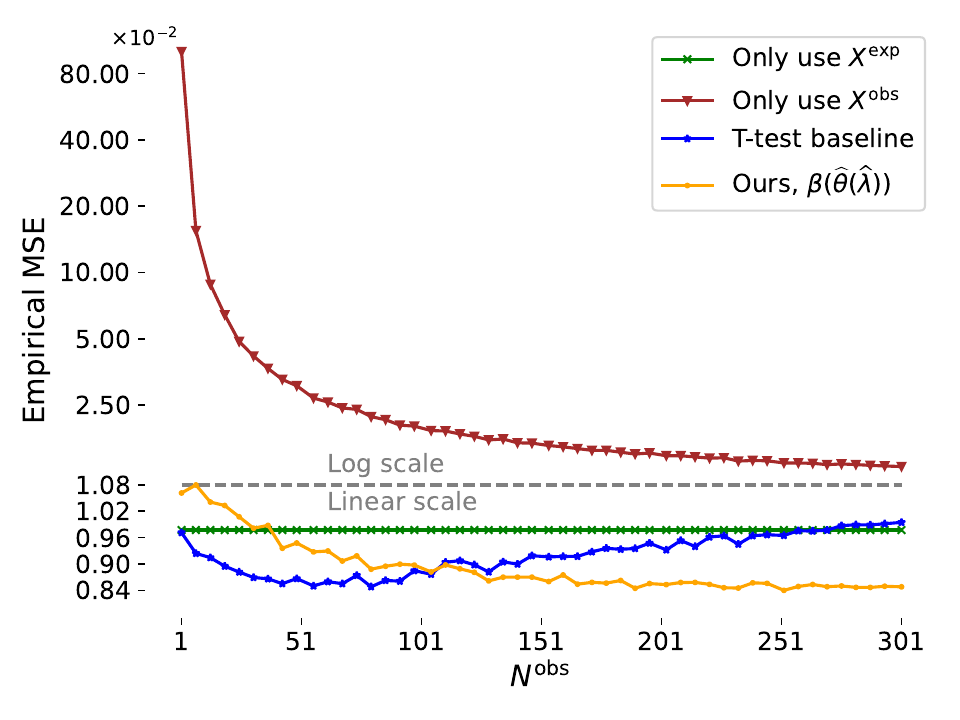}%
        \label{fig:e_mean_mse_n_obs}%
        }
    \caption{No-covariate setting. (a-b) Empirical MSE varying $\varepsilon$. (c) Empirical MSE ratio of the experimental-only estimate to CVCI, and selected $\widehat\lambda$, varying $\varepsilon$. (d) Empirical MSE varying $N\obs$, with a linear–log transformation for visual clarity. The remaining panels of this setting and the experiments with $\sigma^2=100$ are supplemented in Figures~\ref{fig:mean_mse_nocov_extra} and~\ref{fig:mean_mse_var_10}.}
    \label{fig:mean_mse_eps_var_1}%
\end{figure}

As shown in Figures~\ref{fig:a_mean_mse_eps} and \ref{fig:b_mean_mse_eps}, CVCI reliably adapts to varying, unknown values of $\varepsilon$ and outperforms at least one of the single-sourced methods. When $\varepsilon$ is small, it improves upon the $X\expm$-only approach; for intermediate $\varepsilon$, it yields the lowest error; for large $\varepsilon$, it outperforms the $X\obs$-only and t-test approaches while remaining comparable to using $X\expm$ alone. 
As shown in Figure~\ref{fig:c_mean_mse_eps}, our estimator increasingly resembles the experimental estimate as $\varepsilon$ grows, with only minor fluctuations observed before $\widehat{\lambda}$ approaches zero. 
This underscores CVCI's adaptivity to dynamically adjust the reliance on two data sources via cross-validation, without requiring prior knowledge of $\varepsilon$. When observational data are scarce or less reliable, cross-validation leans more heavily on experimental data. 
The above observations hold in both low ($\sigma^2 = 1$) and high ($\sigma^2 = 100$, in Figure~\ref{fig:mean_mse_var_10}) noise settings. Additional results on the impact of noise level are provided in Section~\ref{sec:additional_results_mean}.

\subsection{Linear setting}\label{sec:experiments_linear_setting}
\subsubsection{Settings}
Assume each data sample consists of a tuple of response $\res$, covariates $\covariate$, and binary treatment $\tre$. For experimental data, generate the response as a linear combination of the covariates plus exogenous noise:
$\res = \covariate^\top \theta\expm + \tre \tau^\star + \xi$, where $\xi \sim \cN (0, \sigma^2) \perp \covariate, \tre . $
For observational data, we incorporate a bias $\varepsilon$ to capture unmeasured confounders associated with the treatment:
$\res = \covariate^\top \theta\obs + \tre (\tau^\star + {\varepsilon}) + \xi$, where $\xi \sim \cN (0, \sigma^2) \perp \covariate, \tre.$
Here, $\theta\expm$ and $\theta\obs$ denote the parameters of the respective linear outcome models.
For implementation details see Section~\ref{sec:implementation_linear}.

\subsubsection{Results} \label{sec:linear_results}

We observe trends similar to those in the no-covariate case: our method consistently outperforms at least one of the baselines relying on one data source alone. 
When the bias is moderately low, our method achieves the most accurate causal estimates. 

\begin{figure}[!tb]
    \ifjrssb
    \vspace{0.5mm}
    \fi
    \centering
    \captionsetup[sub]{justification=centering}
    \subfloat[$\theta\expm \neq \theta\obs$, $N\expm=50$, $N\obs=100$.]{%
        \includegraphics[width=0.48\textwidth]{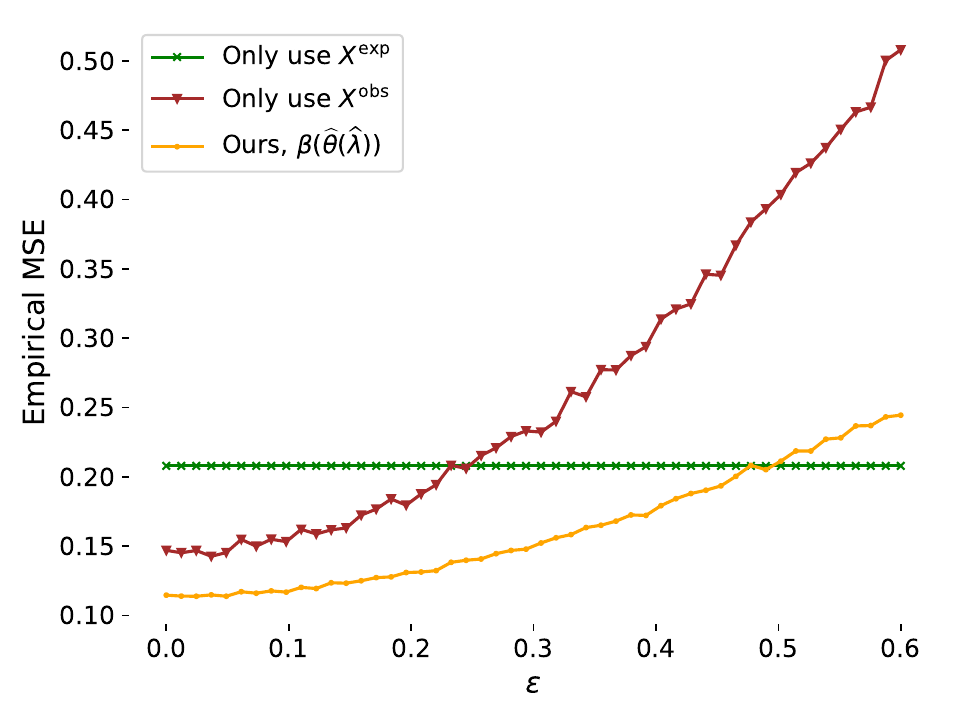}%
        \label{fig:b_linear_mse_eps}%
        }\hfill
    \subfloat[$\theta\expm \neq \theta\obs$, $N\expm=1000$, $N\obs=2000$.]{%
        \includegraphics[width=0.48\textwidth]{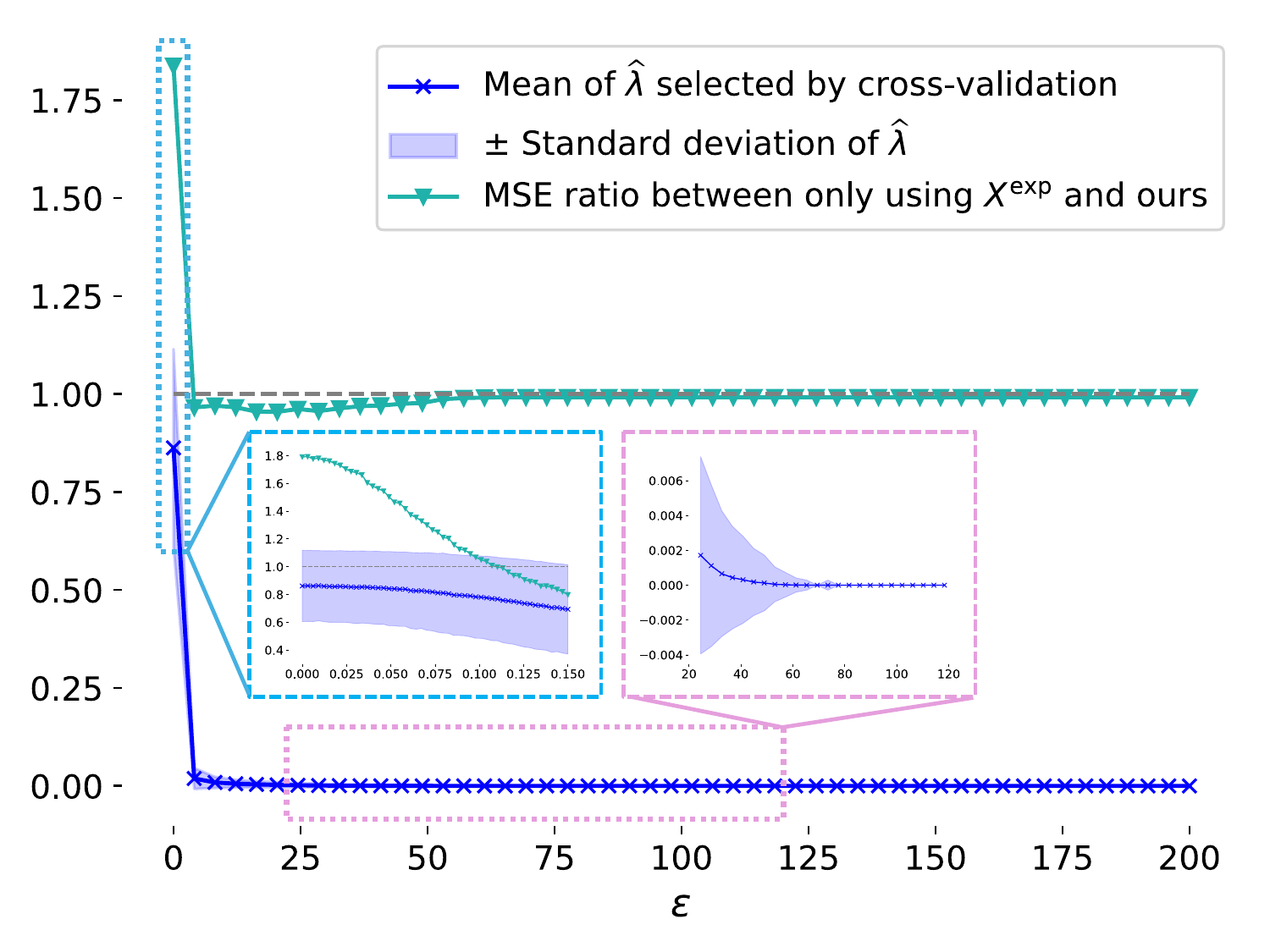}%
        \label{fig:b_linear_mse_ratio}%
        }\\
    \subfloat[$\theta\expm \neq \theta\obs$, $N\expm=50$, $N\obs=500$.]{%
        \includegraphics[width=0.48\textwidth]{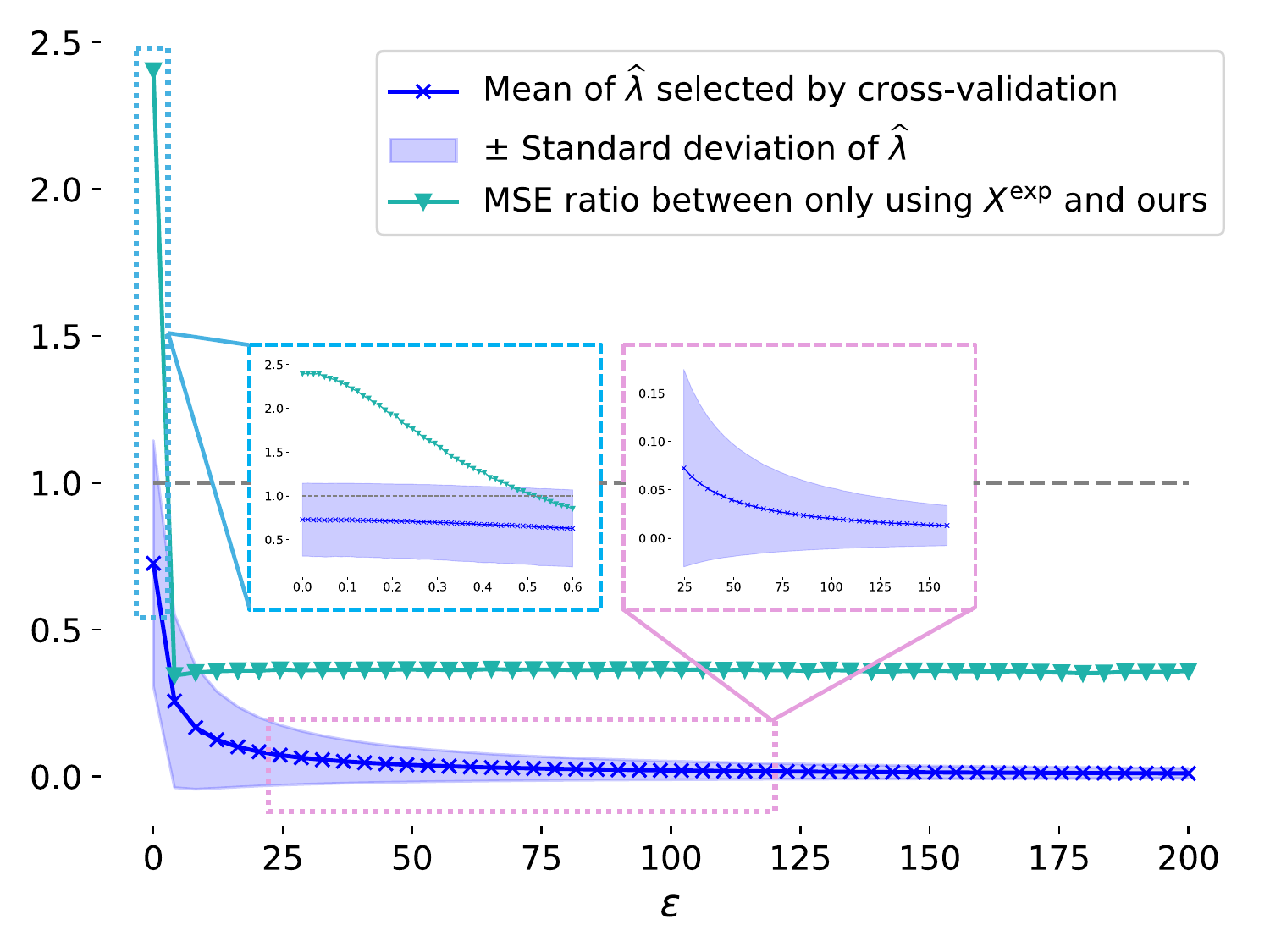}%
        \label{fig:d_linear_mse_ratio}%
        }\hfill
    \subfloat[Histogram of $\widehat\lambda$ for (c).]{%
        \includegraphics[width=0.48\textwidth]{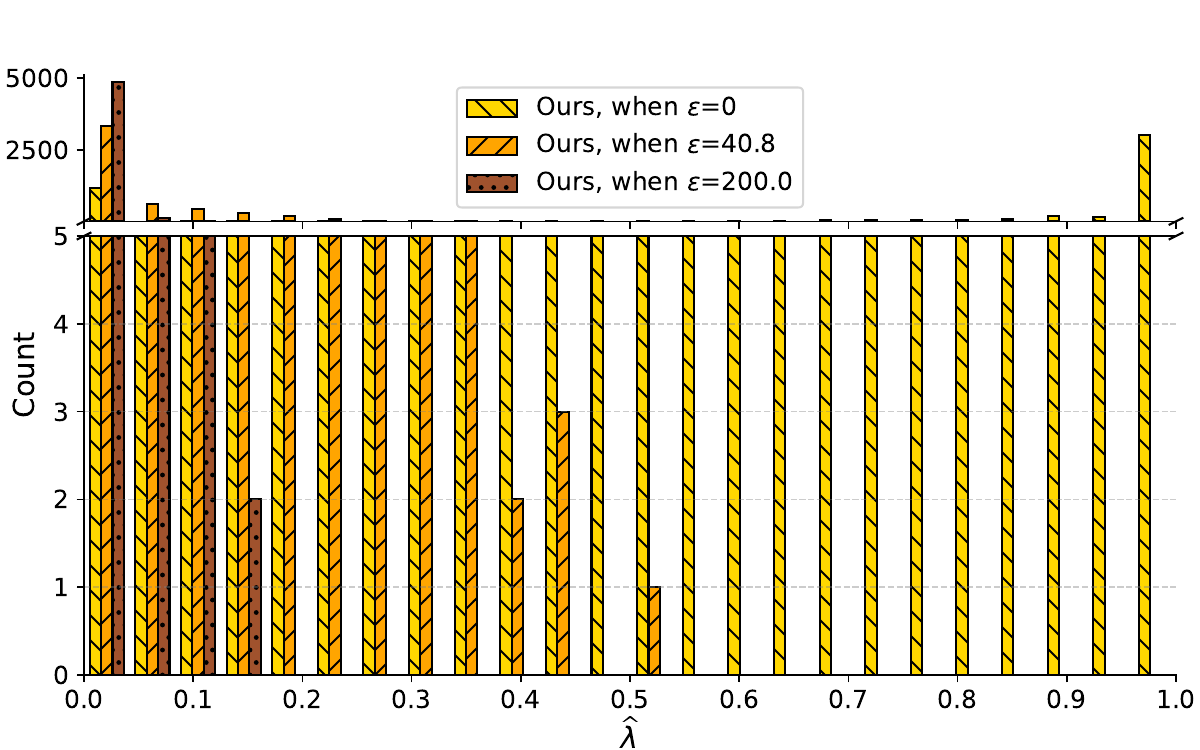}%
        \label{fig:f_linear_mse_lam_dist}%
        }
    \caption{Linear setting. (a) Empirical MSE varying $\varepsilon$. (b-c) Empirical MSE ratio of the experimental-only estimate to CVCI, and selected $\widehat\lambda$, varying $\varepsilon$. (d) Histogram of $\widehat\lambda$ values in the setting of (c); the vertical axis is split into $\leq 5$ and $>5$ counts to show extreme values that inflate the overall MSE. Figures~\ref{fig:linear_mse_eps_extra}--\ref{fig:linear_dist_extra} in the supplementary material give further sample sizes, results varying $N\obs$, the histogram of squared errors, and the $\theta\expm = \theta\obs$ counterparts.}
    \label{fig:linear_mse_eps}
    \label{fig:linear_mse_ratio}%
\end{figure}
We now turn to Figures~\ref{fig:b_linear_mse_ratio}--\ref{fig:f_linear_mse_lam_dist} to examine the behavior when $\varepsilon$ is large. When the experimental data are abundant ($N\expm=1000$, Figure~\ref{fig:b_linear_mse_ratio}), our method reliably converges to the experimental estimates, as cross-validation consistently selects $\widehat\lambda = 0$. This outcome aligns with our expectations: our method appropriately downweights the observational component when the experimental estimates are sufficiently more reliable. In contrast, when the experimental data are limited ($N\expm = 50$, Figure~\ref{fig:d_linear_mse_ratio}), the selected $\widehat\lambda$ generally remains close to zero for large $\varepsilon$, but occasionally small nonzero values are selected (Figure~\ref{fig:f_linear_mse_lam_dist}). When this happens, the resulting squared error can be large because a large $\varepsilon$ would amplify the error for very small $\widehat\lambda$ (supplemented in Figure~\ref{fig:e_linear_mse_se_dist}). 
Under large bias, for example, $\varepsilon = 200$, Figure~\ref{fig:f_linear_mse_lam_dist} shows that among 5000 simulation runs, the selection of $\widehat\lambda > 0.1$ occurred only a handful of times. 
While the method is not immune to error under severe confounding for small $N\expm$, it exhibits robust behavior in the majority of cases. 

Interestingly, the results remain unaffected by whether $\theta\expm$ and $\theta\obs$ are equal or different (supplemented in Figures~\ref{fig:linear_mse_eps_extra}--\ref{fig:linear_dist_extra}). 
This reflects our framework’s design: only estimated treatment effects, not raw covariates or outcome models, are shared across data sources.
Such flexibility is typically not supported by existing methods, which often require stronger assumptions about model alignment or covariate overlap.

\section{Case Study: The LaLonde Dataset}\label{sec:lalonde}
In 1986, Robert LaLonde published a seminal paper that compared the results of a field experiment with the range of estimates an econometrician might have produced using only nonexperimental data, concluding that the nonexperimental methods at that time failed to systematically replicate the trial results \citep{lalonde1986evaluating}. The original study examined the effect on trainee earnings of an employment program implemented through a field experiment.
Discussion and analysis of the LaLonde dataset have led to significant methodological advances in causal inference \citep{heckman1987we, smith2005does, imbens2024lalonde}. 
In our paper, we compare the ATE estimates on the LaLonde dataset from our method with various baselines.

\subsection{Settings} \label{sec:lalonde_settings}
The National Supported Work Demonstration (NSW) was a randomized trial where the treatment is to receive job training between 1975 and 1977. LaLonde and later Dehejia and Wahba analyzed its impact on real earnings (RE) in 1978, with the latter restricting to a smaller subgroup. The resulting NSW dataset contains 185 treated and 260 control samples. We detail the data selection process in Section~\ref{sec:lalonde_data_selection}.


We conduct experiments on both the real and the semi-synthetic data. For real data, the observational control data come from the Panel Study of Income Dynamics (PSID) and Westat's Matched Current Population Survey (CPS). 
They are control-only datasets. We term them \textit{observational control groups}. They are partitioned by pre-intervention variables into subgroups PSID-2, PSID-3, CPS-2, and CPS-3, with the full datasets denoted PSID-1 and CPS-1. We detail the partition procedure in Section~\ref{sec:lalonde_tables_supp}.
To generate semi-synthetic data, we fit linear models on the respective real data sets and re-sample the residuals from Gaussian distributions. This ensures that the experimental estimate is unbiased for the ground-truth effect.

To estimate the ATE, we apply various linear models and will show the following four sets of methods: (i) experimental-only ($\lambda=0$); (ii) CVCI; (iii) observational-only ($\lambda=1$), which uses the NSW treatment group and observational control groups; (iv) pooling, which views all data collectively as observational data, and sets $\lambda=1$.


For real data, we produce Table \ref{tab:lalonde_result_combined} for selected configurations, and Tables~\ref{tab:lalonde_result} and \ref{tab:lalonde_bootstrap} for full configurations, deferred to Section~\ref{sec:lalonde_tables_supp}.
Specifically, Table~\ref{tab:lalonde_result_combined} focuses on two major observational control groups (PSID-1 and CPS-1) and three covariate settings (matching columns 1, 3, and 8 of Tables~\ref{tab:lalonde_result} and \ref{tab:lalonde_bootstrap}). It integrates the point estimates from Table~\ref{tab:lalonde_result} and bootstrapped standard deviations from Table~\ref{tab:lalonde_bootstrap}. For the semi-synthetic data, we produce Table~\ref{tab:lalonde_synthetic}, with an error decomposition in Table~\ref{tab:lalonde_synthetic_decomposition}. 


\subsection{Results}

As a starting point, it is encouraging to see that results from nearly 27 years ago can still be largely replicated precisely today. 
Our reproduced point estimates using a single data source (first and third panels in Tables~\ref{tab:lalonde_result_combined} and \ref{tab:lalonde_result}) match exactly those of Dehejia and Wahba's (columns 1-4 of panels B and C in Table 2, which originally correspond to LaLonde's Table 5 without data selection).
We note, however, that column 5 of their panels B and C was described as controlling for all pre-intervention variables, but simply including all such variables did not allow us to exactly replicate their results. We provide additional discussions of these reproduced results in Section~\ref{sec:reproduced_dehejia}.

\ifjrssb
\else
\makeatletter
\@ifundefined{c@panel}{%
  \newcounter{panel}[table]%
  \renewcommand{\thepanel}{\thetable\Alph{panel}}%
  \renewcommand{\theHpanel}{\thetable.\arabic{panel}}%
}{}
\makeatother
\makeatletter
\@ifundefined{methodrulewidth}{\newlength{\methodrulewidth}\newlength{\methodrulewidthB}}{}
\makeatother
\providecommand{\methodrow}[1]{\multicolumn{4}{l}{\begin{tabular}[t]{@{}>{\raggedright\arraybackslash}p{\methodrulewidth}@{}}#1\\[-2pt]\midrule\end{tabular}}}
\begin{table}[htbp]
    \ifjrssb
    \vspace{0.5mm}
    \fi
    \scriptsize
    \settowidth{\methodrulewidth}{\textbf{($\lambda=1$, $\widetilde X\obs$ only)} \citet{dehejia1999causal}'s setting, $X\obs$:}
    \settowidth{\methodrulewidthB}{\textbf{($\widehat\lambda$, ours)} $\widetilde X\expm+\widetilde X\obs$, $X\expm$: NSW(T+C), $X\obs$:}
    \ifdim\methodrulewidthB>\methodrulewidth \setlength{\methodrulewidth}{\methodrulewidthB}\fi
    \centering
    \caption{Estimates of treatment effect on the LaLonde dataset on selected configurations (Part A), and root mean squared error (RMSE) on semi-synthetic data based on the LaLonde dataset (Part B). 
    Each row: (T) for treatment group, (C) for control group. Each column: estimates by different linear models. For our method, we report the averaged point estimates and averaged selected $\widehat\lambda$ over 5000 runs. In Part A, $\pm 1$ standard deviations are bootstrapped. In Part B, $\widetilde X\expm$ and $\widetilde X\obs$ denote synthetic data generated from the corresponding real data sets. The method with the lowest RMSE for each column is in bold. }
    \begin{tabular}{>{\raggedleft\arraybackslash}m{4.5cm}
    >{\centering\arraybackslash}m{2cm}
    >{\centering\arraybackslash}m{2cm}
    >{\centering\arraybackslash}m{3.4cm}
    }

         \toprule\toprule
         \textbf{Column No.} & 1 & 3 & 8 \\
        \textbf{[Linear setting]} Regress RE78 on: & \{treatment\} & \{treatment, RE75\} &   \{treatment, age, years of schooling, high school dropout status, race, marriage status, RE75, employment status in 1975, RE74, employment status in 1974\} \\
        \toprule

    \multicolumn{4}{l}{\refstepcounter{panel}\label{tab:lalonde_result_combined}\textbf{Part \Alph{panel}: Real data, ATE estimates}} \\ \midrule
    \methodrow{\textbf{($\lambda=0$, $X\expm$ only)}} \\
   NSW(T+C), ATE estimate:  & {1794} $\pm$ 658 & {1750} $\pm$ 657 & {1671} $\pm$ 666 \\\midrule
\methodrow{\textbf{($\widehat\lambda$, ours)} $X\expm+X\obs$, $X\expm$: NSW(T+C), $X\obs$:} \\
NSW(T)+PSID-1(C), ATE estimate: & 1761 $\pm$ 672 & 1511 $\pm$ 721 & 1282 $\pm$ 708 \\
 $\widehat\lambda=$ & (0.0 $\pm$ 0.0) & (0.6 $\pm$ 0.3) & (0.8 $\pm$ 0.3) \\
NSW(T)+CPS-1(C), ATE estimate: & 1740 $\pm$ 673 & 1465 $\pm$ 724 & 1162 $\pm$ 628\\
$\widehat\lambda=$ & (0.3 $\pm$ 0.1) & (0.9 $\pm$ 0.2) & (1.0 $\pm$ 0.2) \\
 \midrule
 \methodrow{\textbf{($\lambda=1$, $X\obs$ only)} \citet{dehejia1999causal}'s setting, $X\obs$:} \\
NSW(T)+PSID-1(C), ATE estimate: & {-15205} $\pm$ 657 & -582 $\pm$ 765 & 4 $\pm$ 842 \\
NSW(T)+CPS-1(C), ATE estimate: & {-8498} $\pm$ 582 & -78 $\pm$ 598 & 1066 $\pm$ 624 \\
\midrule
  \methodrow{\textbf{($\lambda=1$, pool all data as $X\obs$)} $X\obs$:} \\
  NSW(T+C)+PSID-1(C), ATE estimate: & {-13598} $\pm$ 641 & -162 $\pm$ 713 & 741 $\pm$ 666 \\
  NSW(T+C)+CPS-1(C), ATE estimate: & {-8333} $\pm$ 579 & -17 $\pm$ 592 & {1148} $\pm$ 618 \\

    \toprule
    \multicolumn{4}{l}{\refstepcounter{panel}\label{tab:lalonde_synthetic}\textbf{Part \Alph{panel}: Semi-synthetic data, RMSE}} \\ \midrule
    \methodrow{\textbf{($\lambda=0$, $\widetilde X\expm$ only)}} \\
   NSW(T+C), RMSE:  & \textbf{647.7} & \textbf{646.0} & 646.6 \\\midrule
\methodrow{\textbf{($\widehat\lambda$, ours)} $\widetilde X\expm+\widetilde X\obs$, $X\expm$: NSW(T+C), $X\obs$:} \\
NSW(T)+PSID-1(C), RMSE: & 651.5 & 747.7 & 734.1 \\
 $\widehat\lambda=$ & (0.0 $\pm$ 0.0) & (0.6 $\pm$ 0.3) & (0.7 $\pm$ 0.3) \\
NSW(T)+CPS-1(C), RMSE: & 655.2 & 767.7 & \textbf{617.4} \\
 $\widehat\lambda=$ & (0.3 $\pm$ 0.2) & (0.8 $\pm$ 0.2) & (0.9 $\pm$ 0.2) \\
 \midrule
 \methodrow{\textbf{($\lambda=1$, $\widetilde X\obs$ only)} \citet{dehejia1999causal}'s setting, $X\obs$:} \\
NSW(T)+PSID-1(C), RMSE: & 17017.6 & 2469.7 & 1943.6 \\
NSW(T)+CPS-1(C), RMSE: & 10282.3 & 1880.3 & 796.9 \\
\midrule
  \methodrow{\textbf{($\lambda=1$, pool all data as $\widetilde X\obs$)} $X\obs$:} \\
  NSW(T+C)+PSID-1(C), RMSE: & 15409.3 & 2038.9 & 1291.2 \\
  NSW(T+C)+CPS-1(C), RMSE: & 10143.0 & 1848.9 & 773.2 \\

    \bottomrule
    \end{tabular}
    \label{tab:lalonde_result_combined_real_synthetic}
\end{table}
\fi

Importantly, our method gives reliable ATE estimates in the real data setting. 
As shown in Table~\ref{tab:lalonde_result_combined}, our method (the second panel) is more reliable than the approach that depends solely on observational data (the third panel) in all configurations, and in majority of cases 
(30 out of 48 configurations) 
when compared to pooling (the fourth panel). 
In such a case, we confirm what LaLonde found: there is inherent difficulty in accurate nonexperimental modeling due to extreme inter-model variability, even after choosing more suitable subsets. 

Moreover, we identify two trends that align with our intuition: first, inclusion of additional informative covariates leads to a greater weight on the observational component. In Table~\ref{tab:lalonde_result_combined}, $\widehat\lambda$ values are generally small in column 1, where only the treatment is used, and increase in columns 3 and 8, where additional covariates are included. 
Such a trend also presents in Table~\ref{tab:lalonde_result}---columns 6 and 7, which correspond to columns 2 and 4 with the addition of RE74, exhibit noticeably larger $\widehat\lambda$ values. 
Second, for observational control subgroups that are more similar to the NSW control group, the selected $\widehat\lambda$ are generally larger. This agrees with LaLonde’s assertion that subgroups such as PSID-2, PSID-3, CPS-2, and CPS-3 are more comparable to the NSW control
group in distributions of pre-intervention variables.




The semi-synthetic results are generally a callback to the analysis in Section~\ref{sec:linear_results}. Our method achieves the lowest RMSE on CPS, column 8 (Table~\ref{tab:lalonde_synthetic}), corresponding to the regions where it has the lowest error in Figure~\ref{fig:b_linear_mse_eps}. In other cases, we refer back to discussions of Figure~\ref{fig:d_linear_mse_ratio} where it underperforms the experimental estimate, given the substantial bias present in the LaLonde observational data and the small experimental sample size.

\section{Theory}\label{sec:theoretical_results}

\newcommand{\obsmeaninput}{o}
\newcommand{\meancontrast}{d}
\subsection{Improvement of CVCI in finite samples}\label{sec:cv_mean_improvement}

We now give a sufficient condition under which our estimator has strictly smaller MSE than both the experimental and observational sample means, with the proof provided in Section~\ref{sec:proof_prop:cv_mean_improvement}. We illustrate with a setting with no covariates. 
Similar results may be established in the linear setting under proper conditions.

\begin{proposition}[Improvement over both sample means]\label{prop:cv_mean_improvement}
Suppose the experimental and observational samples are independent, with
$\expout_1,\ldots,\expout_{\nexp}\simiid\cN(\truemean,\sigma_{\expshort}^2),$ $
\obsout_1,\ldots,\obsout_{\nobs}\simiid\cN(\truemean+\obserr,\sigma_{\obsshort}^2).
$
Let $s_{\expshort}=\sigma_{\expshort}/\sqrt{\nexp}$ and $s_{\obsshort}=\sigma_{\obsshort}/\sqrt{\nobs}$ be the positive standard errors of the two sample means. Write $\Delta^2=\obserr^2+(s_{\obsshort}-s_{\expshort})^2$, with $\Delta\geq0$. For fixed folds of equal size, it holds that
\begin{align*}
 \E\big[(\EstPar(\estregu)-\truemean)^2\big]
 \leq \big(s_{\expshort}\sqrt{g_{\numfold}}+\Delta\big)^2,
\end{align*}
where the expectation is taken over both samples, with constant $g_{\numfold}\in(1/2,1)$ depending only on the number of folds $\numfold\geq2$. If this bound is below $\min\{s_{\expshort}^2,s_{\obsshort}^2+\obserr^2\}$, CVCI improves on both sample means. 
In particular, when both means have the same standard error $s$, $|\obserr|<(1-\sqrt{g_{\numfold}})s$ suffices. 
If $\numfold=5$, $g_5\approx0.739$, so $|\obserr|\leq 0.14s$ suffices.
\end{proposition}

\subsection{Robustness of CVCI}\label{sec:robustness_theory}


We now present robustness guarantees for CVCI.
  Recall the setup in Section~\ref{sec:problem_formulation}, we consider the squared experimental loss, {\it i.e.}, $\lexp(\beta;\expsam_{\samset}) =(\atefun-\estate(\expsam_{\samset}))^2$ for any set of samples $\expsam_{\samset}$ indexed by $\samset\subseteq[\nexp]$. 
In addition, we make the following two assumptions for our analysis:


\begin{myassumption}{LinATE}{ass:linear_ate}
    Let $\Zfuntilone:\expsamspace\mapsto\R$ be some function satisfying
     $ \E[\Zfuntilone(\expsam_1)]=0$ and $\vecnorm{\Zfuntilone}{\infty} \leq \boundZfuntil$ for some $\boundZfuntil>0$.
     Let $\boundate,\boundatediff, \boundatelin,\boundatenum>0$ be constants.
     For any set $\samset\subseteq[\nexp]$  and any $\delta\in(0,1/2)$ such that $\samsetnum\geq \boundatenum\log(1/\delta)$, 
       with probability at least $1-\delta$, the experimental estimate   $\estate(\expsam_{\samset})$ satisfies:
        \begin{enumerate}
         \item\label{ass:linear_ate:z} 
         $|\trueate|\leq \boundate$,
         \item\label{ass:linear_ate:a} $|\estate(\expsam_{\samset})-\trueate| \leq {\boundatediff\sqrt{\log(1/\delta)}}/{\sqrt{\samsetnum}}$,
         \item\label{ass:linear_ate:b} $\vecnorm{\estate(\expsam_{\samset})-\trueate-{\samsetnum}^{-1}\sum_{j\in\samset}\Zfuntilone(\expsam_j)}{2}\leq {\boundatelin\log(1/\delta)}/{{\samsetnum}}$.
        \end{enumerate}
 \end{myassumption}

\begin{myassumption}{OBS}{ass:obs_ate}
    The observational parameter space $\Parspace\subseteq\R^{\Pardim}$ satisfies $\vecnorm{\Par}{2}\leq \boundobspar$ for all $\Par\in\Parspace$ for some $\boundobspar>0$;
    $\atefun(\Par)\defn \Par_1$ takes the first element of $\Par$ as the estimate of ATE; 
    $\lboundobsh\IdMat\preceq\nabla_\Par^2\lobs(\Par;\dsetobs)
    \preceq \boundobsh\IdMat$ and $\opnorm{\nabla_\Par^3\lobs(\Par;\dsetobs)}\leq \boundobsthr$ for 
    some constants $\boundobsh, \lboundobsh>0,\boundobsthr>0$.
\end{myassumption}

Assumption~\ref{ass:linear_ate} assumes that the ATE estimator $\estate$ based on experimental samples is $\sqrt{\nexp}$-consistent and admits a linear approximation. For example, this is satisfied in our linear setting.
A sufficient condition for Assumption~\ref{ass:linear_ate} is that $\estate$ is derived from some $Z$-estimation problem (Assumption~\ref{ass:zest_as_ate}). We refer to Section~\ref{sec:sufficient_condition_linear_ate} for more details. In Assumption~\ref{ass:obs_ate}, we require the observational loss 
to be strongly convex and have smooth higher-order derivatives. These are standard regularity conditions for analyzing empirical risk minimization. Moreover, the assumption $\atefun(\Par)= \Par_1$ can be generalized to  $\atefun$ being a linear function of $\Par$, as they are equivalent up to a linear transformation on $\Par$. We choose the former for simplicity of presentation.

We use $\polyshort,\polyshortprime>0$ to denote constants that depend polynomially on the parameters in the assumptions (see Section~\ref{sec:additional_notations} for details). 
Our main result holds for any experimental sample size exceeding a threshold. Namely, we assume 
\begin{align}
    \sqrt{\nexp}\geq \polyshort {\numfold}(\log^{1.5}\numfold+\log^{0.5}(1/\delta)), \label{eq:sample_size_condition}
\end{align}
for some constant $\polyshort = \poly>0$ and a user-specified failure probability $\delta \in (0,1/2)$.

\begin{theorem}\label{thm:cv_main}
    Suppose Assumptions~\ref{ass:obs_ate}~and~\ref{ass:linear_ate} hold and the experimental sample size satisfies~\eqref{eq:sample_size_condition}.
     Then there exists some constant $\polyshortprime >0$ such that, with probability at least $1-\delta$,
    \begin{align*}
        (\atefun(\EstPar(\estregu))-\trueate)^2
        \leq \polyshortprime
        \min\Big\{\frac{\log(1/\delta)}{\nexp},1\Big\}.
    \end{align*} 
\end{theorem}

See the proof in Section~\ref{sec:pf_thm_cv_main}. 
A direct consequence of Theorem~\ref{thm:cv_main} is
\begin{corollary}[Robustness of $\atefun(\EstPar(\estregu;\dset))$]
    \label{cor:robustness_atefun_estpar}
    Under Assumptions~\ref{ass:obs_ate}~and~\ref{ass:linear_ate}, there exist some constants $\polyshort,\polyshortprime>0$ such that when $ \numfold\leq  \polyshort\sqrt{\nexp}/\log^{1.5}\nexp$, the estimation error of $\trueate$ is
    \begin{align*}
       \E[(\atefun(\EstPar(\estregu))-\trueate)^2]\leq \frac{\polyshortprime}{\nexp},
    \end{align*} 
    where the expectation is taken over the experimental samples $(\expsam_j)_{j\in[\nexp]}$.
\end{corollary}

Theorem~\ref{thm:cv_main}~and~Corollary~\ref{cor:robustness_atefun_estpar} indicate that our estimator $\atefun(\EstPar(\estregu))$ is robust to the choice of observational samples---it achieves an $\bigO(1/\nexp)$ error rate regardless of the level of bias in observational data. Notably, this $\bigO(1/\nexp)$ rate is known to be optimal and can be attained, for instance, by the AIPW estimator~\citep{hahn1998role} using $\nexp$ experimental samples and no observational data. Moreover, even with a sufficiently large number of observational samples, one cannot achieve a rate faster than $\bigO(1/\nexp)$ without imposing additional assumptions on the observational data. The proof is presented in Section~\ref{sec:pf_cor_robustness_atefun_estpar}. 

We then give a matching minimax lower bound:
Theorem~\ref{thm:mean_estimation_minimax} shows that when taking both experimental and observational data as input, no robust estimator ({\it i.e.}, one with an error rate of order $\bigO(1/\nexp)$ uniformly over  $\obserr\in[-\bounderr,\bounderr]$) can achieve an error rate better than $\bigO(1/\nexp)$, even when $\obserr$ is zero. 
The proof can be found in Section~\ref{sec:pf_thm_mean_estimation_minimax}. 

\begin{theorem}[Minimax lower bound in the  no-covariate setting]
    \label{thm:mean_estimation_minimax}
   Without loss of generality, suppose we are given $\nexp$ experimental samples $\expout_1,\ldots,\expout_{\nexp}\simiid \cN(\truemean,1)$ and $\nobs$  observational samples  $\obsout_1,\ldots,\obsout_{\nobs}\simiid \cN(\truemean+\obserr,1)$ for a mean $\truemean\in[-\boundmean,\boundmean]$ and observational bias $\obserr\in[-\bounderr,\bounderr]$. For any $\meanconsta>0$, define
\begin{align*}
    \minmaxclass_{\meanconsta}
    &\defn
    \{\estmean:\R^{\nexp+\nobs}\to\R\text{ such that }
    \estmean =
\estmean((\expout_i)_{i=1}^{\nexp};
(\obsout_i)_{i=1}^{\nobs})~\text{satisfies}~\\
&~~~~~~~~~~~\E[(\estmean-\truemean)^2]\leq \frac{\meanconsta}{\nexp},
~\text{ for any } \truemean\in [-\boundmean,\boundmean] \text{ and } \obserr\in[-\bounderr,\bounderr].
 \}
\end{align*}
For an absolute constant $\widetilde{c}_1>0$ and any constant $\meanconsta\in[\widetilde{c}_1,\nexp/8]$, 
there is a constant $\meanconstb>0$ depending only on $\meanconsta$ such that
\begin{align*}
   \inf_{\estmean\in\minmaxclass_{\meanconsta}} 
   \sup_{\truemean\in[-\boundmean,\boundmean]}
   \E_{(\expout_i)_{i=1}^{\nexp},(\obsout_i)_{i=1}^{\nobs}~\simiid ~\cN(\truemean,1)}[(\estmean-\truemean)^2]
    \geq \frac{\meanconstb}{\nexp}.
\end{align*}
\end{theorem}

\section{Related Work} \label{sec:related_work}

Our method is inspired by a broader cross-validation-based statistical learning family, which includes stacking \citep{wolpert1992stacked, breiman1996stacked}, aggregation \citep{tsybakov2003optimal, tsybakov2004optimal}, and super learner \citep{van2007super}. 
While the statistical literature usually focuses on predictive fit as the cross-validation objective to aggregate estimators, we design the cross-validation procedure to be explicitly tailored for causal estimation. That is, we have a full nuisance model, but we are mostly interested in its causal parameter (representing the treatment effect) over a specific population (the experimental population). As a result our cross-validation objective is a causal objective that only evaluates the causal parameter, and we split only the experimental data instead of all available data throughout the process. 

A systematic and unified framework to combine experimental and observational data remains largely absent---existing literature is often {\it ad hoc} in nature and hinges on auxiliary assumptions, such as prespecified bias structures \citep{kallus2018removing, van2026adaptive},
 additional model assumptions \citep{yang2020combining, gagnon2023precise}, or covariate similarity \citep{gui2024combining}. We do not assume any of them, as detailed in Section~\ref{sec:problem_formulation}. 
Pooling-based methods treat all data as if they come from one source, often relying on either parametric tests or the assumption that the treatment effect is transportable
\citep{ross2009pooled, gao2023pretest, yang2023elastic, xiong2023federated, armstrong2025adapting, chen2025minimax, dang2025experiment}. 
Shrinkage methods tolerate observational bias but assume that the average effects are equal across data sources within each predefined stratum---a condition that may not hold in practice 
\citep{stein1956inadmissibility, green1991james,  rosenman2023combining}. 
Error-prone estimators balance biased components for each source to cancel out confounding effects~\citep{yang2020combining}. This is an appealing idea, though in practice it can be challenging to construct such estimators. 

\section{Discussion}
We have proposed a simple, general method for integrating experimental and observational data, leveraging cross-validation to adaptively tune their relative contribution. Our approach requires no additional specification or identification assumptions and accommodates a broad range of scenarios beyond the scope of existing methods. We demonstrated its efficacy, adaptivity, and robustness through experiments on both real-world and synthetic datasets. Furthermore, we provided a theoretical analysis showing that it is robust to the bias in observational data, and characterized its improvement and minimax optimality in a no-covariate setting.
In principle, this could be extended from ATE to other causal estimands, such as the conditional average treatment effect. 
The generality of our framework opens opportunities to exploit problem-specific structure, such as the relationship between experimental and observational models, for tailored adaptations in applications.

\if1\blind
\section{Acknowledgments}
\ifjrssb
\else
We thank Dmitry Arkhangelsky, Peter Bickel, Fan Chen, Kevin Chen, Peng Ding, Zeyu Jia, Lihua Lei, Jann Spiess, Ian Waudby-Smith, and Shu Yang  for helpful discussions. 
\fi
We use AI to help write code, process data, and generate content. The authors are responsible for the final content, including materials prepared with AI assistance. 
\fi
\bibliographystyle{abbrvnat}

\bibliography{ref}

\newpage
\ifjrssb
\else
\appendix
\fi
\begin{center}
\vspace*{1mm}
{\large\bf SUPPLEMENTARY MATERIAL}
\end{center}
The supplementary material is organized as follows: Section~\ref{sec:setup_intro_figure} details the setup for Figure~\ref{fig:lalonde_intro}. 
For experiments, implementation details and additional results are provided for the comparison overview (Section~\ref{sec:implementation_overview}), the no-covariate setting (Section~\ref{sec:mean_supp}), the linear setting (Section~\ref{sec:linear_supp}), 
and the LaLonde dataset (Section~\ref{sec:lalonde_specifics}).
Finally, Sections~\ref{sec:proof_method_section} and~\ref{sec:proof_theoretical_results} contain proofs organized by section.

\section{Setup for Figure~\ref{fig:lalonde_intro}}\label{sec:setup_intro_figure}

\begin{figure}[H]
    \centering
        \subfloat[No-covariate setting.]{
        \includegraphics[width=0.45\textwidth]{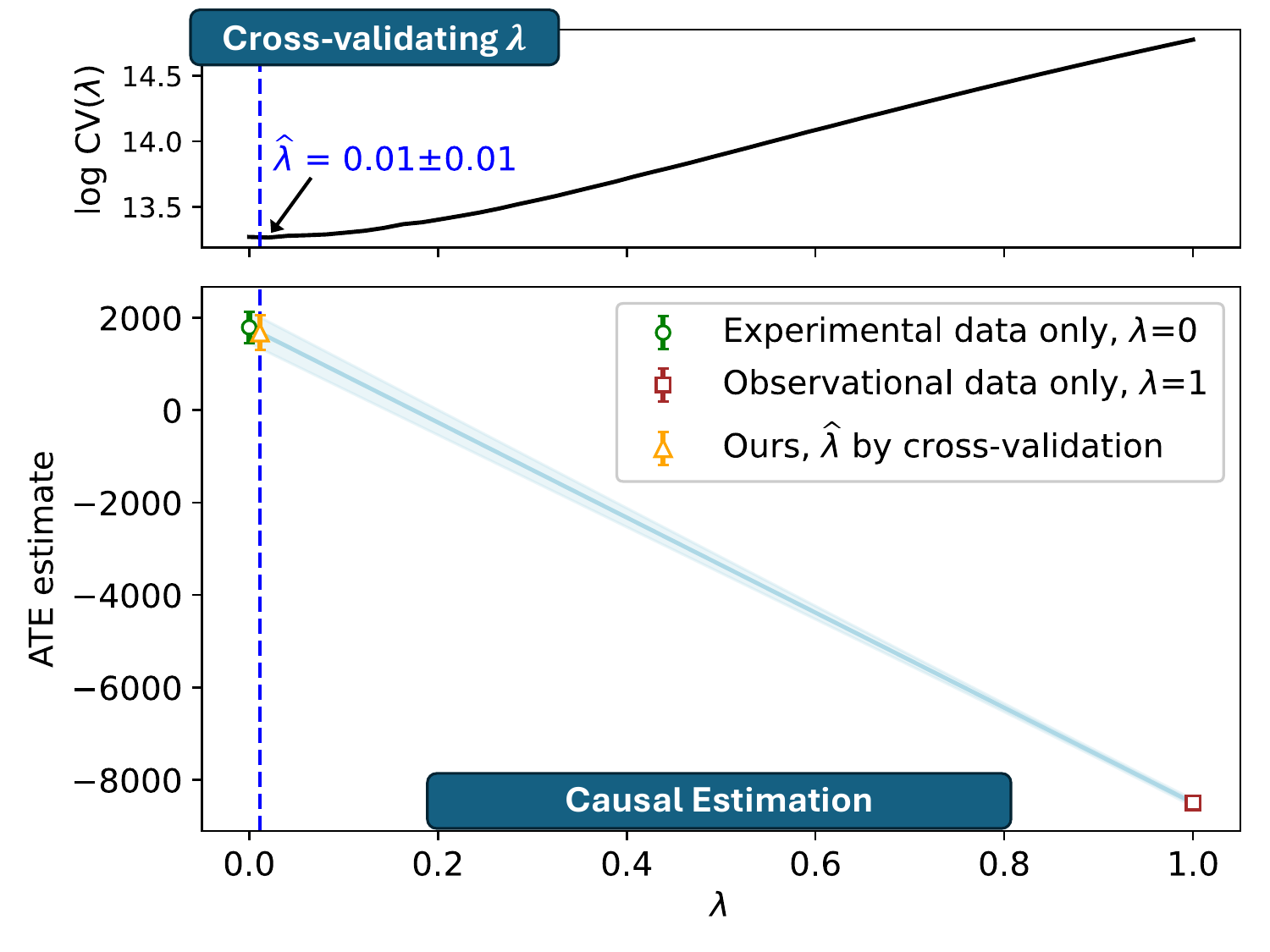}
        \label{fig:lalonde_intro_cps_mean}
        }
    \subfloat[Covariate adjusted linear setting.]{
    \includegraphics[width=0.45\textwidth]{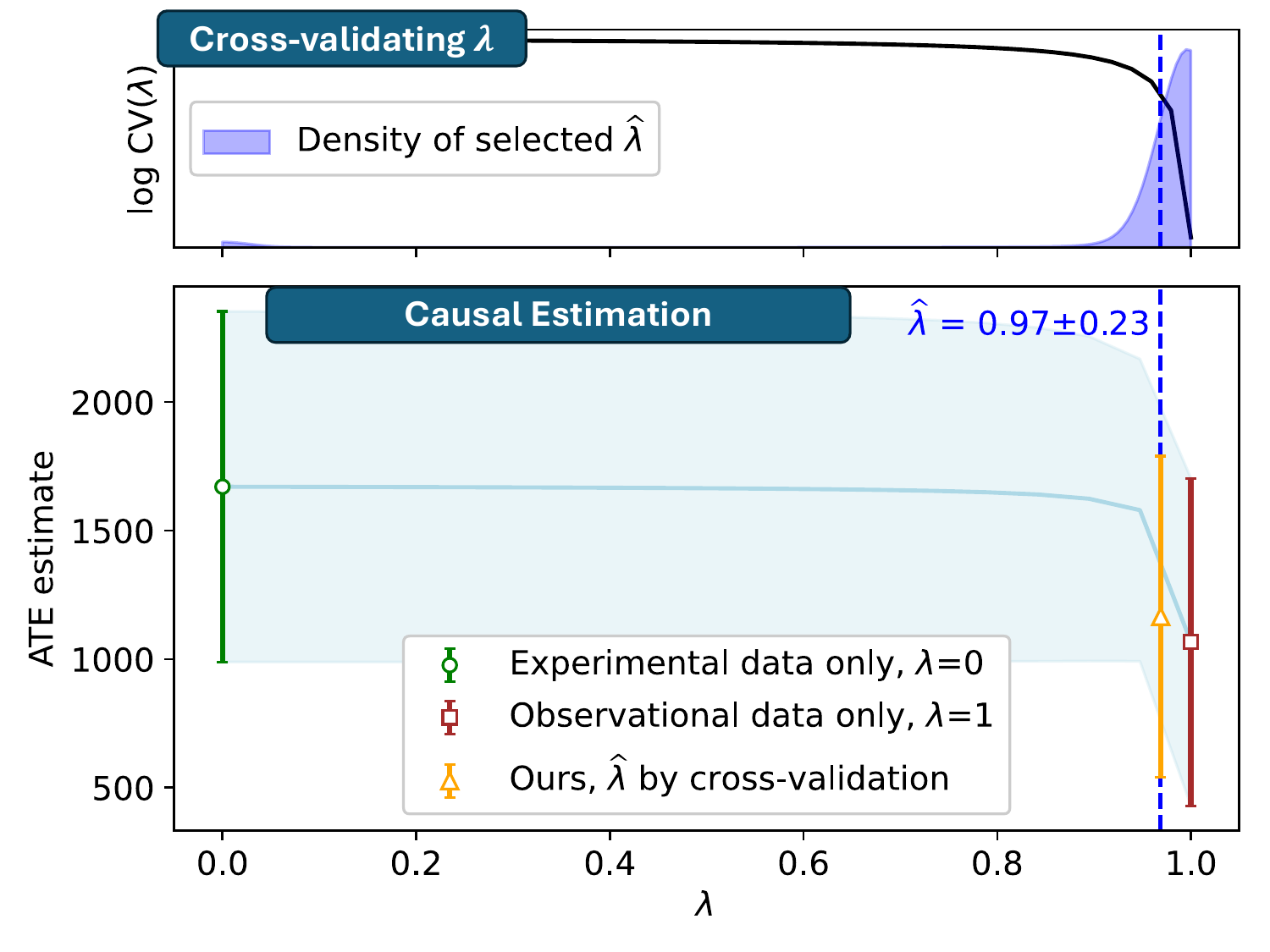}
        \label{fig:lalonde_intro_cps_col8}
        }
    \caption{Cross-validation objective (top) and estimates (bottom) as a function of $\lambda$. CPS control group.} 
    \label{fig:lalonde_intro_cps}
\end{figure}

We supplement Figure \ref{fig:lalonde_intro_cps} for the CPS control group (where Figure~\ref{fig:lalonde_intro} uses the PSID group) and then provide a detailed explanation of both.  
\paragraph{Left versus right}
In both Figures~\ref{fig:lalonde_intro_cps} and \ref{fig:lalonde_intro}, the left subfigures estimate the control mean in the no-covariate case. The right subfigures adjust for all available covariates (that correspond to column 8 in Table~\ref{tab:lalonde_result_combined}).  We employ five-fold cross-validation. Error bars reflect $\pm$1 bootstrapped standard deviations.

\paragraph{Top panel}
The top panels illustrate the process of cross-validating $\lambda$. In each run, experimental data are split into $K$ folds to perform cross-validation, and the $\lambda$ that minimizes the cross-validation objective $\CV(\lambda)$ is selected. The curves in the top panels show $\CV(\lambda)$ averaged over 5000 runs, and the average selected $\widehat\lambda$ is marked by the blue dashed vertical line. 
\paragraph{(Top panel) Why does the average $\widehat\lambda$ not minimize the average $\CV(\lambda)$ in Figures~\ref{fig:lalonde_intro_col8} and \ref{fig:lalonde_intro_cps_col8}?}
As shown by the density plots ({\it i.e.}, how often we select a particular $\lambda$ over 5000 runs), there is a small tail around 0 that skews the average of $\widehat \lambda$ leftwards. 

\paragraph{Bottom panel}
The bottom panels illustrate ATE estimates across different $\lambda$: $\lambda=0$ corresponds to using experimental data alone; $\lambda=1$ to using observational data alone; and the light blue curve in between shows the estimates for $\lambda\in(0,1)$.

\paragraph{(Bottom panel) Why does our method's estimate (the red square) not exactly align with the light blue curve in the back?}
We note that our method’s ATE point estimate (the red square) does not necessarily coincide with this curve at the average $\widehat\lambda$. This discrepancy arises because we may select different $\lambda$ in different runs (due to varying $K$-fold splits), whereas the light blue curve represents the average estimate at a fixed $\lambda$.
Similarly, the bootstrapped standard deviation for our method incorporates uncertainty from both data resampling and cross-validation splitting, while the light blue shaded area reflects data resampling alone at a fixed $\lambda$.

\section{Experiment Overview: Implementation Details}\label{sec:implementation_overview}



\subsection{Experimental and observational estimates}\label{sec:estimates_impl}
 Without covariates, the experimental estimate $\estate$ is the experimental sample mean. With covariates, the experimental estimate $\estate$ is either the AIPW estimate with the known assignment probability and two-fold cross-fitting, or the OLS coefficient on treatment in a regression of the outcome on treatment, the covariates and an intercept. The observational estimate $\widehat\tau\obs$ is the sample mean without covariates, or the OLS treatment coefficient on the observational covariates. 

\subsection{Baselines}
Baselines take the experimental and observational estimates with their estimated variances, denoted as $(\estate, \widehat v_{\mathrm{e}})$ and $(\widehat\tau\obs, \widehat v_{\mathrm{o}})$. Write
\begin{align*}
    V = \widehat v_{\mathrm{e}} + \widehat v_{\mathrm{o}}, \qquad w = \frac{\widehat v_{\mathrm{e}}}{V}, \qquad \widehat\Delta = \widehat\tau\obs - \estate, \qquad t = \frac{\widehat\Delta}{\sqrt{V}}.
\end{align*}
The baselines include:

\begin{itemize}[leftmargin=*, itemsep=2pt]
    \item \emph{Pre-test}:
    \begin{align*}
        \widehat\tau_{\mathrm{PT}} = \begin{cases} \estate & \text{if } |t| > 1.96, \\ \estate + w\,\widehat\Delta & \text{otherwise.} \end{cases}
    \end{align*}
    \item \emph{Soft threshold}:
    \begin{align*}
        \widehat\tau_{\mathrm{ST}}(c) = \begin{cases} \estate + c\,w\sqrt{V}\,\mathrm{sign}(\widehat\Delta) & \text{if } |t| > c, \\ \estate + w\,\widehat\Delta & \text{otherwise,} \end{cases} \qquad c = 1.96;
    \end{align*}
    the pooled discrepancy $w\,\widehat\Delta$ is capped at $c\,w\sqrt{V}$ in absolute value, so the estimate moves from the precision-weighted combination toward $\estate$ as the discrepancy grows.
    \item \emph{Adaptive soft threshold (AKS adaptive)} \citep{armstrong2025adapting}: $\widehat\tau_{\mathrm{ST}}(c^\star)$, where the cutoff $c^\star$ is chosen to minimize the worst-case regret in MSE relative to an oracle that knows the size of the bias, uniformly over bias sizes; $c^\star$ depends only on the correlation $-\sqrt{\widehat v_{\mathrm{e}}/V}$ between the experimental estimate and the discrepancy, and we take it from the authors' table and interpolation (for example $c^\star = 0.64$ at a correlation of $-0.77$).
    \item \emph{Error-prone estimator} \citep{yang2020combining}: a consistent estimate $\widehat\tau_2$ from a validation sample is improved by the difference of two error-prone estimates, $\widehat\tau_{2,\mathrm{ep}}$ on the validation sample and $\widehat\tau_{1,\mathrm{ep}}$ on the main sample, whose limit is zero when both samples come from one population:
    \begin{align*}
        \widehat\tau_{\mathrm{EP}} = \widehat\tau_2 - \widehat\Gamma\,\widehat V_{\mathrm{ep}}^{-1}\,(\widehat\tau_{2,\mathrm{ep}} - \widehat\tau_{1,\mathrm{ep}}),
    \end{align*}
    with $\Gamma=\mathrm{Cov}(\widehat\tau_2, \widehat\tau_{2,\mathrm{ep}} - \widehat\tau_{1,\mathrm{ep}})$, and $V_{\mathrm{ep}}=\mathrm{Var}( \widehat\tau_{2,\mathrm{ep}} - \widehat\tau_{1,\mathrm{ep}})$.  We take the experimental sample as the validation sample ($\widehat\tau_2 = \estate$), the observational sample as the main sample, and, following their Remark~3, the OLS treatment coefficient on the shared covariates as the error-prone procedure. If the two error-prone limits differ by $b$ the estimator is biased by $(\Gamma/V_{\mathrm{ep}})\,b$. 
    Columns 12, 24 and 25 of Figure~\ref{fig:bias_level_map} (missing covariates with confounded assignment, IHDP and STAR with omitted covariates) reproduce the setting of their paper: both samples come from one population and only the recorded covariates differ, so the premise holds, the error-prone estimator is unbiased, and its MSE is smaller than CVCI's by a factor of $1.4$ to $4.4$. In the other designs the experimental error-prone estimate is unbiased by randomization while the observational one carries the observational bias, so the premise holds only at zero bias. 
    \item \emph{Precision weighting}: $\widehat\tau_{\mathrm{PW}} = \estate + w\,\widehat\Delta$.
    \item \emph{Pooled OLS}: the coefficient $\widehat\tau$ in the regression $Y_i = \alpha + \tau\,W_i + Z_i^\top\theta_Z + \delta\,S_i + e_i$ on the stacked samples, with $S_i$ the study indicator and $Z_i$ the covariates present in both samples.
    \item \emph{CVCI}: without covariates we use Eq.~\eqref{eq:mean_close_form} with leave-one-out cross-validation. With covariates we use the closed form of Eq.~\eqref{eq:linear_close_form}. $\lambda$ is chosen on a grid of $50$ values in $[0, 1]$ by $5$-fold cross-validation stratified by treatment. 
    
\end{itemize}

\subsection{Synthetic designs in Figure~\ref{fig:bias_level_map}}\label{sec:synthetic_designs_impl}
The true ATE $\tau^\star = 0.5$. 
For the no-covariate setting,
    \begin{align*}
        \res\expm &= \tau^\star + \xi, \qquad \xi \sim \cN(0, 1)\\
        \res\obs &= \tau^\star + \varepsilon + \xi.
    \end{align*}
     In the constant bias setting, $\varepsilon$ is constant. The heavy-tailed variant replaces  $\varepsilon$ by $\varepsilon\,(1 + T)$ with a Student's $t$ distribution $T \sim t_3$.
     
   For the linear setting, each sample draws covariates $\covariate \sim \cN(0, \IdMat_5)$, a coefficient vector $\theta \sim \cN(0, \IdMat_5)$ (afresh in every run, independently for the two samples unless stated), noise $\xi \sim \cN(0, 1)$, and experimental treatment $\tre\expm \sim \mathrm{Bern}(0.5)$ unless otherwise specified. Bias designs are as follows:

    \begin{itemize}[leftmargin=*, itemsep=2pt]
    \item \emph{Constant bias}: $\tre\obs \sim \mathrm{Bern}(0.2)$ and
    \begin{align*}
        \res\expm &= \covariate^\top\theta\expm + \tre\expm\,\tau^\star + \xi, \\
        \res\obs &= \covariate^\top\theta\obs + \tre\obs\,(\tau^\star + \varepsilon) + \xi;
    \end{align*}
    the \emph{same-coefficient} variant sets $\theta\obs = \theta\expm$.
    \item \emph{Heterogeneous bias}: the treatment effect is heterogeneous (with ATE still being $\tau^\star$), and the bias is heterogeneous and dependent on a covariate $Z_2$, 
    \begin{align*}
        \res\expm &= \covariate^\top\theta\expm + \tre\expm\,(\tau^\star + 0.5\,Z_1) + \xi, \\
        \res\obs &= \covariate^\top\theta\obs + \tre\obs\,\bigl(\tau^\star + 0.5\,Z_1 + \varepsilon\,(1 + Z_2)\bigr) + \xi.
    \end{align*}
    
    \item \emph{Heavy-tailed bias}: the bias is sampled from a Student's $t$ distribution,
    \begin{align*}
        \res\expm &= \covariate^\top\theta\expm + \tre\expm\,\tau^\star + \xi,\\
        \res\obs &= \covariate^\top\theta\obs + \tre\obs\,\bigl(\tau^\star + \varepsilon\,(1 + T)\bigr) + \xi, \qquad T \sim t_3.
    \end{align*}
    \item \emph{Misspecified observational model}: 
    \begin{align*}
        \tre\obs \sim \mathrm{Bern}(\mathrm{sigmoid}(-1 + 1.5\,Z_1)).
    \end{align*} In both samples,
    \begin{align*}
        \res &= \covariate^\top\theta + \tre\,\tau^\star + m\,(Z_1^2 - 1) + \xi
    \end{align*}
    with their own $\theta\expm$ and $\theta\obs$.
    The linear working model omits the curvature, which biases $\widehat\tau\obs$ but not $\estate$.
    \item \emph{Missing covariates}: in both samples
    \begin{align*}
        \res &= \covariate^\top\theta + \tre\,\tau^\star + \xi
    \end{align*}
    with their own $\theta\expm$ and $\theta\obs$.
    In the observational sample,
    \begin{align*}
        \tre\obs \sim \mathrm{Bern}\bigl(\mathrm{sigmoid}(\eta\,(Z_4 + Z_5))\bigr)
    \end{align*}
    and the observational sample records only $Z_1, Z_2, Z_3$. The sign of the bias changes with $\theta\obs$ across runs, so the Monte Carlo bias averages to zero and the bias cells of this design are blank in Figure~\ref{fig:bias_level_map}.
    \item \emph{Missing covariates, confounded assignment}: with a shared $\theta$, in both samples,  
    \begin{align*}
     \res &= \covariate^\top\theta + \tre\tau^\star + \xi,    \\
        \tre &\sim \mathrm{Bern}\bigl(\mathrm{sigmoid}(\eta\,(Z_4 + Z_5))\bigr).
    \end{align*}
    The experimental sample records all five covariates, and the observational sample records only $Z_1, Z_2, Z_3$.
    \item \emph{Unmeasured confounding}: an unobserved $U \sim \cN(0, 1)$ enters the outcome of both samples and the observational treatment,
    \begin{align*}
        \res &= \covariate^\top\theta + \tre\,\tau^\star + U + \xi, \\
        \tre\obs &\sim \mathrm{Bern}\bigl(\mathrm{sigmoid}(-1 + \gamma U)\bigr),
    \end{align*}
    with the confounding strength $\gamma$ on $21$ equally spaced values in $[0, 1.17]$ at $(N\expm, N\obs) = (50, 500)$ and in $[0, 0.195]$ at $(N\expm, N\obs) = (1000, 2000)$.
\end{itemize}
The population bias of each setting is computed once by Monte Carlo. 
We repeat 5000 runs for each setting of each design.

\begin{figure}[!tbp]
    \ifjrssb
    \vspace{0.5mm}
    \fi
    \centering
    \includegraphics[width=\textwidth, height=0.74\textheight, keepaspectratio]{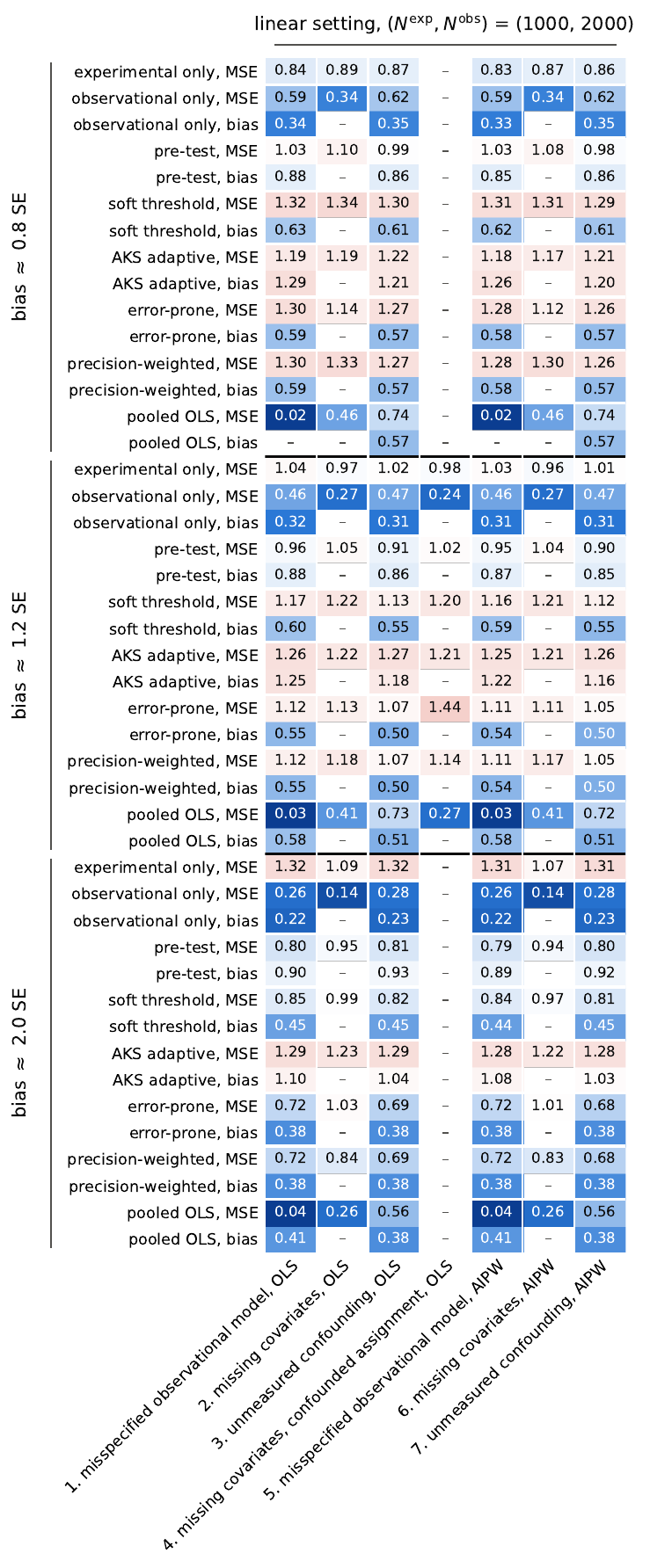}
    \caption{Ratios of CVCI to the row estimator for the linear designs with $(N\expm, N\obs) = (1000, 2000)$, in the layout of the synthetic columns of Figure~\ref{fig:bias_level_map}.}
    \label{fig:bias_level_map_large}
\end{figure}

\subsection{Real-data designs}
For IHDP \citep{hill2011bayesian}, the trial has $747$ children, $25$ covariates and $18.6\%$ treated, with Hill's nonlinear response surface for outcomes. For STAR \citep{krueger1999experimental}, the trial has $3730$ kindergarten students in small or regular classes, $7$ covariates and $46\%$ treated, with the sum of the reading and mathematics scores as the  outcome. 
Designs use residual sampling: fit a linear model $Y_i = \alpha + \tau\,W_i + Z_i^\top\theta_Z + e_i$ on the whole trial, obtain the residuals with fitted values $\widehat f_w(Z_i) = \widehat\alpha + \widehat\tau\,w + Z_i^\top\widehat\theta_Z$. Denote the residual standard deviation as $s$. Each run redraws the outcomes with covariates fixed. Then $\tau^\star = \widehat\tau$. The designs include:
\begin{itemize}[leftmargin=*, itemsep=2pt]
    \item \emph{Confounding by selection on the outcome residual} (columns with $\gamma \in \{0.1, 0.2, 0.3\}$): 20\% of the trial keeps its randomized treatment and forms the experimental sample with
    \begin{align*}
        \res\expm_i &= \widehat f_{\tre_i}(Z_i) + s\,\xi_i, \qquad \xi_i \sim \cN(0, 1). 
    \end{align*}
    For the rest, a standard normal $u_i$ drives both the treatment and the outcome to form the observational sample
    \begin{align*}
        \tre\obs_i &\sim \mathrm{Bern}\bigl(\mathrm{sigmoid}(c + \gamma\,u_i)\bigr), \\
        \res\obs_i &= \widehat f_{\tre\obs_i}(Z_i) + s\,u_i,
    \end{align*}
    with $c$ the logit of the trial's treated share.
    \item \emph{Omitted covariates}: select $k$ covariates with the largest standardized outcome coefficients ($k = 5$ for IHDP, $2$ for STAR). Denote $u_i$ as a standardized one-dimensional summary of the contribution of those $k$ covariates to the outcome. In both samples,
    \begin{align*}
        \tre_i &\sim \mathrm{Bern}\bigl(\mathrm{sigmoid}(c + \gamma\,u_i)\bigr), \\
        Y_i &= \widehat f_{\tre_i}(Z_i) + s\,\xi_i.
    \end{align*}
 The experimental sample records all covariates. The observational sample records $Z$ without the $k$ omitted ones. We set $\gamma \in \{0.2, 0.4, 0.8\}$.
\end{itemize}


\section{No-covariate Experiments: Implementation Details and Additional Results} \label{sec:mean_supp}

\subsection{Implementation details} \label{sec:implementation_mean}
 We utilize the closed-form solution in Eq.~\eqref{eq:mean_close_form}. We set $\tau^\star=0.5$, where the specific value chosen would not affect the qualitative results. For cross-validation, we set $K=N\expm$ and conduct a grid search over candidate values of $\lambda \in [0, 1]$ in 50 linearly spaced bins. The t-test baseline is as follows: the null hypothesis is that the two populations have the same mean, while the alternative is that their means differ. If it rejects the null hypothesis, we set $\lambda = 0$ to rely solely on experimental samples. Otherwise, we set $\lambda = N\obs / (N\expm+N\obs)$ to incorporate both sources. For experiments varying $N\obs$ (or $N\expm$), we generate a large observational (or experimental) dataset and draw random subsets of the desired size for each run. We repeat 5000 runs for each experiment.
 
 For figure production, the insets display zoomed-in views of the plots over small values of $\varepsilon$ to highlight the performance gains in that region, and over an intermediate range to provide a closer examination of the model's behavior: $\varepsilon \in [0, 0.2]$ and $\varepsilon \in [0.53, 1.47]$ in Figures~\ref{fig:c_mean_mse_eps} and \ref{fig:d_mean_mse_eps} ($\sigma^2=1$), and $\varepsilon \in [0, 2]$ and $\varepsilon \in [5.3, 14.7]$ in Figures~\ref{fig:c_mean_mse_eps_var_10} and \ref{fig:d_mean_mse_eps_var_10} ($\sigma^2=100$). In Figures~\ref{fig:e_mean_mse_n_obs}, \ref{fig:f_mean_mse_n_exp}, \ref{fig:e_mean_mse_n_obs_var_10}, and \ref{fig:f_mean_mse_n_exp_var_10}, we apply a continuous piecewise transformation to the vertical axis to improve visual clarity. Specifically, values below a threshold $b$ are scaled linearly, while values above $b$ are log-transformed relative to the threshold. This transformation takes the form  
\[
\text{stretch}(y) = 
\begin{cases}
a \cdot \frac{y}{b}, & y \leq b \\
a + \log\left(\frac{y}{b}\right), & y > b
\end{cases}
\]
where $a$ controls the intensity of the stretch and ensures continuity at the transition point $y = b$. This approach preserves detail for small values while compressing the dynamic range of larger values, making trends and comparisons more visually accessible. The transformation is invertible, allowing us to recover the original values on the vertical axis.
We set $b$ to be the maximum of our method's empirical MSE, and $a$ to be $5$.

\subsection{Additional results} \label{sec:additional_results_mean}
\begin{figure}[H]
    \ifjrssb
    \vspace{0.5mm}
    \fi
    \centering
    \captionsetup[sub]{justification=centering}
    \subfloat[Same settings as Figure~\ref{fig:b_mean_mse_eps}. Inset: Zoom in.]{%
        \includegraphics[width=0.48\textwidth]{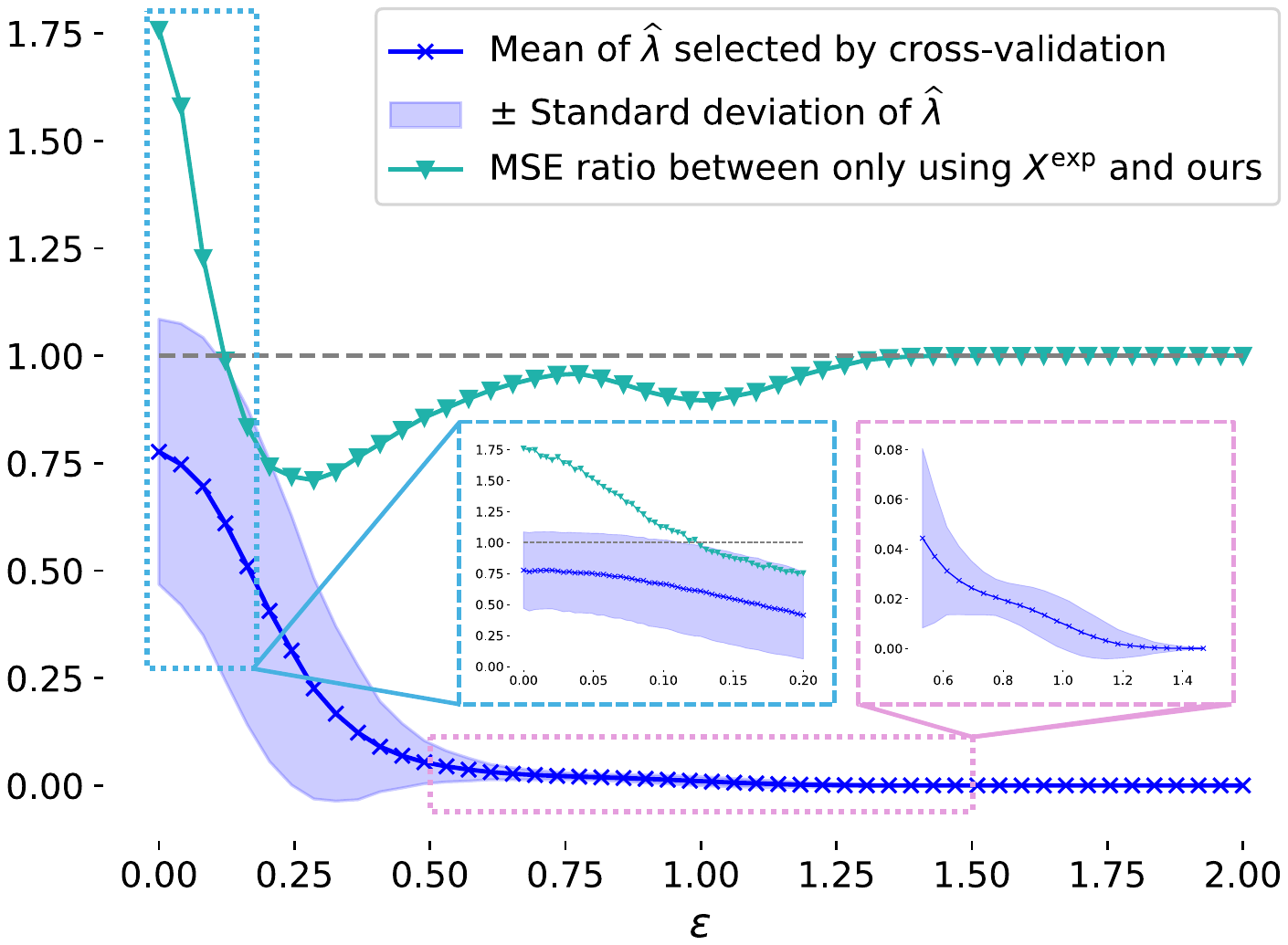}%
        \label{fig:d_mean_mse_eps}%
        }\hfill
    \subfloat[$N\obs=150$, $\varepsilon=0.1$, $\sigma^2=1$.]{%
        \includegraphics[width=0.48\textwidth]{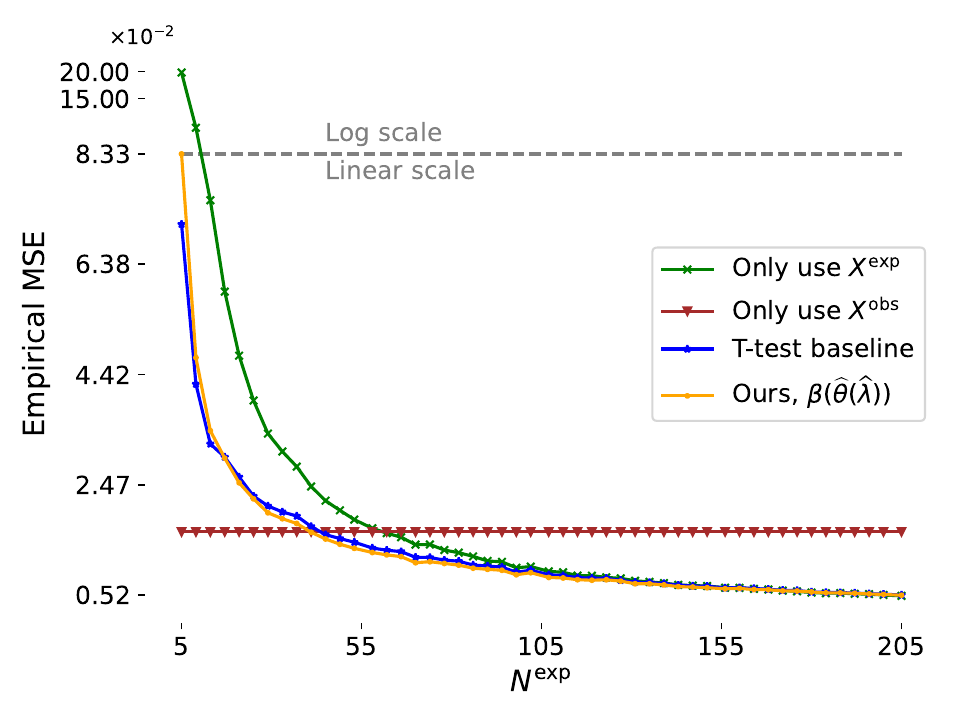}%
        \label{fig:f_mean_mse_n_exp}%
        }
    \caption{No-covariate setting, continued from Figure~\ref{fig:mean_mse_eps_var_1}. (a) Empirical MSE ratio of the experimental-only estimate to CVCI, and selected $\widehat\lambda$, varying $\varepsilon$ at $N\obs=200$. (b) Empirical MSE varying $N\expm$ (with a linear–log transformation for visual clarity), where we see that CVCI consistently outperforms the $X\expm$-only baseline when the bias is small.}
    \label{fig:mean_mse_nocov_extra}%
\end{figure}

\begin{figure}[H]
    \ifjrssb
    \vspace{0.5mm}
    \fi
    \centering
    \captionsetup[sub]{justification=centering}
        \subfloat[$N\expm=100$, $N\obs=5000$, $\sigma^2=100$. ]{%
        \includegraphics[width=0.48\textwidth]{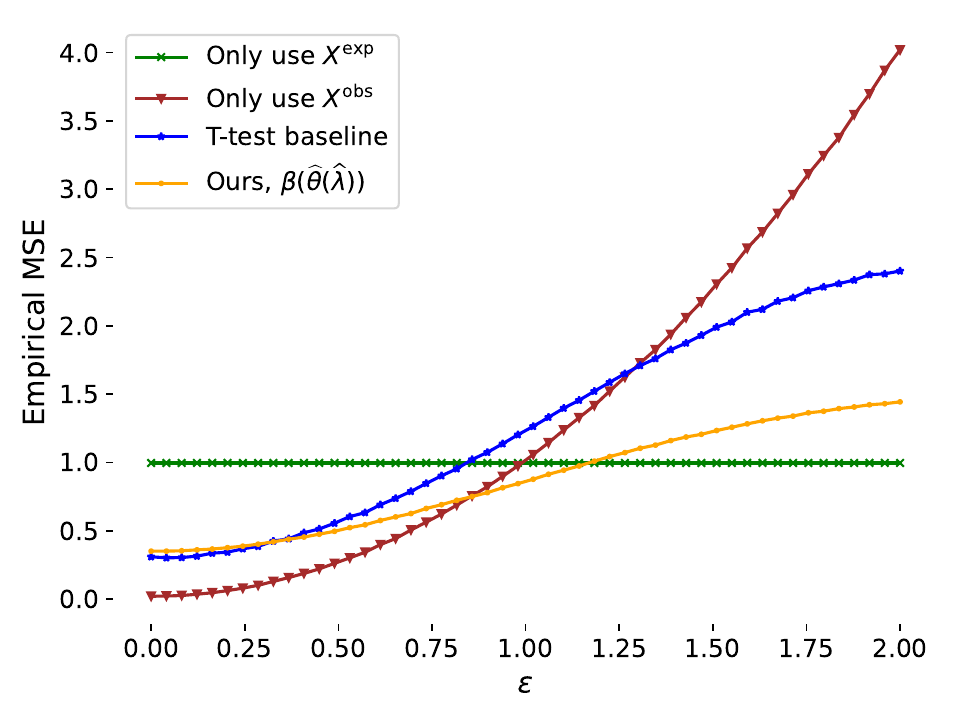}%
        \label{fig:a_mean_mse_eps_var_10}%
        }\hfill
    \subfloat[$N\expm=100$, $N\obs=200$, $\sigma^2=100$.]{%
        \includegraphics[width=0.48\textwidth]{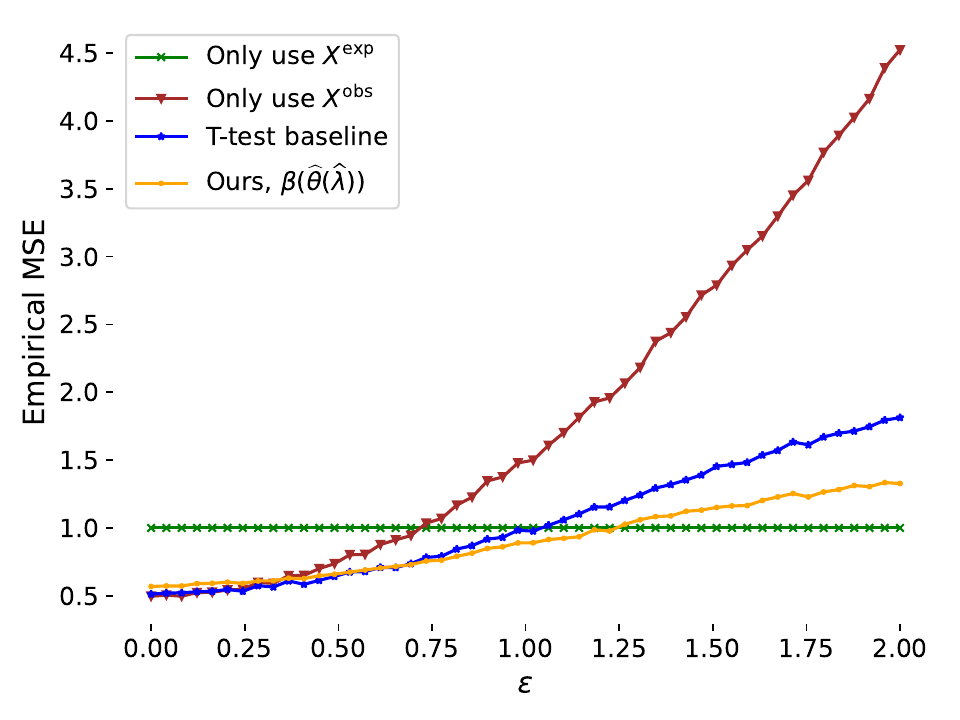}%
        \label{fig:b_mean_mse_eps_var_10}%
        }\\
    \subfloat[Same settings as (a). Inset: Zoom in.]{%
        \includegraphics[width=0.48\textwidth]{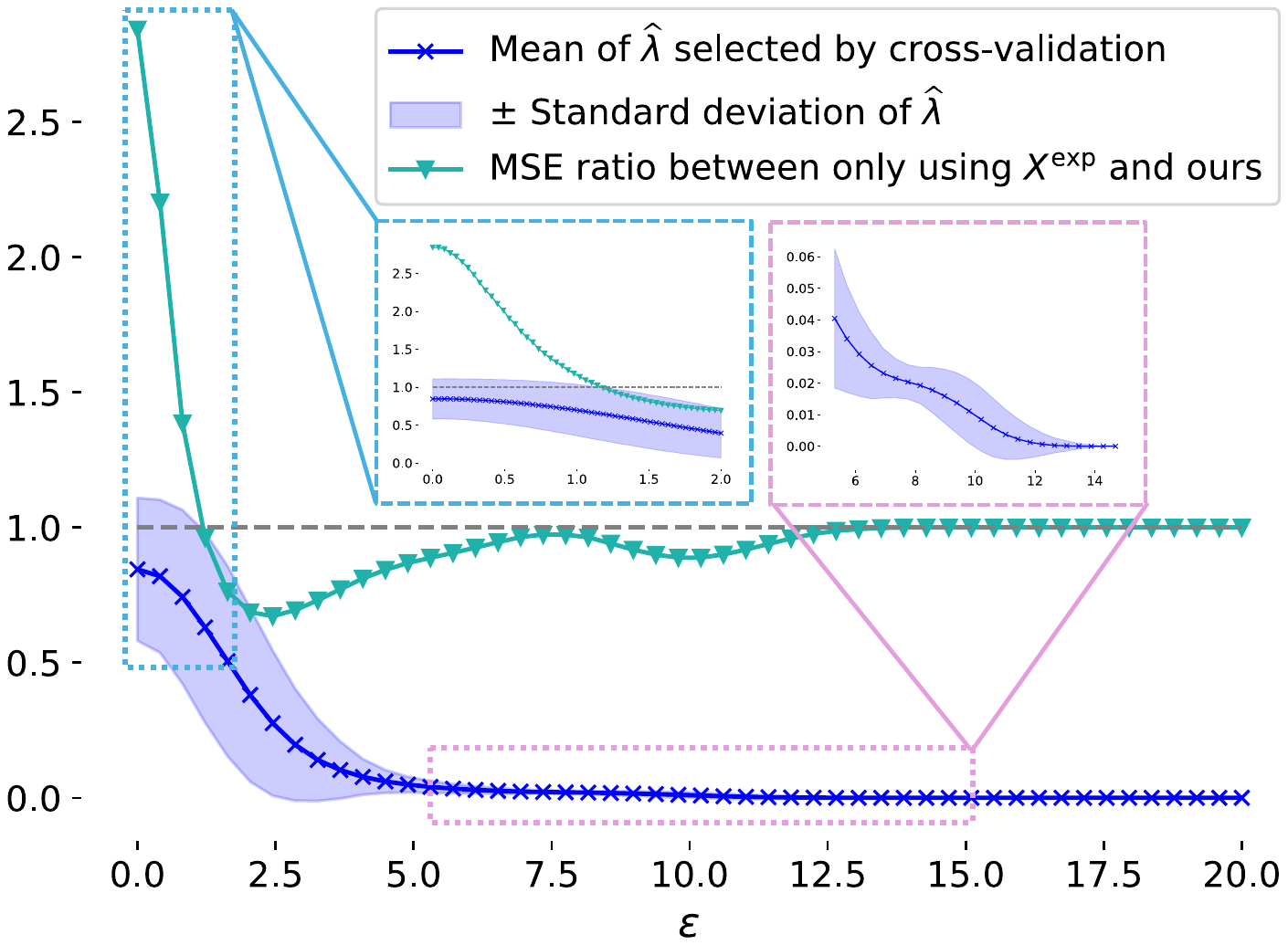}%
        \label{fig:c_mean_mse_eps_var_10}%
        }\hfill
    \subfloat[Same settings as (b). Inset: Zoom in.]{%
        \includegraphics[width=0.48\textwidth]{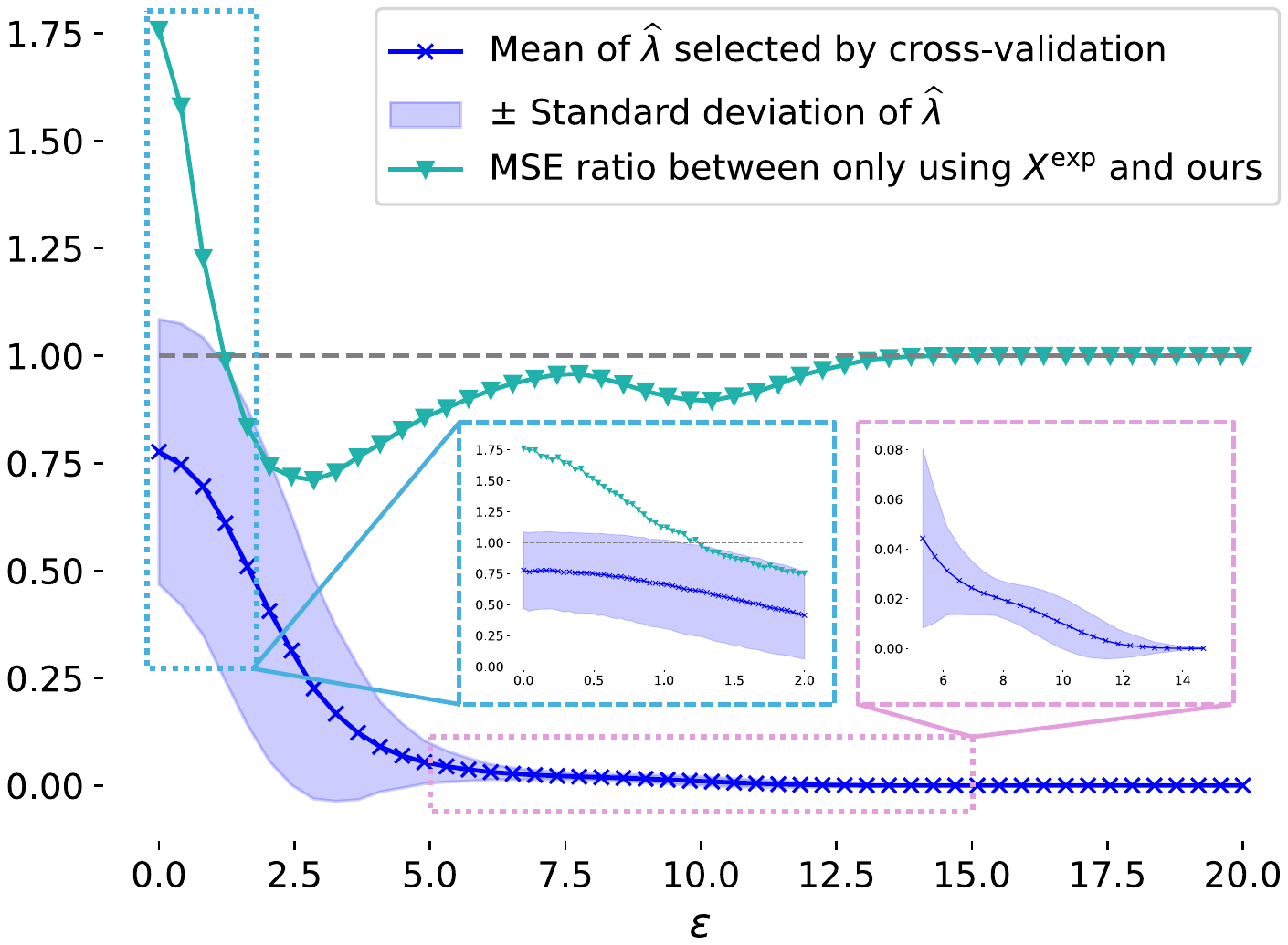}%
        \label{fig:d_mean_mse_eps_var_10}%
        }\\
     \subfloat[$N\expm=100$, $\varepsilon=1$, $\sigma^2=100$. ]{%
        \includegraphics[width=0.48\textwidth]{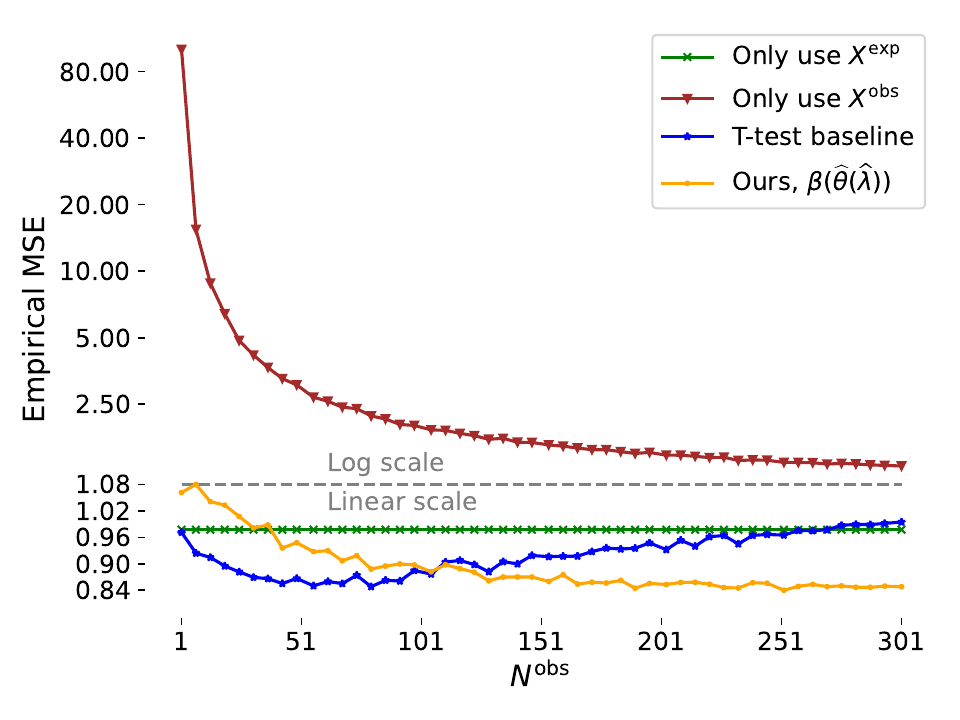}%
        \label{fig:e_mean_mse_n_obs_var_10}%
        }\hfill
    \subfloat[$N\obs=150$, $\varepsilon=1$, $\sigma^2=100$. ]{%
        \includegraphics[width=0.48\textwidth]{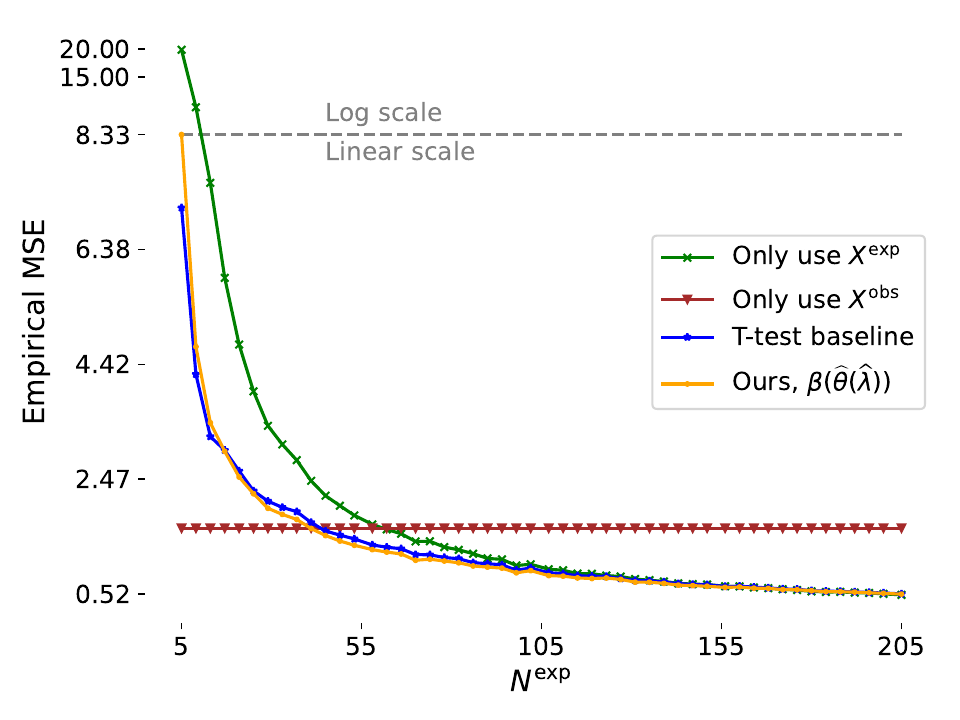}%
        \label{fig:f_mean_mse_n_exp_var_10}%
        }
    \caption{No-covariate setting. Same setup as Figures~\ref{fig:mean_mse_eps_var_1} and~\ref{fig:mean_mse_nocov_extra}, but with $\sigma^2=100$. Panels (a-c) and (e) correspond to Figures~\ref{fig:a_mean_mse_eps}--\ref{fig:c_mean_mse_eps} and~\ref{fig:e_mean_mse_n_obs}; panels (d) and (f) correspond to Figures~\ref{fig:d_mean_mse_eps} and~\ref{fig:f_mean_mse_n_exp}.}
    \label{fig:mean_mse_var_10}%
\end{figure}

Figure~\ref{fig:mean_mse_nocov_extra} gives the two remaining panels of the $\sigma^2=1$ setting: the MSE ratio at $N\obs=200$, the counterpart of Figure~\ref{fig:c_mean_mse_eps}, and the MSE as $N\expm$ varies.
Raising the noise level $\sigma^2$ from 1 to 100, we observe that each sub-figure in Figure~\ref{fig:mean_mse_var_10} mirrors its counterpart in Figures~\ref{fig:mean_mse_eps_var_1} and~\ref{fig:mean_mse_nocov_extra}. While the overall behaviors remain qualitatively unchanged, the MSEs scale up by a factor of roughly 100. This is due to the bias-variance decomposition of MSE, where the variance component dominates as the noise level increases. The scaling also shifts the threshold of $\varepsilon$ beyond which biased observational data lose their utility: from $\varepsilon \approx 0.125$ in Figures~\ref{fig:a_mean_mse_eps} and \ref{fig:b_mean_mse_eps} to $\varepsilon \approx 1.25$ in Figures~\ref{fig:a_mean_mse_eps_var_10} and \ref{fig:b_mean_mse_eps_var_10}. 

\section{Linear Setting: Implementation Details and Additional Results}\label{sec:linear_supp}
\subsection{Implementation details} \label{sec:implementation_linear}
We utilize the closed-form solution in Eq.~\eqref{eq:linear_close_form}.  For each experimental and observational sample, we independently generate the covariates $ \covariate \sim \cN (0, \sigma^2 \IdMat)$, the treatment $\tre \sim \mathrm{Bern}(0.5)$ for experimental data and $\mathrm{Bern}(0.2)$ for observational data, and an exogenous noise $\xi \sim \cN (0, \sigma^2) \perp \covariate, \tre$. We set $\sigma^2=1$. For experimental samples, the response is generated as
$\res = \covariate^\top \theta\expm + \tre \tau^\star + \xi$. We set $\tau^\star=0.5$.
For observational samples, we introduce the bias via $\res = \covariate^\top \theta\obs + \tre (\tau^\star + \varepsilon) + \xi$. 
The weights of $\theta\expm$ and $\theta\obs$ are sampled from a multivariate normal distribution $\cN(0, \IdMat)$. We then append a 1 to each $Z$ and a $0$ to $\theta\expm$ and $\theta\obs$ to account for the intercept term. The dimensions of $\theta\expm$ and $\theta\obs$ are set to 6 (including the intercept). For the experiments varying $\varepsilon$, weights are sampled independently for each simulation. For experiments varying $N\obs$, weights are sampled once to generate a large observational dataset, from which random subsets of the desired size are drawn in each run. For cross-validation, we set $K=5$ and use 50 linearly spaced bins for candidate values of $\lambda$.
To calculate the experimental estimate $\estate$, we employ the average of AIPW estimates with a known propensity score (0.5). A linear outcome model is fit on half of the experimental data, and the AIPW estimates are computed using the remaining half.
When splitting the data either for computing the AIPW estimate or for cross-validation, we stratify by treatment assignment, resulting in each fold containing approximately 50\% treated and 50\% control samples. We repeat 5000 runs for each experiment.

For figure production, the insets in Figures~\ref{fig:b_linear_mse_ratio}, \ref{fig:d_linear_mse_ratio} and \ref{fig:a_linear_mse_ratio}--\ref{fig:b_linear_mse_ratio_app} provide zoomed-in views: over small values of $\varepsilon$ to highlight performance gains, and over the range $\varepsilon \in [24.49, 118.37]$ to enable a closer examination of the model's behavior. We apply the same linear-log transformation described in Section~\ref{sec:implementation_mean} to Figure~\ref{fig:linear_mse_n_obs_extra}, where $N\obs$ varies. The threshold $b$ is set to the maximum MSE of our method. The transformation intensity parameter $a$ is set to 3 in Figures~\ref{fig:e_linear_mse_n_obs} and \ref{fig:c_linear_mse_n_obs_same-cov}, and to 5 in Figures~\ref{fig:f_linear_mse_n_obs} and \ref{fig:d_linear_mse_n_obs_same-cov}.

\subsection{Additional results}
\label{sec:additional_results_linear}

\begin{figure}[H]
    \ifjrssb
    \vspace{0.5mm}
    \fi
    \centering
    \captionsetup[sub]{justification=centering}
    \subfloat[$\theta\expm \neq \theta\obs$, $N\expm=50$, $N\obs=500$.]{%
        \includegraphics[width=0.48\textwidth]{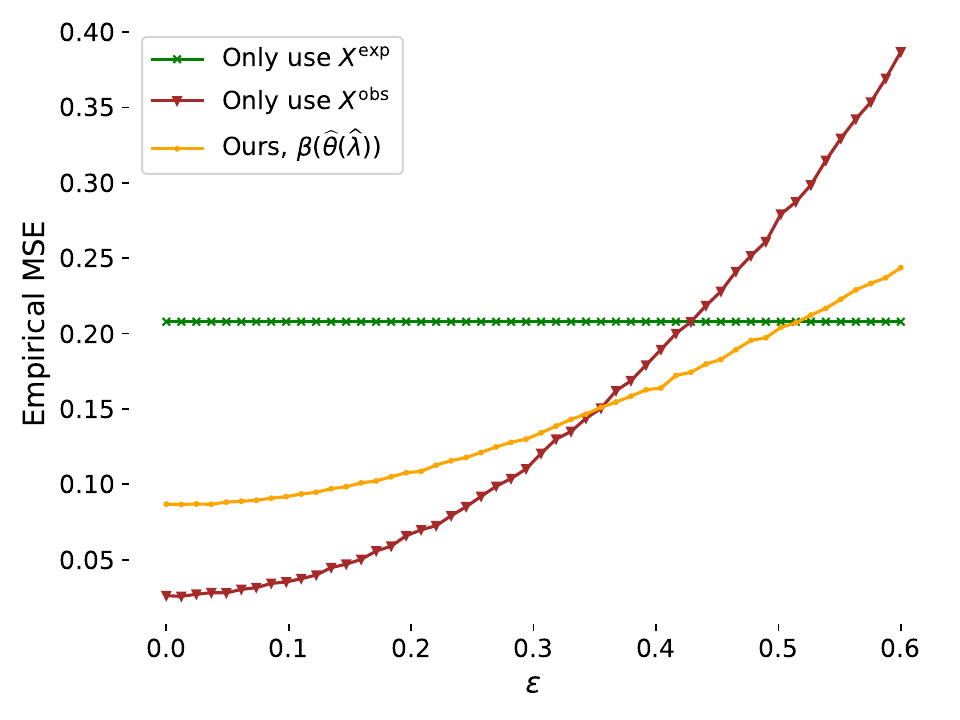}%
        \label{fig:c_linear_mse_eps}%
        }\hfill
    \subfloat[$\theta\expm = \theta\obs$, $N\expm=50$, $N\obs=500$.]{%
        \includegraphics[width=0.48\textwidth]{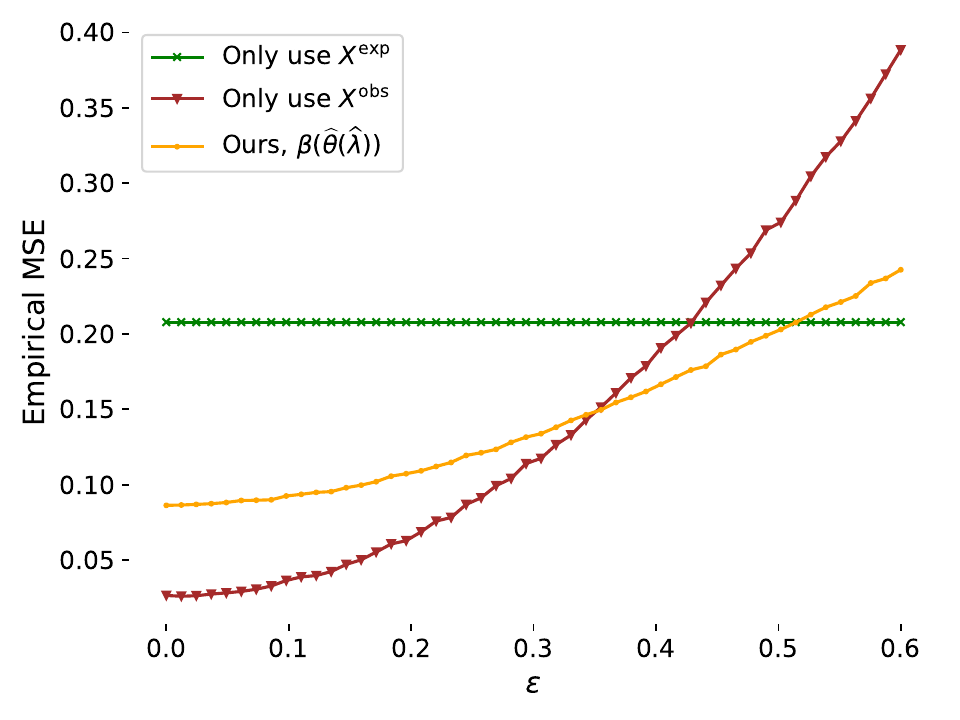}%
        \label{fig:a_linear_mse_eps_same-cov}%
        }\\
    \subfloat[$\theta\expm \neq \theta\obs$, $N\expm=1000$, $N\obs=2000$.]{%
        \includegraphics[width=0.48\textwidth]{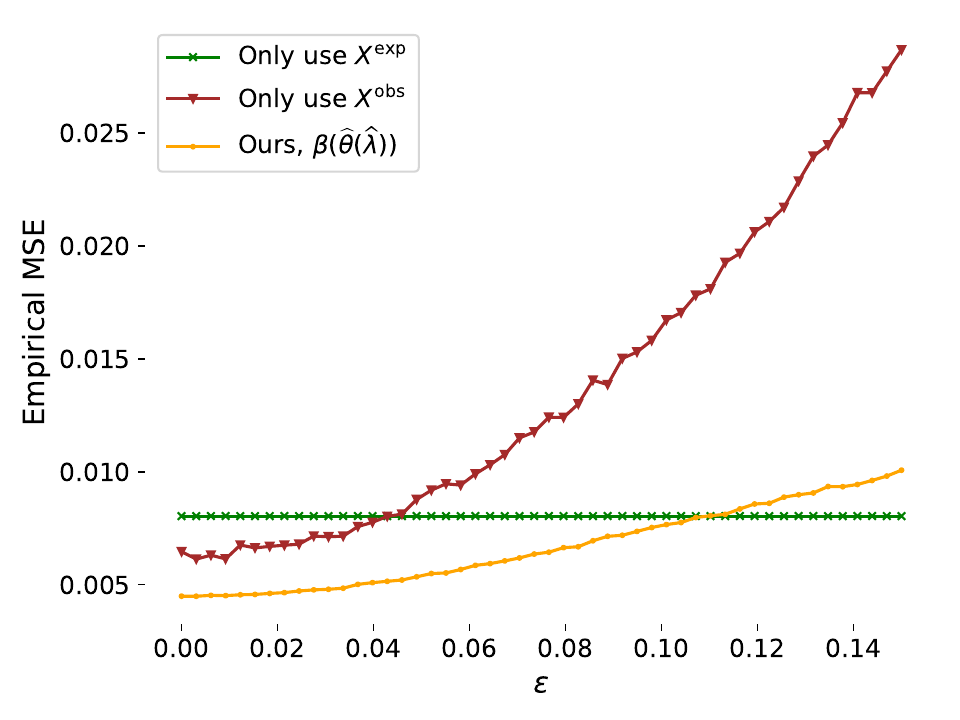}%
        \label{fig:d_linear_mse_eps}%
        }\hfill
    \subfloat[$\theta\expm = \theta\obs$, $N\expm=1000$, $N\obs=2000$.]{%
        \includegraphics[width=0.48\textwidth]{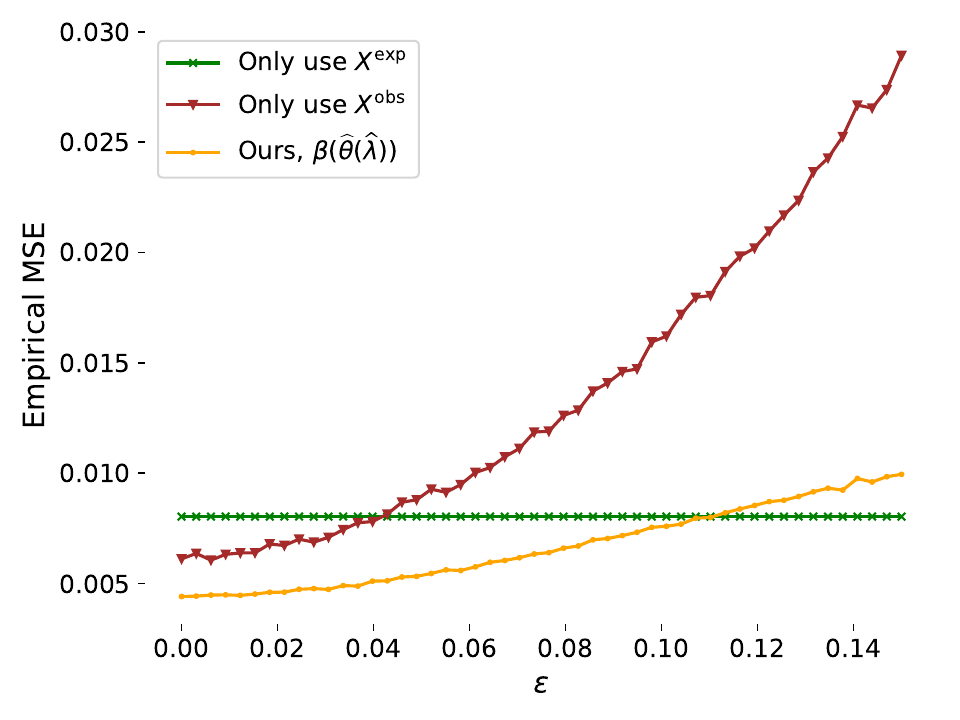}%
        \label{fig:b_linear_mse_eps_same-cov}%
        }\\
    \subfloat[$\theta\expm = \theta\obs$, $N\expm=50$, $N\obs=100$.]{%
        \includegraphics[width=0.48\textwidth]{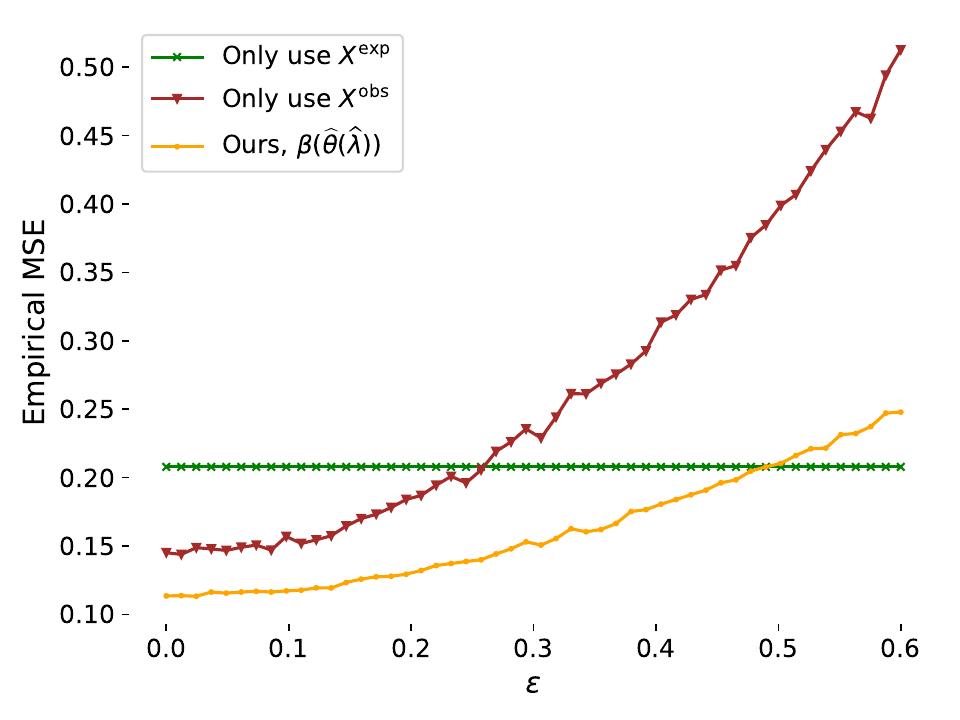}%
        \label{fig:a_linear_mse_eps}%
        }
    \caption{Linear setting, empirical MSE varying $\varepsilon$. Panels (b) and (d) are the $\theta\expm = \theta\obs$ counterparts of (a) and (c); panel (e) is the $\theta\expm = \theta\obs$ counterpart of Figure~\ref{fig:b_linear_mse_eps}.}
    \label{fig:linear_mse_eps_extra}%
\end{figure}

\begin{figure}[H]
    \ifjrssb
    \vspace{0.5mm}
    \fi
    \centering
    \captionsetup[sub]{justification=centering}
    \subfloat[$\theta\expm \neq \theta\obs$, $N\expm=50$, $\varepsilon=0.05$.]{%
        \includegraphics[width=0.48\textwidth]{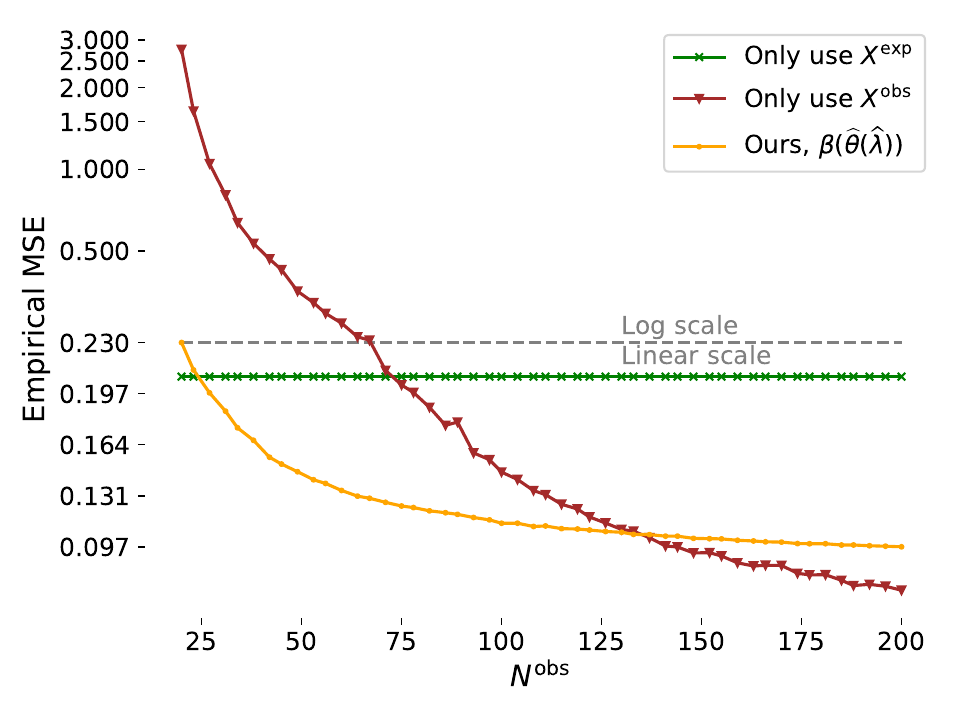}%
        \label{fig:e_linear_mse_n_obs}%
        }\hfill
    \subfloat[$\theta\expm \neq \theta\obs$, $N\expm=1000$, $\varepsilon=0.05$.]{%
        \includegraphics[width=0.48\textwidth]{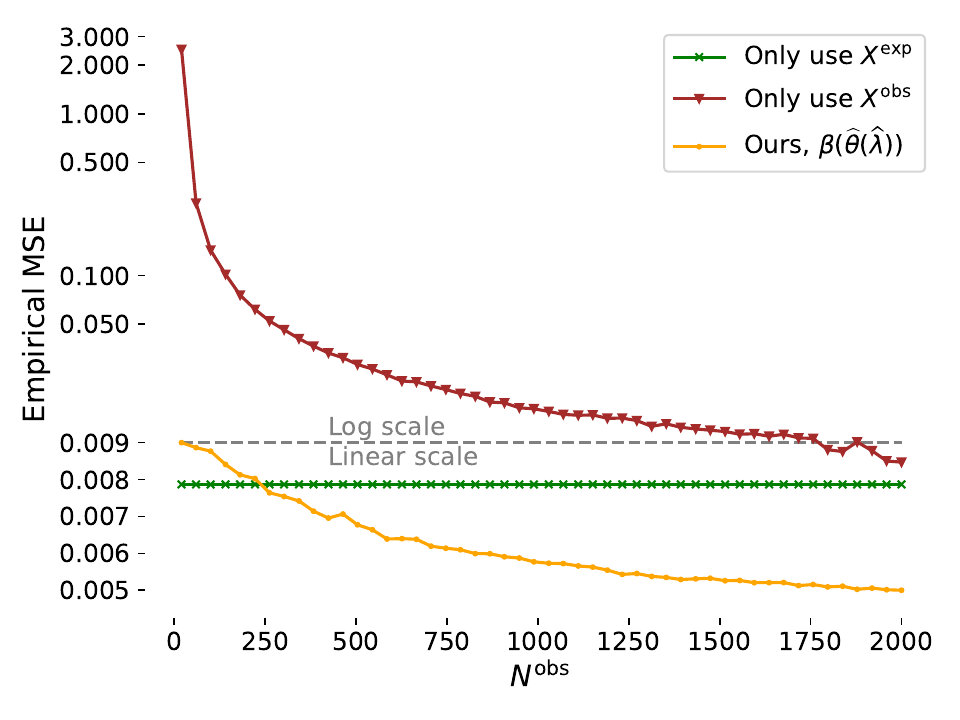}%
        \label{fig:f_linear_mse_n_obs}%
        }\\
    \subfloat[$\theta\expm = \theta\obs$, $N\expm=50$, $\varepsilon=0.05$.]{%
        \includegraphics[width=0.48\textwidth]{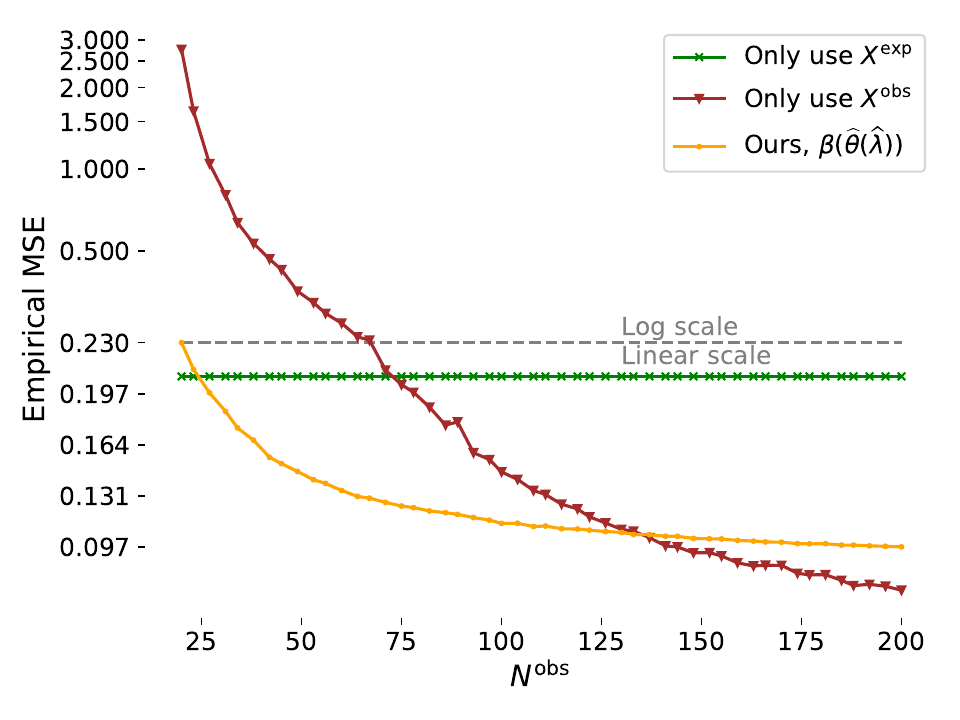}%
        \label{fig:c_linear_mse_n_obs_same-cov}%
        }\hfill
    \subfloat[$\theta\expm = \theta\obs$, $N\expm=1000$, $\varepsilon=0.05$.]{%
        \includegraphics[width=0.48\textwidth]{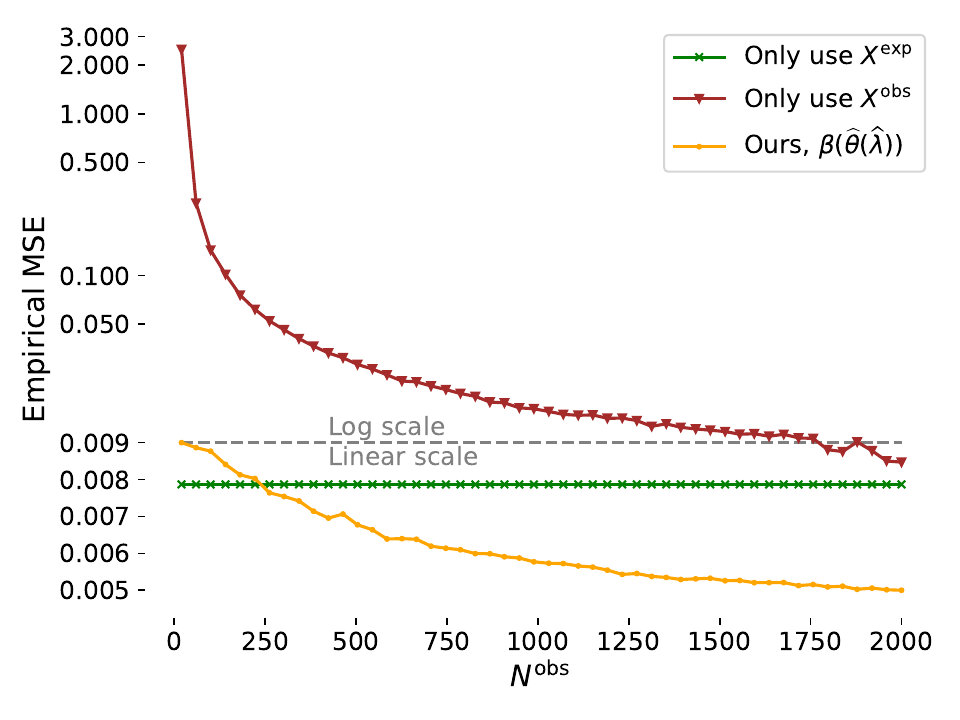}%
        \label{fig:d_linear_mse_n_obs_same-cov}%
        }
    \caption{Linear setting, empirical MSE varying $N\obs$ at $\varepsilon=0.05$; we apply a linear–log transformation for visual clarity. Panels (c-d) are the $\theta\expm = \theta\obs$ counterparts of (a-b).}
    \label{fig:linear_mse_n_obs_extra}%
\end{figure}

\begin{figure}[H]
    \ifjrssb
    \vspace{0.5mm}
    \fi
    \centering
    \captionsetup[sub]{justification=centering}
    \subfloat[$\theta\expm = \theta\obs$, $N\expm=1000$, $N\obs=2000$.]{%
        \includegraphics[width=0.48\textwidth]{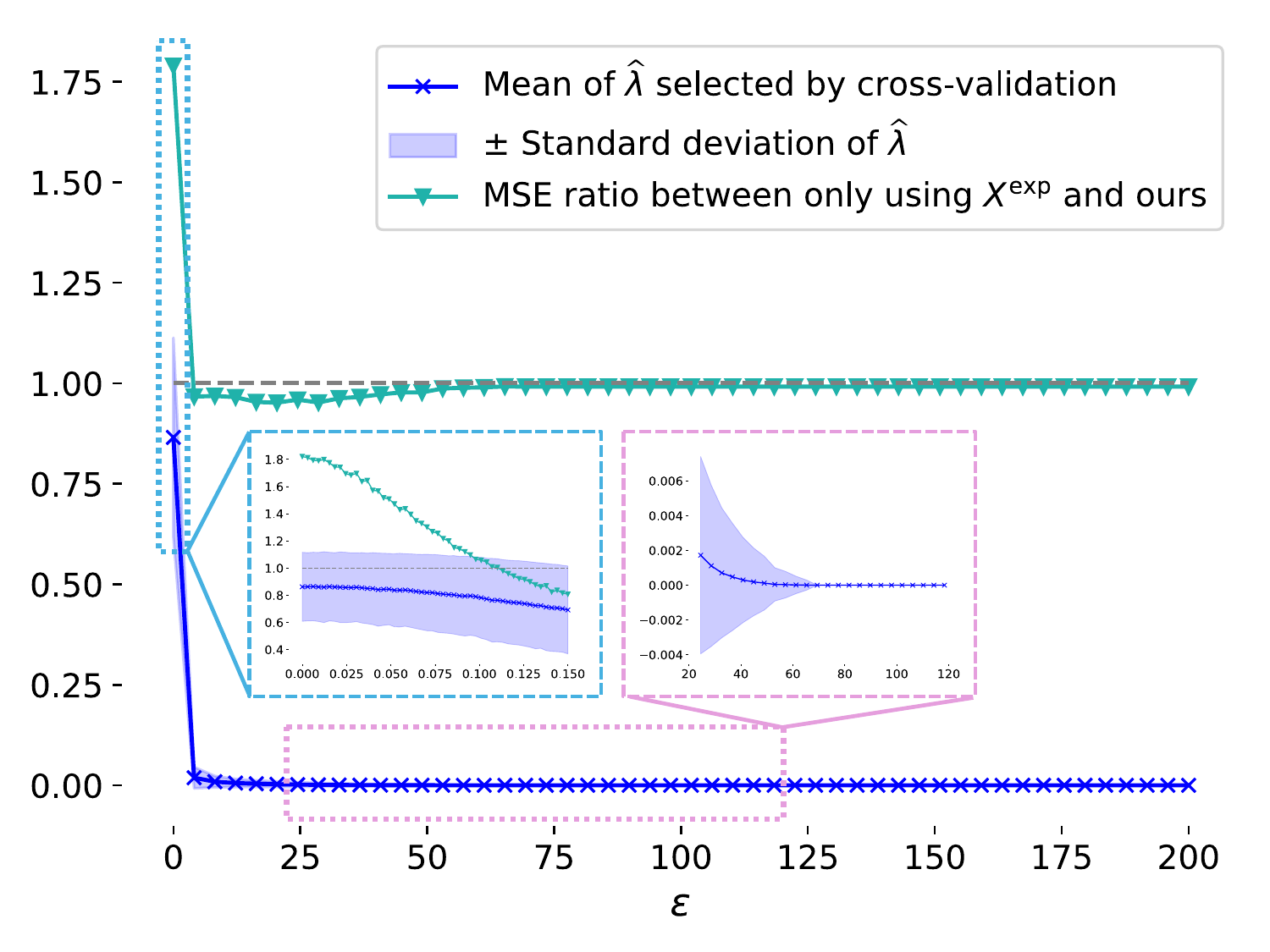}%
        \label{fig:a_linear_mse_ratio}%
        }\hfill
    \subfloat[$\theta\expm = \theta\obs$, $N\expm=50$, $N\obs=500$.]{%
        \includegraphics[width=0.48\textwidth]{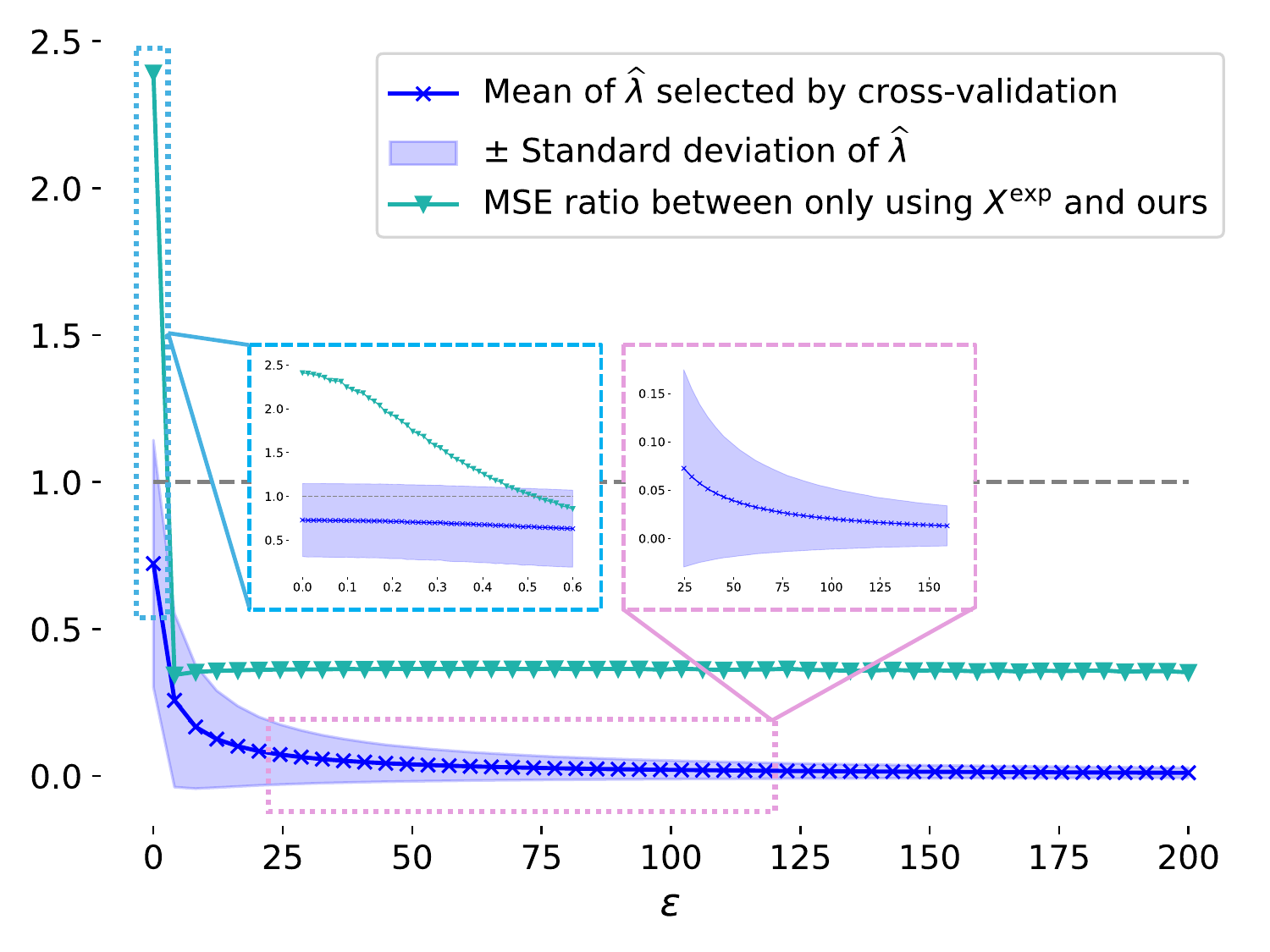}%
        \label{fig:c_linear_mse_ratio}%
        }\\
    \subfloat[$\theta\expm = \theta\obs$, $N\expm=500$, $N\obs=5000$.]{%
        \includegraphics[width=0.48\textwidth]{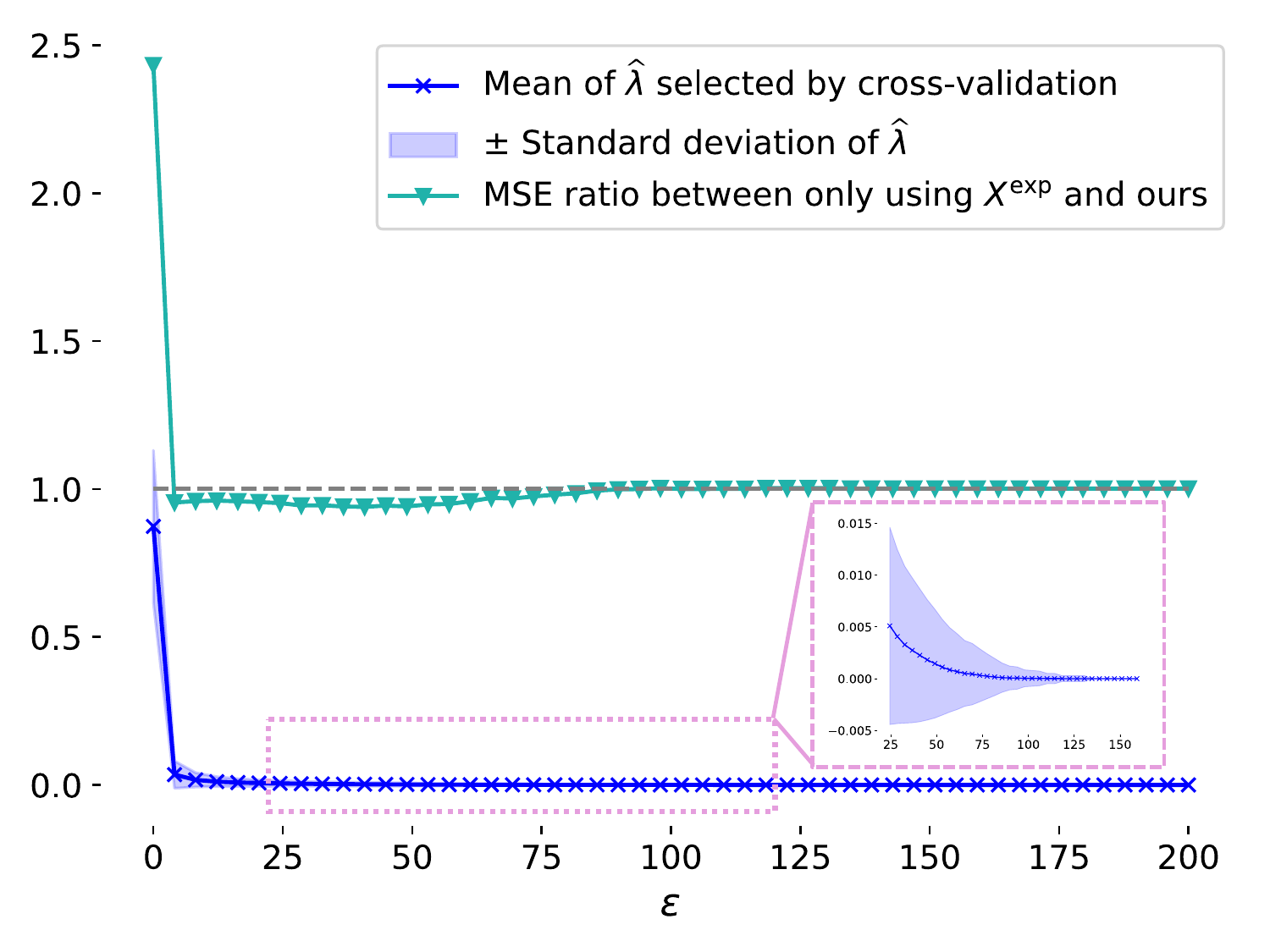}%
        \label{fig:a_linear_mse_ratio_app}%
        }\hfill
    \subfloat[$\theta\expm \neq \theta\obs$, $N\expm=500$, $N\obs=5000$.]{%
        \includegraphics[width=0.48\textwidth]{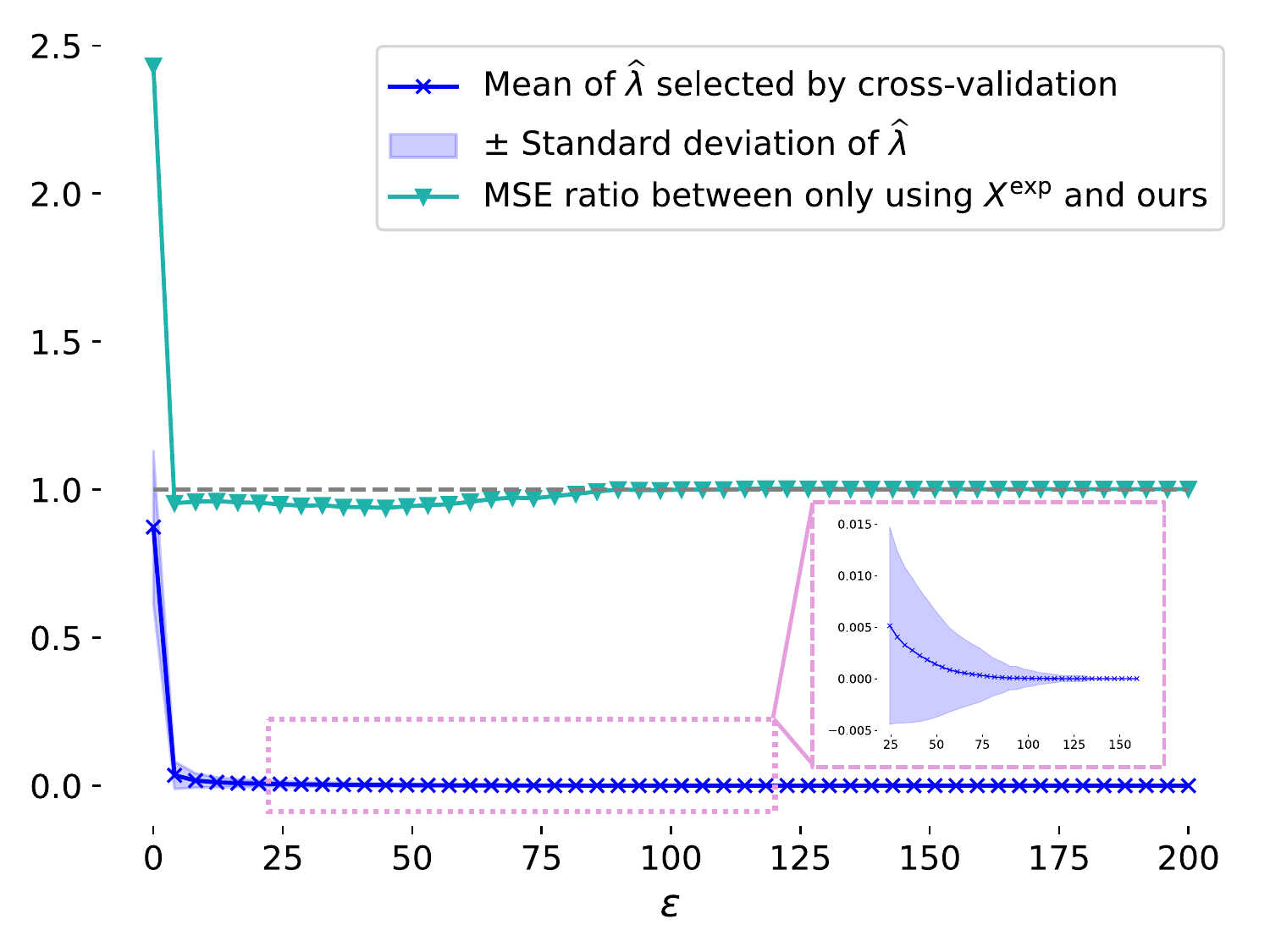}%
        \label{fig:b_linear_mse_ratio_app}%
        }
    \caption{Linear setting, large-bias results. Empirical MSE ratio of the experimental-only estimate to CVCI, and selected $\widehat\lambda$, varying $\varepsilon$. Panels (a) and (b) are the $\theta\expm = \theta\obs$ counterparts of Figures~\ref{fig:b_linear_mse_ratio} and~\ref{fig:d_linear_mse_ratio}; panels (c) and (d) use $(N\expm, N\obs)=(500, 5000)$.}
    \label{fig:linear_mse_ratio_extra}%
\end{figure}

\begin{figure}[H]
    \ifjrssb
    \vspace{0.5mm}
    \fi
    \centering
    \captionsetup[sub]{justification=centering}
    \subfloat[Squared errors, $\theta\expm \neq \theta\obs$.]{%
        \includegraphics[width=0.48\textwidth]{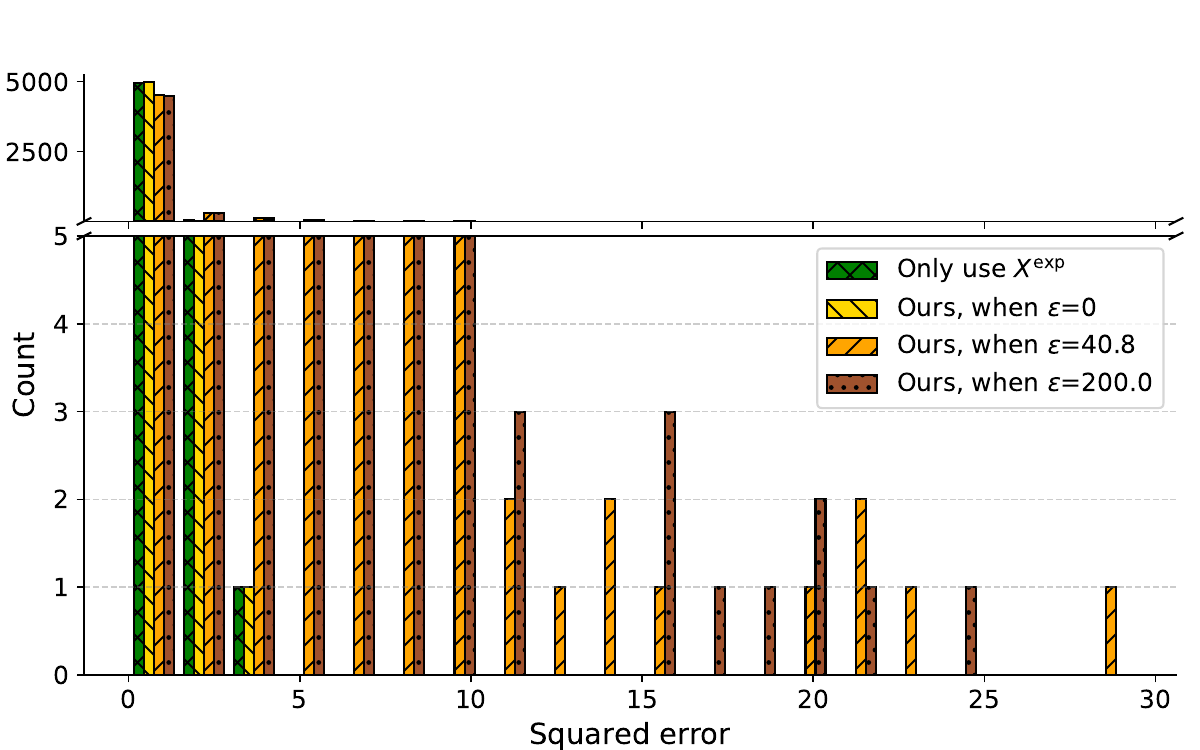}%
        \label{fig:e_linear_mse_se_dist}%
        }\hfill
    \subfloat[Squared errors, $\theta\expm = \theta\obs$.]{%
        \includegraphics[width=0.48\textwidth]{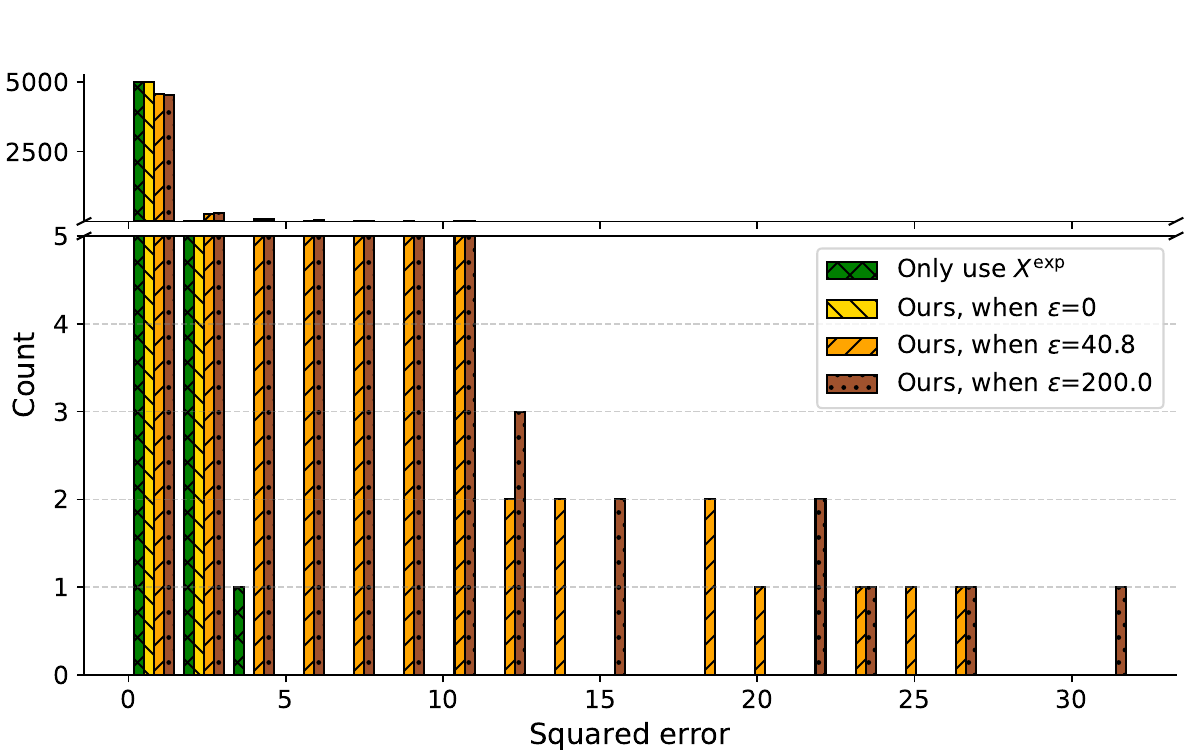}%
        \label{fig:e_linear_mse_se_dist_same-cov}%
        }\\
    \subfloat[$\widehat\lambda$, $\theta\expm = \theta\obs$.]{%
        \includegraphics[width=0.48\textwidth]{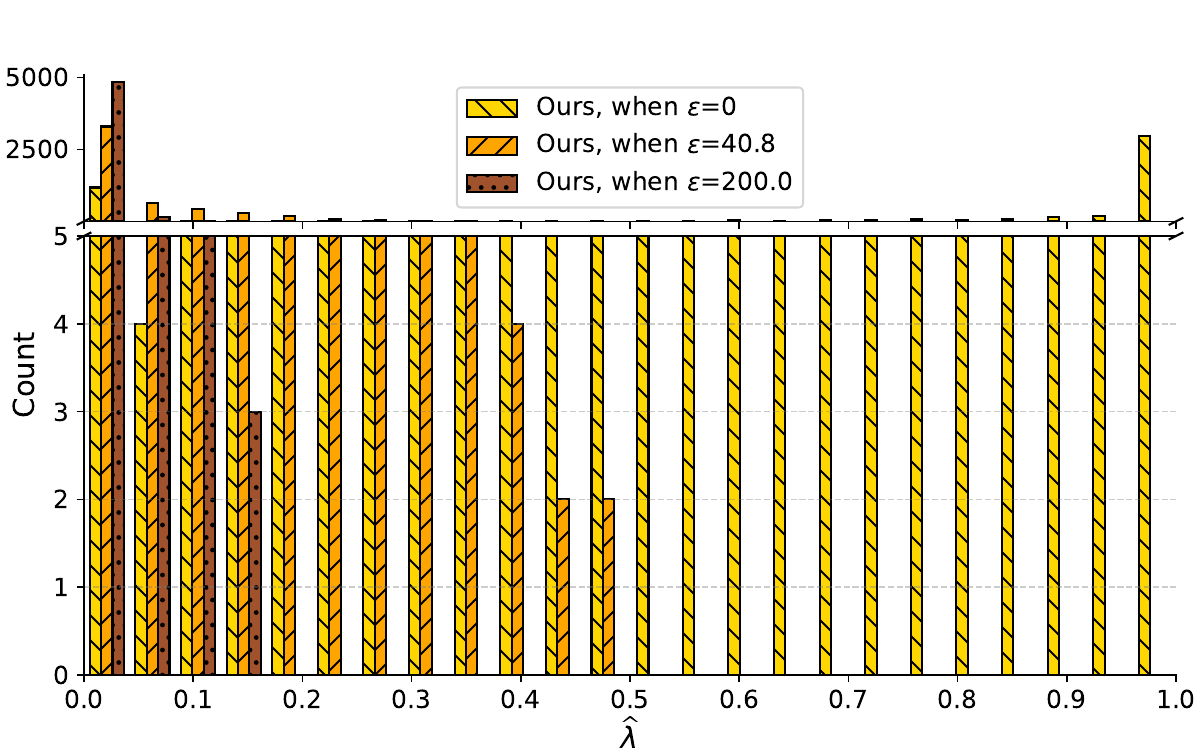}%
        \label{fig:f_linear_mse_lam_dist_same-cov}%
        }
    \caption{Linear setting, histograms of squared errors and $\widehat\lambda$ values over 5000 runs with $N\expm=50$, $N\obs=500$. Panel (a) is the histogram of squared errors in the setting of Figure~\ref{fig:d_linear_mse_ratio}; panels (b-c) are the $\theta\expm = \theta\obs$ counterparts of (a) and Figure~\ref{fig:f_linear_mse_lam_dist}. The vertical axis is split into $\leq 5$ and $>5$ counts to show extreme values that inflate the overall MSE.}
    \label{fig:linear_dist_extra}%
\end{figure}

Figures~\ref{fig:linear_mse_eps_extra}--\ref{fig:linear_dist_extra} complement Figure~\ref{fig:linear_mse_eps}. Figure~\ref{fig:linear_mse_eps_extra} repeats the comparison of Figure~\ref{fig:b_linear_mse_eps} at further sample sizes and with $\theta\expm = \theta\obs$: CVCI improves on the experimental-only estimate when $\varepsilon$ is small and on the observational-only estimate when $\varepsilon$ is large, and the range of $\varepsilon$ over which it improves on the experimental-only estimate narrows from about $0.5$ at $N\expm=50$ to about $0.1$ at $N\expm=1000$. Figure~\ref{fig:linear_mse_n_obs_extra} fixes $\varepsilon=0.05$ and varies $N\obs$: the MSE of CVCI decreases as observational samples are added and falls below that of the experimental-only estimate. Figure~\ref{fig:linear_mse_ratio_extra} extends the large-bias results of Figures~\ref{fig:b_linear_mse_ratio} and~\ref{fig:d_linear_mse_ratio}: for $N\expm=500$ and $N\expm=1000$ the MSE ratio returns to one as $\widehat\lambda$ goes to zero, whereas for $N\expm=50$ it stays below one. Figure~\ref{fig:linear_dist_extra} gives the run-level histograms for the latter case: for large $\varepsilon$, $\widehat\lambda$ is near zero in nearly all runs, and a small number of runs have large squared errors. In each figure, the panels with $\theta\expm = \theta\obs$ closely match their $\theta\expm \neq \theta\obs$ counterparts.

\section{LaLonde Dataset} \label{sec:lalonde_specifics}

\subsection{Data selection}\label{sec:lalonde_data_selection}

\begin{figure}[H]
\centering
\includegraphics[width=0.6\textwidth]{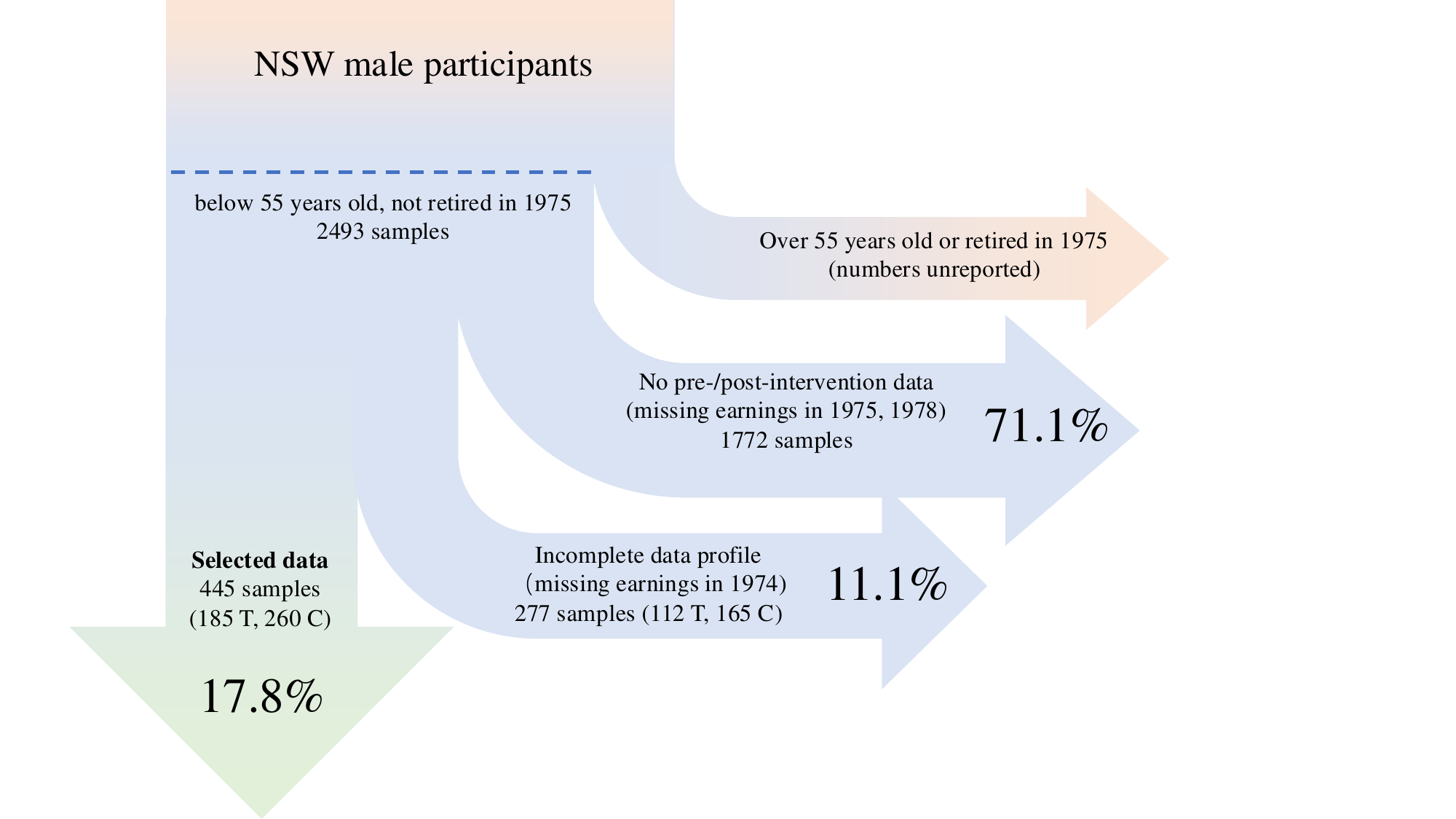} 
\caption{Illustration of data selection process. T and C refer to the number of samples in treatment and control groups, respectively. The size of the arrows does not reflect the actual percentage.}
\label{fig:data_selection}
\end{figure}

\begin{table}[H]
\centering
\caption{Proportions of binary true values in treatment and control groups in NSW data post selection.}
\begin{tabular}{@{}lcc@{}}
\toprule
Variable & Treatment    & Control     \\ \midrule
Black    & 0.8432   & 0.8269   \\
Hispanic     & 0.0595   & 0.1077   \\
Married  & 0.1892   & 0.1538   \\
No degree   & 0.7081   & 0.8346   \\
Unemployed in 1974     & 0.7081   & 0.7500  \\
Unemployed in 1975     & 0.6000   & 0.6846   \\
\hline
Overall counts & 185   & 260   \\ \bottomrule
\end{tabular}
\label{table:proportions}
\end{table}

Dehejia and Wahba's paper narrowed the focus to male participants who were under 55 years of age at the time of the program's initiation. This specific subgroup was chosen because it allows for a consistent examination of labor outcomes and eliminates the potential impact of retirement. The selection criteria were further refined to those individuals who had earnings data available for both 1975 (pre-intervention) and 1978 (post-intervention). This subset comprised 297 treated and 425 control participants. However, to enhance the analysis's robustness and to focus on those with more complete data profiles, the dataset was further narrowed to those who also had earnings data available for 1974, reducing the sample to 185 treated and 260 control participants. As illustrated in Figure \ref{fig:data_selection}, the subset selection is based on only pre-intervention variables. 
Table \ref{table:proportions} shows the proportion of each pre-intervention variable in treatment and control groups, for the sake of assessing internal validity.

\subsection{Discussion on the reproduction of Dehejia and Wahba’s results} \label{sec:reproduced_dehejia}
We now discuss our reproduced results from Dehejia and Wahba in the first and third panels in Table~\ref{tab:lalonde_result}. The correspondence is as follows: point estimates in our column 1 align exactly with their panels B(1) and C(1), 2 with B(2), 3 with B(3) and C(3), 4 with B(4), 6 with C(2), and 7 with C(4). 
Interestingly, our column 5 (which regresses on all covariates excluding RE74 and 1974 employment status) yields the most accurate statistically significant ATE estimate (\$1167). This result was not highlighted by either Dehejia and Wahba or LaLonde, suggesting a potentially overlooked finding. Notably, this improvement is achieved using CPS-1, which is not considered a specially selected subgroup. This outcome contradicts LaLonde’s assertion that subgroups such as PSID-2, PSID-3, CPS-2, and CPS-3 are more comparable to the NSW control group.

We then critically assess Dehejia and Wahba’s claim that 1974 earnings (RE74) are a valuable predictor in estimating treatment effects.
In our reproduced results, the impact of RE74 could be mixed. For example, we compare results in columns 2 and 6, where they share the same set of covariates but column 6 additionally includes RE74. Column 2 has four statistically significant estimates of treatment effect, none of which are statistically significant in column 6. The column 2 estimates deviate a large amount from those of the NSW data, implying that the training had a negative impact on future earnings. In column 6, incorporating RE74 gives less negative results, but at the cost of losing statistical significance. 

Interestingly, RE74 appears to add value in a specific setting: the linear model in column 7, applied to the CPS-3 subgroup, produces the best near-significant estimate ($1,326$) across all configurations with a p-value of 0.09. Although this does not meet our predetermined 0.05 significance threshold, it still indicates marginal significance. This suggests that CPS-3 may serve as a promising comparison group. Additionally, we note that caution is necessary when interpreting the CPS-3 results. The selection criteria for CPS-3 excluded individuals with 1975 incomes above the poverty line, whereas NSW participants were not restricted in this way. Specifically, the ATE may differ depending on whether individuals' incomes in 1975 were above or below the poverty threshold. The result may be attributable to good luck rather than a meaningful underlying effect.

\subsection{Tables \ref{tab:lalonde_result} and \ref{tab:lalonde_bootstrap}: additional configurations}\label{sec:lalonde_tables_supp}
We now present the full configurations.
The observational data are partitioned into six subgroups: 
\begin{enumerate}
    \item PSID-1, CPS-1: full datasets;
    \item PSID-2: PSID-1 subjects who were unemployed when surveyed in 1976;
    \item PSID-3: PSID-2 subjects who were unemployed in 1975;
    \item CPS-2: CPS-1 subjects who were unemployed when surveyed in 1976;
    \item CPS-3: CPS-2 subjects whose income in 1975 was lower than the poverty level.
\end{enumerate}

Each column represents the estimated effect of treatment according to a specific linear setting as follows:

\begin{enumerate}
    \item Regress RE78 on treatment;
    \item Regress RE78 on treatment, age, $\text{age}^2$, years of schooling, high school dropout status, and race;
    \item Regress RE78 on treatment and RE75;
    \item Regress RE78 on treatment, age, $\text{age}^2$, years of schooling, high school dropout status, race, and RE75;
    \item Regress RE78 on treatment, age, years of schooling, high school dropout status, race, marriage status, RE75 and employment status in 1975;
    \item Regress RE78 on treatment, age, $\text{age}^2$, years of schooling, high school dropout status, race, and RE74;
    \item Regress RE78 on treatment, age, $\text{age}^2$, years of schooling, high school dropout status, race, RE75, and RE74;
    \item Regress RE78 on treatment, age, years of schooling, high school dropout status, race, marriage status, RE75, employment status in 1975, RE74, and employment status in 1974.
\end{enumerate}

In the following Table~\ref{tab:lalonde_result}, each panel contains results for different methods detailed in Section \ref{sec:lalonde_settings}. Each row represents the data configuration with (T) for treatment group and (C) for control group. For the second panel, we report the estimated treatment effect with $\pm 1$ standard deviation over 5000 runs, followed by $\widehat\lambda$ in parentheses selected by five-fold cross-validation. For the other panels, the p-values (in parentheses) come from testing the null hypothesis that the treatment coefficient is zero.  Statistically significant results (under $0.05$) are in \textbf{bold}.

We note that in Table~\ref{tab:lalonde_result}, the reported standard deviations of our method come from random $K$-fold splits in each run. In contrast, Table~\ref{tab:lalonde_bootstrap} presents bootstrap standard deviations: for our method, this captures uncertainty from both data resampling and cross-validation splitting, while for the other methods, it reflects uncertainty from data resampling alone. 

\newpage
\thispagestyle{empty} 
\begin{table}[H]
    \tiny
    \centering  
    \caption{Estimates of treatment effects on the LaLonde dataset. Full configurations.  
    } 
    \begin{tabular}{ r c c c c c c c c}
         \toprule\toprule
  & 1 & 2 & 3 & 4 & 5 & 6 & 7 & 8 \\ \midrule
    \textbf{($\lambda=0$, $X\expm$ only)}    &  &  &  &  &  & & & \\\cmidrule(r){1-1}
   NSW(T+C)  & \textbf{1794} & \textbf{1672} & \textbf{1750} & \textbf{1631} & \textbf{1610} & \textbf{1688} & \textbf{1672} & \textbf{1671} \\
p-value & (0.0048) & (0.009) & (0.0059) & (0.0108) & (0.0122) & (0.0082) &(0.0091) & (0.0095) \\\midrule 
\textbf{($\widehat\lambda$, ours)}  $X\expm+X\obs$,   &  &  &  &  &  & & & \\
$X\expm$: NSW(T+C), $X\obs:$   &   &  &  &  &  & & & \\\cmidrule(r){1-1}
NSW(T)+PSID-1(C) & 1761$\pm$24 & 1595$\pm$96 & 1511$\pm$163 & 1345$\pm$220 & 1161$\pm$294 & 1453$\pm$186 & 1303$\pm$264 & 1282$\pm$270 \\
 $\widehat\lambda=$ & (0.0$\pm$0.0) & (0.1$\pm$0.1) & (0.6$\pm$0.2) & (0.6$\pm$0.3) & (0.8$\pm$0.2) & (0.5$\pm$0.3) & (0.8$\pm$0.3) & (0.8$\pm$0.3) \\
NSW(T)+PSID-2(C) &  1692$\pm$70 & 1544$\pm$127 & 1281$\pm$268 & 1243$\pm$272 & 1381$\pm$25 & 1340$\pm$246 & 1157$\pm$243 & 1142$\pm$195 \\
$\widehat\lambda=$ & (0.1$\pm$0.0) & (0.1$\pm$0.1) & (0.7$\pm$0.2) & (0.6$\pm$0.3) & (1.0$\pm$0.1) & (0.6$\pm$0.3) & (0.9$\pm$0.2) & (0.9$\pm$0.2) \\
NSW(T)+PSID-3(C) & 1279$\pm$209 & 1358$\pm$234 & 1388$\pm$58 & 1256$\pm$266 & 1375$\pm$27 & 1176$\pm$267 & 1162$\pm$268 & 1159$\pm$172 \\
$\widehat\lambda=$ & (0.9$\pm$0.2) & (0.5$\pm$0.3) & (1.0$\pm$0.1) & (0.6$\pm$0.3) & (1.0$\pm$0.1) & (0.8$\pm$0.2) & (0.8$\pm$0.2) & (0.9$\pm$0.2) \\
NSW(T)+CPS-1(C) & 1740$\pm$37 & 1571$\pm$111 & 1465$\pm$181 & 1219$\pm$335 & 1202$\pm$105 & 1381$\pm$211 & 1187$\pm$344 & 1162$\pm$187 \\
$\widehat\lambda=$ & (0.3$\pm$0.1) & (0.4$\pm$0.3) & (0.9$\pm$0.2) & (0.9$\pm$0.2) & (1.0$\pm$0.1) & (0.9$\pm$0.3) & (0.9$\pm$0.2) & (1.0$\pm$0.1) \\
NSW(T)+CPS-2(C) & 1695$\pm$68 & 1528$\pm$137 & 1478$\pm$183 & 1227$\pm$280 & 1090$\pm$227 & 1223$\pm$290 & 1158$\pm$257 & 1122$\pm$246 \\
$\widehat\lambda=$ & (0.2$\pm$0.1) & (0.4$\pm$0.2) & (0.6$\pm$0.2) & (0.8$\pm$0.2) & (0.9$\pm$0.2) & (0.9$\pm$0.2) & (0.9$\pm$0.2) & (0.9$\pm$0.2) \\
NSW(T)+CPS-3(C) & 1569$\pm$150 & 1288$\pm$269 & 1454$\pm$196 & 1122$\pm$249 & 1179$\pm$112 & 1299$\pm$82 & 1343$\pm$59 & 1120$\pm$251 \\
$\widehat\lambda=$ & (0.3$\pm$0.1) & (0.7$\pm$0.3) & (0.4$\pm$0.2) & (0.9$\pm$0.2) & (1.0$\pm$0.1) & (1.0$\pm$0.1) & (1.0$\pm$0.1) & (0.9$\pm$0.2) \\
 \midrule
 \textbf{($\lambda=1$, $X\obs$ only)}     &  &  &  &  &  & & & \\
 \citet{dehejia1999causal}, &  &  &  &  & & & &  \\
 $X\obs$:  &  &  &  &  & & & &  \\ \cmidrule(r){1-1}
NSW(T)+PSID-1(C) & \textbf{-15205} & \textbf{-7741} & -582 & -265 & 428 & -879 & 218 & 4 \\ 
p-value & ($<$.0001) & ($<$.0001) & (0.4892) & (0.7633) & (0.6613) & (0.3451) & (0.8014) & (0.9967) \\
NSW(T)+PSID-2(C) & \textbf{-3647} & \textbf{-2810} & 721 & 297 & 1377 & 94 & 907 & 999 \\
p-value & (0.0002) & (0.0097) & (0.4167) & (0.7678) & (0.204) & (0.9281) & (0.3669) & (0.3753) \\
NSW(T)+PSID-3(C) & 1070 & 35 & 1370 & 243 & 1371 & 821 & 822 & 1049 \\
p-value & (0.2353) & (0.9743) & (0.1277) & (0.8254) & (0.2414) & (0.4558) & (0.456) & (0.3902) \\
NSW(T)+CPS-1(C) & \textbf{-8498} & \textbf{-4417} & -78 & 525 & \textbf{1167} & -8 & 739 & 1066 \\
p-value & ($<$.0001) & ($<$.0001) & (0.8849) & (0.3459) & (0.0373) & (0.989) & (0.1769) & (0.0541) \\
NSW(T)+CPS-2(C) & \textbf{-3822} & \textbf{-2208} & -263 & 371 & 885 & 615 & 879 & 891 \\
p-value & ($<$.0001) & (0.0031) & (0.6467) & (0.5752) & (0.183) & (0.3595) & (0.1789) & (0.1778) \\
NSW(T)+CPS-3(C) & -635 & 375 & -91 & 844 & 1129 & 1270 & 1326 & 866 \\
p-value & (0.3342) & (0.6483) & (0.8875) & (0.2961) & (0.1597) & (0.1122) & (0.0965) & (0.2797)  \\
\midrule
  \textbf{($\lambda=1$, $X\obs$ only)}  &  &  &  &  & & & &  \\ 
  Pooling,
  &  &  &  &  & & & &  \\ 
   view all data as $X\obs$: &  &  &  &  & & & &  \\ \cmidrule(r){1-1}
  NSW(T+C)+PSID-1(C) & \textbf{-13598} & \textbf{-5303} & -162 & 326 & 767 & -99 & 683 & 741 \\ 
  p-value & ($<$.0001) & ($<$.0001) & (0.8394) & (0.6878) & (0.3589) & (0.9084) & (0.392) & (0.3749) \\ 
  NSW(T+C)+PSID-2(C) & -889 & -58 & 1101 & 969 & 1264 & 964 & 1163 & \textbf{1368} \\ 
  p-value & (0.2417) & (0.9363) & (0.0959) & (0.1375) & (0.0557) & (0.1526) & (0.0731) & (0.038) \\ 
  NSW(T+C)+PSID-3(C) & \textbf{1555} & \textbf{1353} & \textbf{1599} & \textbf{1366} & \textbf{1570} & \textbf{1528} & \textbf{1522} & \textbf{1710} \\ 
   p-value & (0.0114) & (0.0272) & (0.0087) & (0.0251) & (0.0108) & (0.0116) & (0.0121) & (0.0055) \\ 
  NSW(T+C)+CPS-1(C) & \textbf{-8333} & \textbf{-3594} & -17 & 714 & \textbf{1202} & 277 & 911 & \textbf{1148} \\ 
   p-value & ($<$.0001) & ($<$.0001) & (0.9745) & (0.1943) & (0.0293) & (0.6239) & (0.0919) & (0.0349) \\ 
  NSW(T+C)+CPS-2(C) & \textbf{-3267} & -683 & -26 & 923 & \textbf{1188} & 1078 & \textbf{1229} & \textbf{1265} \\ 
  p-value & ($<$.0001) & (0.3116) & (0.9633) & (0.122) & (0.0468) & (0.0755) & (0.0372) & (0.0323) \\ 
  NSW(T+C)+CPS-3(C) & 282 & \textbf{1268} & 524 & \textbf{1354} & \textbf{1430} & \textbf{1611} & \textbf{1587} & \textbf{1521} \\ 
  p-value & (0.6278) & (0.0344) & (0.3545) & (0.0216) & (0.0151) & (0.0058) & (0.0065) & (0.009) \\ 
   \bottomrule
    \end{tabular}
    \label{tab:lalonde_result}
\end{table}

\newpage
\thispagestyle{empty} 
\begin{table}[H]
\tiny 
    \centering
    \caption{Bootstrap standard deviations of estimated treatment effects on the LaLonde dataset. Full configurations. } 
    \begin{tabular}{ r c c c c c c c c}
         \toprule\toprule
  & 1 & 2 & 3 & 4 & 5 & 6 & 7 & 8 \\ \midrule
    \textbf{($\lambda=0$, $X\expm$ only)}    &  &  &  &  &  & & & \\\cmidrule(r){1-1}
   NSW(T+C)  & 658 & 656 & 657 & 659 & 657 & 656 & 661 & 666 \\\midrule
\textbf{($\widehat\lambda$, ours)}  $X\expm+X\obs$,   &  &  &  &  &  & & & \\
$X\expm$: NSW(T+C), $X\obs:$   &   &  &  &  &  & & & \\\cmidrule(r){1-1}
NSW(T)+PSID-1(C) & 672 & 681 & 721 & 723 & 674 & 725 & 701 & 708 \\
 $\widehat\lambda=$ & (0.0) & (0.1) & (0.3) & (0.3) & (0.3) & (0.3) & (0.3) & (0.3) \\
NSW(T)+PSID-2(C) & 681 & 696 & 659 & 694 & 657 & 694 & 661 & 666 \\
$\widehat\lambda=$ & (0.1) & (0.2) & (0.3) & (0.3) & (0.3) & (0.3) & (0.3) & (0.3) \\
NSW(T)+PSID-3(C) & 643 & 699 & 631 & 695 & 662 & 665 & 672 & 668 \\
$\widehat\lambda=$ & (0.3) & (0.3) & (0.3) & (0.3) & (0.3) & (0.3) & (0.3) & (0.3) \\
NSW(T)+CPS-1(C) & 673 & 685 & 724 & 680 & 609 & 721 & 665 & 628 \\
$\widehat\lambda=$ & (0.1) & (0.3) & (0.2) & (0.3) & (0.2) & (0.3) & (0.3) & (0.2) \\
NSW(T)+CPS-2(C) & 680 & 703 & 725 & 686 & 624 & 660 & 638 & 632 \\
$\widehat\lambda=$ & (0.1) & (0.3) & (0.3) & (0.3) & (0.2) & (0.3) & (0.3) & (0.3) \\
NSW(T)+CPS-3(C) & 729 & 686 & 719 & 642 & 618 & 614 & 615 & 639 \\
$\widehat\lambda=$ & (0.2) & (0.3) & (0.3) & (0.3) & (0.3) & (0.3) & (0.3) & (0.3) \\
 \midrule
 \textbf{($\lambda=1$, $X\obs$ only)}     &  &  &  &  &  & & & \\
 \citet{dehejia1999causal}, $X\obs$:  &  &  &  &  & & & &  \\ \cmidrule(r){1-1}
NSW(T)+PSID-1(C) & 657 & 784 & 765 & 778 & 896 & 782 & 764 & 842 \\
NSW(T)+PSID-2(C) & 900 & 929 & 828 & 932 & 1020 & 1060 & 935 & 984 \\ 
NSW(T)+PSID-3(C) & 890 & 1033 & 873 & 1016 & 1078 & 1018 & 1018 & 1074 \\  
NSW(T)+CPS-1(C) & 582 & 614 & 598 & 610 & 625 & 630 & 617 & 624 \\ 
NSW(T)+CPS-2(C) & 604 & 696 & 610 & 657 & 668 & 689 & 668 & 666 \\ 
NSW(T)+CPS-3(C) & 670 & 736 & 673 & 719 & 743 & 747 & 739 & 731 \\ 
\midrule
  \textbf{($\lambda=1$, $X\obs$ only)}  &  &  &  &  & & & &  \\ 
  Pooling,
  &  &  &  &  & & & &  \\ 
   view all data as $X\obs$: &  &  &  &  & & & &  \\ \cmidrule(r){1-1}
   NSW(T+C)+PSID-1(C) & 641 & 726 & 713 & 694 & 701 & 709 & 691 & 666 \\ 
  NSW(T+C)+PSID-2(C) & 690 & 659 & 663 & 648 & 657 & 659 & 646 & 630 \\ 
  NSW(T+C)+PSID-3(C) & 662 & 648 & 659 & 648 & 657 & 645 & 648 & 635 \\ 
  NSW(T+C)+CPS-1(C) & 579 & 602 & 592 & 602 & 615 & 629 & 612 & 618 \\ 
  NSW(T+C)+CPS-2(C) & 602 & 634 & 607 & 618 & 626 & 636 & 624 & 621 \\ 
  NSW(T+C)+CPS-3(C) & 633 & 619 & 634 & 618 & 625 & 620 & 621 & 611 \\ 
   \bottomrule
    \end{tabular}
    \label{tab:lalonde_bootstrap}
\end{table}

\newpage
\subsection{Table~\ref{tab:lalonde_synthetic}: error decomposition}
\begin{table}[H]
   \scriptsize 
    \centering  
    \caption{Root Mean Squared Error (RMSE) and its decomposition using LaLonde semi-synthetic data. $\widetilde X\expm$: synthetic based on $X\expm$. $\widetilde X\obs$: synthetic based on $X\obs$. Selected configurations.
    } 
    \begin{tabular}   
    { >{\raggedleft\arraybackslash}m{4cm}
     >{\centering\arraybackslash}m{1cm}
      >{\centering\arraybackslash}m{1cm}
       >{\centering\arraybackslash}m{1cm}
       >{\centering\arraybackslash}m{1cm}
       >{\centering\arraybackslash}m{1cm}
        >{\centering\arraybackslash}m{1cm}}
         \toprule\toprule
    Column No. & \multicolumn{2}{c}{1} & \multicolumn{2}{c}{3} & \multicolumn{2}{c}{8} \\ 
      Regress RE78 on: & \multicolumn{2}{c}{\{treatment\}} & \multicolumn{2}{c}{\{treatment, RE75\}} &   \multicolumn{2}{c}{\{treatment, all covariates\}} \\ 
        \toprule

    \textbf{($\lambda=0$, $\widetilde X\expm$ only) } NSW(T+C)   &  & & &  & \\ \cmidrule(r){1-1}
RMSE   & \multicolumn{2}{c}{647.7} & \multicolumn{2}{c}{646.0} & \multicolumn{2}{c}{646.6}  \\
bias    & \multicolumn{2}{c}{$-$9.6} & \multicolumn{2}{c}{$-$9.1} & \multicolumn{2}{c}{$-$10.4}  \\
standard deviation    & \multicolumn{2}{c}{647.7} & \multicolumn{2}{c}{645.9} & \multicolumn{2}{c}{646.5}  \\
\midrule 
\textbf{($\widehat\lambda$, ours)}  $\widetilde  X\expm+\widetilde  X\obs$, $X\expm$: NSW(T+C),  &  & & &  & \\
 $X\obs$ includes NSW(T) and: & PSID & CPS &  PSID & CPS & PSID & CPS \\ \cmidrule(r){1-1}
RMSE & 651.5 & 655.2 & 747.7 & 767.7 & 734.1 & 617.4 \\ 
bias & $-$50.9 & $-$67.0 & $-$237.3 & $-$271.9 & $-$251.0 & $-$176.4 \\ 
standard deviation & 649.6  & 651.8  & 709.1  & 717.9  & 689.9  & 591.7  \\ 
$\widehat\lambda=$ & 0.0 $\pm$ 0.0 & 0.3 $\pm$ 0.2 & 0.6 $\pm$ 0.3   & 0.8$\pm$ 0.2 & 0.7 $\pm$ 0.3 &  0.9 $\pm$ 0.2 \\
 \midrule
\textbf{($\lambda=1$, $\widetilde  X\obs$ only)} \citet{dehejia1999causal},   &  &  &  &  & \\
 $X\obs$ includes  NSW(T) and: & PSID & CPS & PSID & CPS  & PSID & CPS  \\  \cmidrule(r){1-1}
RMSE  & 17017.6 & 10282.3 & 2469.7 & 1880.3 & 1943.6 & 796.9 \\
bias & $-$16977.8 & $-$10257.6 & $-$2318.4 & $-$1802.2 & $-$1655.8 & $-$574.8 \\ 
standard deviation & 1162.3  & 712.5  & 851.1 & 536.3  & 1017.8  & 552.0  \\ \midrule
\textbf{($\lambda=1$, pool all data as $\widetilde  X\obs$),} 
&  &  &  &  & \\
 $X\obs$ includes NSW(T+C) and & PSID & CPS & PSID & CPS  & PSID & CPS  \\  \cmidrule(r){1-1}
RMSE  & 15409.3 & 10143.0 & 2038.9 & 1848.9 & 1291.2 & 773.2 \\
bias & $-$15398.7 & $-$10130.6 & $-$1956.7 & $-$1779.3 & $-$1141.9 & $-$572.6 \\ 
standard deviation &  570.2  &  501.6   & 573.1 & 502.5   &  602.7  & 519.6    \\ 
\midrule
   \bottomrule
    \end{tabular}
    \label{tab:lalonde_synthetic_decomposition}
\end{table}

\newpage

\section{Proofs in Section~\ref{sec:method}}\label{sec:proof_method_section}

\subsection{Closed-form solution for the no-covariate setting: Deriving Eq.~\eqref{eq:mean_close_form} and additional discussion} \label{proof:mean_close_form}
We now derive Eq.~\eqref{eq:mean_close_form}, which is  
\[  \widehat\theta(\lambda) = \arg \min_\theta  \sum_{i = 1}^{N\expm} (1 - \lambda) (Y_i\expm - \theta)^2 + \lambda (\overline{Y}\obs - \theta)^2 
    = (1 - \lambda) \overline{Y}\expm + \lambda \overline{Y}\obs. \]
The result follows from the following calculation:
\begin{align*}
    &\arg \min_\theta  (1 - \lambda) (\overline{Y}\expm - \theta)^2 + \lambda (\overline{Y}\obs - \theta)^2 \\
    &\qquad = \arg \min_\theta (1 - \lambda) \theta^2 - 2 (1 - \lambda) \overline{Y}\expm \theta + \lambda \theta^2 - 2 \lambda \overline{Y}\obs \theta\\
    &\qquad = \arg \min_\theta \theta^2 - 2 \Big( (1 - \lambda) \overline{Y}\expm + \lambda \overline{Y}\obs \Big) \theta\\
    &\qquad = (1 - \lambda)    \overline{Y}\expm +  \lambda \overline{Y}\obs.
\end{align*}

Moreover, we note that the following four minimizers are equivalent:
\begin{align*}
    \widehat\theta(\lambda) &= \arg \min_\theta  (1 - \lambda) (\overline{Y}\expm - \theta)^2 + \lambda (\overline{Y}\obs - \theta)^2 \\
    &= \arg \min_\theta  \frac{1 - \lambda}{N\expm} \Big( \sum_{i = 1}^{N\expm}  (Y_i\expm - \theta)^2 \Big) + \lambda (\overline{Y}\obs - \theta)^2\\
    &= \arg \min_\theta  \frac{1 - \lambda}{N\expm} \Big( \sum_{i = 1}^{N\expm}  (Y_i\expm - \theta)^2 \Big) + \frac{\lambda}{N\obs} \Big( \sum_{i = 1}^{N\obs}  (Y_i\obs- \theta)^2 \Big)\\
    &= \arg \min_\theta  (1 - \lambda) (\overline{Y}\expm - \theta)^2  + \frac{\lambda}{N\obs} \Big( \sum_{i = 1}^{N\obs}  (Y_i\obs- \theta)^2 \Big).
\end{align*}
The equivalence of these formulations follows directly from Lemma~\ref{lem:additive_mse}. Specifically, the first and second terms in each formulation resemble $ (1 - \lambda) \theta^2 - 2 (1 - \lambda) \overline{Y}\expm  \theta$ and $\lambda \theta^2 - 2  \lambda \overline{Y}\obs \theta$, respectively, up to additive constants that do not affect the minimizer. This equivalence implies that aggregate- and unit-level losses yield the same minimizer, reflecting their alignment in the underlying principle across granularity.

\subsection{Additive structure of the quadratic experimental loss }\label{sec:proof_additive_mse}

\begin{lemma}\label{lem:additive_mse}
    For a scalar-valued function $f$, a fixed input sequence $x_1, \ldots, x_N$, and a scalar $t$, it holds that 
    \begin{align*}
        \Big(t - \frac{1}{N} \sum_i f (x_i)\Big)^2 \propto_{t} \frac{1}{N}\sum_{i} \Big(t - f (x_i)\Big)^2,
    \end{align*}
    where $\propto_t$ denotes proportional to with respect to $t$ up to constants.
\end{lemma} 
Informally, treating the experimental data as fixed, the squared error between a given scalar (e.g., the causal parameter) and the average experimental estimate is proportional to the average squared error between that scalar and each individual estimate, up to constants. 

We prove the additive property for squared loss:
\begin{proof} We have
    \begin{align*}
 \MoveEqLeft{ \Big(t - \frac{1}{N} \sum_if (x_i)\Big)^2}
 \\
   &=  \frac{1}{N}\sum_{i}\Big(t- \frac{1}{N} \sum_j f (x_j)\Big)^2 \\ 
   &= \frac{1}{N}\sum_{i}\Big(t - f (x_i) - \Big( \frac{1}{N} \sum_j f (x_j) - f (x_i)\Big)\Big)^2\\
   &= \frac{1}{N}\sum_{i} \Big[\Big(t - f (x_i)\Big)^2 + \Big( \frac{1}{N} \sum_j f (x_j) - f (x_i)\Big)^2   \\
   &\qquad - 2 \Big(t - f(x_i) \Big) \Big( \frac{1}{N} \sum_j f (x_j) - f (x_i)\Big)  \Big]\\
   &\propto_{t} \frac{1}{N}\sum_{i} \Big(t - f (x_i)\Big)^2 - \frac{1}{N}\sum_{i} 2 \Big(t - f (x_i) \Big) \Big( \frac{1}{N} \sum_j f (x_j) - f (x_i)\Big).
\end{align*}
The second term vanishes as
\begin{align*}
   \sum_{i} \Big(t - f (x_i) \Big) \Big( \frac{1}{N} \sum_j f (x_j) - f (x_i) \Big)  
    \propto_{t}\ t  \sum_i \Big( \frac{1}{N} \sum_j f (x_j) - f (x_i)\Big),
\end{align*}
 concluding the proof.
\end{proof}

\subsection{Closed-form solution for the linear setting: Deriving Eq.~\eqref{eq:linear_close_form}} \label{proof:linear_close_form}
We now derive the closed-form solution for the linear setting. We are interested in 
\begin{align*}
   \widehat\theta(\lambda) = \arg\min_{\theta} (1-\lambda) (\beta(\theta) - \estate)^2 + \frac{\lambda}{N\obs}\sum_{i=1}^{N\obs} \Big(Y_i\obs - \theta^\top  \begin{bmatrix} \tre_i\obs\\ \covariate_i\obs \end{bmatrix}\Big)^2.
\end{align*}
We can write $\beta(\theta) = e_1^\top \theta$, where $e_1^\top = ( 1 \quad 0  \cdots 0 ) $. Then we have 
\begin{align*}
    \widehat\theta(\lambda) 
    &= \arg\min_{\theta} (1 - \lambda) (e_1^\top \theta)^2 - 2 (1 - \lambda) \estate e_1^\top \theta \\
    &\qquad + \frac{\lambda}{N\obs} \sum_{i=1}^{N\obs} \Big( \Big( \begin{bmatrix} \tre_i\obs \\ \covariate_i\obs \end{bmatrix}^\top \theta \Big)^2 - 2 \res_i\obs \begin{bmatrix} \tre_i\obs \\ \covariate_i\obs \end{bmatrix}^\top \theta \Big)\\
    &= 
    \arg\min_{\theta} \theta^\top \Big( (1 - \lambda) e_1 e_1^\top + \frac{\lambda}{N\obs} \sum_{i=1}^{N\obs} \begin{bmatrix} \tre_i\obs 
        \\ 
        \covariate_i\obs \end{bmatrix} \begin{bmatrix} \tre_i\obs 
            \\
             \covariate_i\obs \end{bmatrix}^\top  \Big) \theta 
             \\&\qquad
             -
              \Big( 2 (1 - \lambda) \estate e_1^\top + \frac{2\lambda}{N\obs} \sum_{i=1}^{N\obs}  Y_i\obs \begin{bmatrix} \tre_i\obs \\ \covariate_i\obs \end{bmatrix}^\top \Big) \theta.
\end{align*}
We take the gradient with respect to $\theta$ and set it to 0:
\begin{align*}
     \Big( (1 - \lambda) e_1 e_1^\top + \frac{\lambda}{N\obs} \begin{bmatrix} \tre\obs \\ \covariate\obs \end{bmatrix} \begin{bmatrix} \tre\obs \\ \covariate\obs \end{bmatrix}^\top  \Big) \theta \\
     \qquad -  \Big( (1 - \lambda) \estate e_1 + \frac{\lambda}{N\obs}  \begin{bmatrix} \tre\obs \\ \covariate\obs \end{bmatrix} Y\obs \Big) &= 0.
\end{align*}
Solving this linear system gives the desired minimizer. When $\lambda = 0$, the minimizer may not be unique, but every solution must satisfy $\beta(\theta)=\estate$, thereby matching the experimental estimate. When $\lambda=1$, the objective reduces to ordinary least squares on observational data, yielding the observational estimate.

\section{Proofs in Section~\ref{sec:theoretical_results}}\label{sec:proof_theoretical_results}

\subsection{Proof of Proposition~\ref{prop:cv_mean_improvement}}\label{sec:proof_prop:cv_mean_improvement}
Write $A=\overline{Y}\expm$ for the full experimental mean and $A_k=\overline{Y}_{\fold_k}\expm$ for the held-out mean in fold $k$. Since the folds have equal size, $\sum_j A_j=\numfold A$. The training mean $A_{-k}$ averages the other $\numfold-1$ fold means, so
\[
 A_{-k}=\frac{\sum_{j\ne k}A_j}{\numfold-1}
       =\frac{\numfold A-A_k}{\numfold-1}
       =A-\frac{A_k-A}{\numfold-1}.
\]
For a given observational mean $\obsmeaninput$, write $\meancontrast=\obsmeaninput-A$. By~\eqref{eq:mean_close_form}, the combined training estimate in fold $k$ is $(1-\regu)A_{-k}+\regu\obsmeaninput$. Subtracting it from the held-out mean gives
\[
 A_k-\big[(1-\regu)A_{-k}+\regu\obsmeaninput\big]
 =\frac{\numfold-\regu}{\numfold-1}(A_k-A)-\regu\meancontrast.
\]
The CV criterion averages these squared differences, and the cross term vanishes because $\sum_k(A_k-A)=0$. Thus, with $S=\numfold^{-1}\sum_k(A_k-A)^2$, we have
\[
 \CV(\regu)=S\left(\frac{\numfold-\regu}{\numfold-1}\right)^2+\regu^2\meancontrast^2.
\]
Minimizing over $\regu\in[0,1]$ gives $\EstPar(\estregu)=T(\overline{Y}\obs)$, where
\[
 T(\obsmeaninput)=A+w(\meancontrast)\meancontrast,\qquad w(\meancontrast)=\min\!\left\{1,\frac{\numfold S}{(\numfold-1)^2\meancontrast^2+S}\right\}.
\]
Since $T$ is continuous and piecewise differentiable, with $T'(\obsmeaninput)=1$ when $w=1$ and $T'(\obsmeaninput)=w(2w/\numfold-1)\in[-1,0]$ when $w<1$, it is $1$-Lipschitz.

Write $\overline{Y}\obs=\truemean+\obserr+s_{\obsshort}U$ and $O_0=\truemean+s_{\expshort}U$ using the same standard normal $U$ independent of the experimental samples. Then the midpoint $M=(A+O_0)/2$ has mean $\truemean$, variance $s_{\expshort}^2/2$, and is independent of $D=O_0-A$ and $S$ by Gaussianity. Hence, writing $w_0=w(D)$, we have
\begin{align*}
 g_{\numfold}&\defn\frac{\E[(T(O_0)-\truemean)^2]}{s_{\expshort}^2}
 =\frac{\E\big[(M-\truemean+(w_0-1/2)D)^2\big]}{s_{\expshort}^2}\\
 &=\frac12+\frac{\E[(w_0-1/2)^2D^2]}{s_{\expshort}^2}
 =1-\frac{\E[w_0(1-w_0)D^2]}{s_{\expshort}^2}.
\end{align*}
Since $D/s_{\expshort}\sim\cN(0,2)$ and $S/s_{\expshort}^2\sim\chi^2_{\numfold-1}$ are independent, $g_{\numfold}$ depends only on $\numfold$.  Finally, the Lipschitz continuity of $T$ gives $\E[(T(\overline{Y}\obs)-T(O_0))^2]\leq\E[(\overline{Y}\obs-O_0)^2]=\Delta^2$. Applying the $L^2$-triangle inequality to $T(\overline{Y}\obs)-\truemean=(T(O_0)-\truemean)+(T(\overline{Y}\obs)-T(O_0))$ yields the bound.

\subsection{Preliminaries on the proofs for Section~\ref{sec:robustness_theory}}

\subsubsection{A sufficient condition for Assumption~\ref{ass:linear_ate}}\label{sec:sufficient_condition_linear_ate}
By Lemma~\ref{lm:linear_expansion_estate},
a sufficient condition for Assumption~\ref{ass:linear_ate} is the following condition assuming that $\estate(\expsam_{\samset})$ is derived from some $Z$-estimation problem. In this case, $(\boundZfuntil,\boundate,\boundatediff,\boundatelin,\boundatenum)$ can be chosen as constants that depend polynomially on the parameters $(\datepar,1/\strongcvx,\boundatepar,\bZfunzero,\bZfunone,\bZfuntwo)$ in Assumption~\ref{ass:zest_as_ate}.

\begin{myassumption}{$Z$-est}{ass:zest_as_ate}
Let $\ateparspace\subseteq\R^{\datepar}$ be some open convex set.
        For a set of i.i.d.   experimental samples $\expsam_{\samset}\defn(\expsam_{j})_{j\in\samset}$,  we define 
        $\estate(\expsam_{\samset})\defn\estatepar_1$, where $\estatepar_1$ is the first coordinate of  $\estatepar\in\ateparspace$, the solution to the following estimating equation:
        \begin{align*}
            \sum_{j\in\samset} \Zfun(\expsam_j;\estatepar)=\bzero
        \end{align*}
        for some $Z$-function $\Zfun:\expsamspace\times\R^{\datepar}\to\R^{\datepar}$. Define $\PopZfun(\atepar)\defn \E[\Zfun(\expsam_1;\atepar)]$ for any $\atepar\in\ateparspace$.
         Moreover, assume that 
         \begin{enumerate}
            \item\label{ass:zest_as_ate:a}
         $\PopZfun(\trueatepar)=\bzero$ for some $\trueatepar\in\ateparspace$ such that $\trueatepar_1=\trueate$; there exists some constant $\boundatepar>0$ such that  $\vecnorm{\atepar}{2}\leq \boundatepar$ for all $\atepar\in\ateparspace$.
         \item\label{ass:zest_as_ate:c} 
         $\Zfun$ is twice continuously differentiable. There exist some constants $\bZfunzero,\bZfunone,\bZfuntwo>0$ such that 
         \begin{align*}             \sup_{\sam\in\expsamspace,\atepar\in\ateparspace} \vecnorm{\Zfun(\sam;\atepar)}{2}&\leq \bZfunzero,\\         \sup_{\sam\in\expsamspace,\atepar\in\ateparspace} \opnorm{\nabla\Zfun(\sam;\atepar)}&\leq \bZfunone,\\            \sup_{\sam\in\expsamspace,\atepar\in\ateparspace} \opnorm{\nabla^2\Zfun(\sam;\atepar)}
         &\leq \bZfuntwo.
         \end{align*} 
         \item\label{ass:zest_as_ate:b}
          $\sigma_{\min}(\nabla\PopZfun(\trueatepar))\geq \strongcvx$ for some constant $\strongcvx>0$. There exist some constants $\polyshort,\polyshortprime>0$ such that for any $\delta\in(0,1/2)$ and any index set $\samset$ with  $\samsetnum\geq \polyshortprime\log(1/\delta)$, with probability at least $1-\delta$, 
         $\vecnorm{\estatepar-\trueatepar}{2}\leq \frac{\polyshort\sqrt{\log(1/\delta)}}{\sqrt{\samsetnum}}$.
        Here, the constants $\polyshort,\polyshortprime$ depend polynomially on the parameters $(\datepar,1/\strongcvx,\boundatepar,\bZfunzero,\bZfunone,\bZfuntwo)$.
        \end{enumerate}
    \end{myassumption}

In Assumption~\ref{ass:zest_as_ate}, we posit that the ATE estimator $\estate$ is given by the first coordinate of some $Z$-estimator. 
Specifically, Assumption~\ref{ass:zest_as_ate:a} assumes that the true ATE $\trueate$ equals the  first coordinate of the true parameter $\trueatepar$ of the $Z$-estimation problem. Note that this can be generalized to any linear function of $\trueatepar$ by a simple change of variables.
Assumption~\ref{ass:zest_as_ate:c} imposes standard smoothness conditions on the $Z$-function and its derivatives. Assumption~\ref{ass:zest_as_ate:b} assumes $\sqrt{\nexp}$-convergence of the $Z$-estimator. This is satisfied when e.g., the $Z$-function is the gradient of some convex loss. In fact, a sufficient condition for Assumption~\ref{ass:zest_as_ate:b} is the following convexity condition~\ref{ass:convexity_atepar}.
 We refer to Lemma~\ref{lm:convexity_atepar} for more details.
\begin{myassumption}{Con}{ass:convexity_atepar}
 $\nabla\PopZfun(\atepar)\succeq \bzero$ for any $\atepar\in\ateparspace$ and $\nabla\PopZfun(\trueatepar)\succeq \gamma \IdMat$ for some constant $\gamma>0$.
\end{myassumption}

It is readily verified that the ordinary least squares (OLS) estimator satisfies Assumption~\ref{ass:zest_as_ate} when 
the observed outcome $\expres_i$ is linear in the covariates $\expcov_i$ and the treatment assignment $\exptre_i$. 
Additionally, under proper conditions, the inverse propensity weighted (IPW) estimator~\citep{horvitz1952generalization} satisfies Assumption~\ref{ass:zest_as_ate} when the true propensity score $\prop(\expcov_i)\defn \P(\exptre_i=1|\expcov_i)$ follows a logistic model, {\it i.e.}, $\prop(\expcov_i) = \exp(\covariate_i^{\expshort\top }\omega^\star)/(1+\exp(\covariate_i^{\expshort\top } \omega^\star))$ for some $\omega^\star\in\R^{\dexp}$, and is estimated via logistic regression (see Example~3 in~\citet{lin2024worthwhile}).

\subsubsection{Notation}\label{sec:additional_notations}
We now restate and clarify the notation. For any set $\samset \subseteq [\nexp]$, we define $\expsam_\samset \defn (\expsam_i)_{i \in \samset}$ as the subset of experimental samples indexed by $\samset$. In particular, recall that  $\expsam_{\fold_i}$ denotes the set of experimental samples in the $i$-th fold, for $i \in [\numfold]$. We write $\expsam_{[\nexp]} = \expsam$ and $\obssam_{[\nobs]} = \obssam$ to denote the full set of experimental and observational samples, respectively. With this notation, the full dataset is $\dset = (\expsam, \obssam) = (\expsam_{[\nexp]}, \obssam_{[\nobs]})$, and the dataset excluding the $i$-th experimental fold is $\dset_{-i} = (\expsam_{-\fold_i}, \obssam) = (\expsam_{[\nexp] \setminus \fold_i} , \obssam_{[\nobs]})$, for $i \in [\numfold]$. We  write $\EstPar(\regu)=\EstPar(\regu;\dset)$ to specify the dependence of $\EstPar(\regu)$ on $\dset.$
We also define $\dsetobs \defn \obssam = \obssam_{[\nobs]}$.

For each subset of experimental samples $\expsam_{\samset}=(\expsam_j)_{j\in[\samset]}$, 
we write the experimental loss $\lexp(\atefun(\Par); \expsam_{\samset}) = (\atefun(\Par) - \estate(\expsam_{\samset}))^2$, where $\estate(\expsam_{\samset})$ denotes an estimate of the average treatment effect (ATE) based on the samples indexed by $\samset$. We also write $\lexp(\atefun(\Par); \distexp) = (\atefun(\Par) - \ate)^2$ for the population loss. In addition, for any function $f$, with slight abuse of notation, 
we let $\empmean{\samset}[f(\expsam)] \defn \frac{1}{\samsetnum} \sum_{j \in \samset} f(\expsam_j)$ denote the empirical average over a subset $\samset$ of the experimental samples.

We use $\vecnorm{\cdot}{2}$ to denote the Euclidean norm for vectors and $\opnorm{\cdot}$ to denote the spectral norm (or operator norm) for matrices and third-order tensors. Concretely, for a third-order tensor $\mathcal{T} \in \mathbb{R}^{d_1 \times d_2 \times d_3}$, its spectral norm (or operator norm) is defined as
\[
\opnorm{\mathcal{T}} \defn \sup_{\vecnorm{x}{2} = \vecnorm{y}{2} = \vecnorm{z}{2} = 1} \sum_{i=1}^{d_1} \sum_{j=1}^{d_2} \sum_{k=1}^{d_3} \mathcal{T}_{ijk} x_i y_j z_k.
\]

Throughout the  proofs, we use $\polyshort,\polyshortprime>0$ to denote  constants that depend polynomially on the parameters in the assumptions. We allow their values to change from place to place. More specifically, when Assumptions~\ref{ass:obs_ate}~and~\ref{ass:linear_ate} hold, 
the constants $\polyshort=\poly$ (or $\polyshortprime=\polyprime$) depends polynomially on the parameters $(\boundobspar,1/\lboundobsh,\boundobsh,\boundobsthr;
\boundZfuntil,\boundate,\\\boundatediff,\boundatelin,\boundatenum)$. Alternatively, when Assumption~\ref{ass:obs_ate}~and~\ref{ass:zest_as_ate} hold, the constants $\polyshort=\polyz$ (or $\polyshortprime=\polyprimez$) depend polynomially on the parameters $(\boundobspar,1/\lboundobsh,\boundobsh,\boundobsthr;\datepar,1/\strongcvx,\boundatepar,\bZfunzero,\bZfunone,\bZfuntwo)$. The set of parameters the constants $\polyshort, \polyshortprime$ depend on should be clear from context, as it only depends on what assumptions are made. We therefore omit the explicit dependence in the notation.

\subsection{Proof of Theorem~\ref{thm:cv_main}}\label{sec:pf_thm_cv_main}

Under Assumptions~\ref{ass:obs_ate}~and~\ref{ass:linear_ate} and the sample size condition $\sqrt{\nexp}\geq \polyshortprime {\numfold}(\log^{1.5}\numfold+\log^{0.5}(1/\delta))$ in Eq.~\eqref{eq:sample_size_condition}, we will show the following two lemmas.
\begin{lemma}\label{lm:lexp_finite}
    For any $\delta\in(0,1/2)$,  we have with probability at least $1-\delta$ that, for all $\regu\in[0,1]$,
\begin{align*}
      &\quad\Big|\frac{1}{\numfold} \sum_{i=1}^\numfold \lexp(\atefun(\EstPar(\regu; \dset_{-i})); \distexp) -\frac{1}{\numfold} \sum_{i=1}^\numfold \lexp(\atefun(\EstPar(\regu; \dset_{-i})); \expsam_{\fold_i} )
      \\&\qquad+
      \frac{1}{\numfold} \sum_{i=1}^\numfold (\estate(\expsam_{\fold_i})-\ate)^2\Big|\\
      &\leq 
      \polyshort \frac{\log(1/\delta)}{\nexp}
      + 
      \polyshort\frac{\sqrt{\log(1/\delta)}}{\sqrt{\nexp}}
      \cdot 
      \sqrt{\lexp(\atefun(\EstPar(\regu; \dset)); \distexp).
      }
  \end{align*}
\end{lemma}

See the proof in Section~\ref{sec:pf_lexp_finite}.

\begin{lemma}\label{lm:lexp_cv}
    For any $\delta\in(0,1/2)$,  we have with probability at least $1-\delta$ that, for all $\regu\in[0,1]$,
\begin{align*}
      &\quad\Big|\frac{1}{\numfold} \sum_{i=1}^\numfold \lexp(\atefun(\EstPar(\regu; \dset_{-i})); \distexp) -
      \lexp(\atefun(\EstPar(\regu; \dset)); \distexp)
      \Big|\\
      &\leq
      \polyshort \frac{\log(1/\delta)}{\nexp}
  + 
  \polyshort\frac{\sqrt{\log(1/\delta)}}{\sqrt{\nexp}}
  \cdot 
  \sqrt{\lexp(\atefun(\EstPar(\regu; \dset)); \distexp)
  }.
      \end{align*}
\end{lemma}

See the proof in Section~\ref{sec:pf_lexp_cv}.

With the two lemmas at hand, we are ready to prove Theorem~\ref{thm:cv_main}.
Let 
\begin{align*}
\trueregu\defn\argmin_{\regu\in[0,1]}\lexp(\atefun(\EstPar(\regu;\dset));\distexp)
=
\argmin_{\regu\in[0,1]}(\atefun(\EstPar(\trueregu;\dset))-\trueate)^2
\end{align*}
be the optimal regularization parameter that minimizes the estimation error 
given the dataset $\dset$. Since $\EstPar(0;\dset)=\estate(\expsam_{[\nexp]})$ satisfies $|\estate(\expsam_{[\nexp]})-\trueate|\leq \boundatediff\sqrt{\log(1/\delta)}/\sqrt{\nexp}$  with probability at least $1-\delta$ by Assumption~\ref{ass:linear_ate:a}, we have
\begin{subequations}
\begin{align}
\lexp(\atefun(\EstPar(\trueregu;\dset));\distexp)
\leq 
\lexp(\atefun(\EstPar(0;\dset));\distexp)
\leq 
\polyshort\frac{\log(1/\delta)}{\nexp}
\label{eq:main_pf_eq_0}
\end{align}
with probability at least $1-\delta$.

Let $\regucst$ denote the averaged squared error $\sum_{i=1}^\numfold (\estate(\expsam_{\fold_i})-\ate)^2/\numfold$ independent of $\regu$. Therefore, combining Lemmas~\ref{lm:lexp_finite}and~\ref{lm:lexp_cv}, and applying a triangle inequality, we obtain
\begin{align}
   &\quad~ \Big|
    \frac{1}{\numfold} \sum_{i=1}^\numfold \lexp(\atefun(\EstPar(\regu; \dset_{-i})); \expsam_{\fold_i} )
      -
    \regucst
      - 
      \lexp(\atefun(\EstPar(\regu;\dset));\distexp)
      \Big|\notag\\
      &
      \leq
      \polyshort\frac{\log(1/\delta)}{\nexp}
      + 
      \polyshort\frac{\sqrt{\log(1/\delta)}}{\sqrt{\nexp}}
      \cdot 
      \sqrt{\lexp(\atefun(\EstPar(\regu; \dset)); \distexp)
      }\label{eq:main_pf_eq_1}
\end{align}
for all $\regu\in[0,1]$ with probability at least $1-\delta$. 
\end{subequations}

Consequently,  on the event where Eqs.~\eqref{eq:main_pf_eq_0}~and~\eqref{eq:main_pf_eq_1} hold, we have 
\begin{subequations}
\begin{align}
    &\quad
\frac{1}{\numfold} \sum_{i=1}^\numfold \lexp(\atefun(\EstPar(\estregu; \dset_{-i})); \expsam_{\fold_i} )
-
\regucst\notag
\\
&\geq
\lexp(\atefun(\EstPar(\estregu;\dset));\distexp)
-
\Big(
\polyshort\frac{\log(1/\delta)}{\nexp}
+ 
\polyshort\frac{\sqrt{\log(1/\delta)}}{\sqrt{\nexp}}
\cdot 
\sqrt{\lexp(\atefun(\EstPar(\estregu; \dset)); \distexp)
}
\Big),\label{eq:main_pf_eq_2}
\\
&\quad
\frac{1}{\numfold} \sum_{i=1}^\numfold \lexp(\atefun(\EstPar(\estregu; \dset_{-i})); \expsam_{\fold_i} )
-
\regucst\notag
\\
&\leq
\frac{1}{\numfold} \sum_{i=1}^\numfold \lexp(\atefun(\EstPar(\trueregu; \dset_{-i})); \expsam_{\fold_i} )
-
\regucst \notag
\\
&\leq
\polyshort\frac{\log(1/\delta)}{\nexp}
+ 
\polyshort\frac{\sqrt{\log(1/\delta)}}{\sqrt{\nexp}}
\cdot 
\sqrt{\lexp(\atefun(\EstPar(\trueregu; \dset)); \distexp)
}\notag
\\
&\leq 
\polyshort\frac{\log(1/\delta)}{\nexp}
\label{eq:main_pf_eq_3}.
\end{align}
Combining Eqs.~\eqref{eq:main_pf_eq_2}~and~\eqref{eq:main_pf_eq_3} and solving a quadratic inequality yields 
\begin{align*}
    \lexp(\atefun(\EstPar(\estregu;\dset));\distexp)\leq \polyshort\frac{\log(1/\delta)}{\nexp}
\end{align*}
with probability at least $1-\delta$.  The proof is completed by noting that 
\begin{align*}
    \lexp(\atefun(\EstPar(\estregu;\dset));\distexp)
=(\atefun(\EstPar(\estregu;\dset))-\trueate)^2
\leq (\boundate+\boundobspar)^2\leq \polyshort
\end{align*}
almost surely.

\end{subequations}

\newpage

\newpage

\newpage

\subsubsection{Proof of Lemma~\ref{lm:lexp_finite}}\label{sec:pf_lexp_finite}
Adopt the shorthands $\estate_i=\estate(\expsam_{\fold_i}),\estate_{-i}=\estate(\{\expsam_{\fold_j},j\neq i \}),$ \\$
\estate=\estate(\{\expsam_{\fold_j},j\in[\numfold] \})$.
Also define 
\begin{align}
    \EstPar(\regu;\dsetobs)\defn 
    \arg\min_{\Par\in\Parspace} \Big\{ (1 - \regu) \lexp(\atefun(\Par); \distexp) + \regu \lobs(\theta; \dsetobs) \Big\}.\label{eq:estpar_dsetobs_def}
\end{align}
  By some basic algebra, we have
  \begin{align*}
     \MoveEqLeft{ \frac{1}{\numfold} \sum_{i=1}^\numfold \lexp(\atefun(\EstPar(\regu; \dset_{-i})); \distexp) -\frac{1}{\numfold} \sum_{i=1}^\numfold \lexp(\atefun(\EstPar(\regu; \dset_{-i})); \expsam_{\fold_i} ) }\\
      &\qquad +
        \frac{1}{\numfold} \sum_{i=1}^\numfold (\estate_i-\ate)^2
\\
&=     
\frac{2}{\numfold}\sum_{i=1}^\numfold (\estate_i-\ate)(\atefun(\EstPar(\regu;\dset_{-i}))-\ate)\\
&=
\underbrace{\frac{2}{\numfold}\sum_{i=1}^\numfold (\estate_i-\ate)(\atefun(\EstPar(\regu;\dset_{-i}))-\atefun(\EstPar(\regu;\dsetobs)))}_{\revdef\remainderterm_1}\\
&\qquad +
\underbrace{\frac{2}{\numfold}\sum_{i=1}^\numfold (\estate_i-\ate)\cdot(\atefun(\EstPar(\regu;\dsetobs))-\ate)}_{\revdef\remainderterm_2}.
  \end{align*}
We make the following claims which will be shown at the end of the proof: 
\begin{subequations}
\begin{itemize}
    \item[1.] When $\regu\in(0,1]$, we have
\begin{align}
   &|\atefun(\EstPar(\regu;\dset_{-i}))-\atefun(\EstPar(\regu;\dsetobs))
   - 
   2(1-\regu)(\estate_{-i}-\ate)\onehot_1^\top\hesshort(\regu)^{-1} \onehot_1
   |\nonumber\\
   &
   \qquad \leq \polyshort\frac{\log(\numfold/\delta)}{{\nexp}},
   \label{eq:temp_claim_1}
    \end{align}
    for all $i\in[\numfold]$ 
    for some $\polyshort=\poly>0$ with probability at least $1-\delta$,
    where 
    \begin{align*}
        \hesshort(\regu)\defn  \regu  \nabla_\Par^2 \lobs(\EstPar(\regu;\dsetobs);\dsetobs)
+ (1-\regu)\nabla_\Par^2 \lexp(\atefun(\EstPar(\regu;\dsetobs));\distexp).
    \end{align*} 
    Moreover, when $\regu=0$, we have 
    $
        |\atefun(\EstPar(0;\dset_{-i}))-\atefun(\EstPar(0;\dsetobs))|
        =| \estate_{-i} - \trueate|.$
        \item[2.]
       Similarly,  when $\regu\in(0,1]$, we have
\begin{align}
   &|\atefun(\EstPar(\regu;\dset))-\atefun(\EstPar(\regu;\dsetobs))
   - 
   2(1-\regu)(\estate-\ate)\onehot_1^\top\hesshort(\regu)^{-1} \onehot_1
   |\nonumber\\
   &\qquad\leq \polyshort\frac{\log(1/\delta)}{{\nexp}},
   \label{eq:temp_claim_15}
    \end{align}
    for some $\polyshort=\poly>0$ with probability at least $1-\delta$.
  In addition, when $\regu=0$, we have 
    $
        |\atefun(\EstPar(0;\dset))-\atefun(\EstPar(0;\dsetobs))|
        =| \estate - \trueate|.
        $
        \item[3.]
         There exists some $\polyshort=\poly>0$ such that
        \begin{align}
           \sup_{\regu\in(0,1]} (1-\regu)\onehot_1^\top\hesshort(\regu)^{-1}\onehot_1\leq \polyshort.
            \label{eq:temp_claim_2}
        \end{align}
\end{itemize}
\end{subequations}
By claim~\eqref{eq:temp_claim_1}, when $\regu>0$, we have
\begin{align*}
  \remainderterm_1  &
   =
  \frac{4(1-\regu)\onehot_1^\top\hesshort(\regu)^{-1}\onehot_1}{\numfold}\sum_{i=1}^\numfold (\estate_i-\ate)(\estate_{-i}-\ate)
+\remainder_1
\end{align*}
for some $\remainder_1$  such that $|\remainder_1|\leq \polyshort\log^{1.5}(\numfold/\delta)/\nexp/\sqrt{\nexp/\numfold}\leq\polyshort\log(1/\delta)/\nexp$
with probability at least $1-\delta$.  Moreover, we have by Eq.~\eqref{eq:concentration_estate_2} in Lemma~\ref{lm:concentration_estate} that
\begin{align*}
    \frac{1}{\numfold}\sum_{i=1}^\numfold (\estate_{-i}-\ate)(\estate_{i}-\ate)
    &\leq \polyshort\frac{\log(1/\delta)}{\nexp}
    \end{align*}
    with probability at least $1-\delta$. Combining the last two bounds and using claim~\eqref{eq:temp_claim_2} yields
    \begin{align*}
        \remainderterm_1
        &\leq \polyshort\frac{\log(1/\delta)}{\nexp}
        \end{align*}
        for all $\regu\in(0,1]$ for some $\polyshort=\poly>0$ with probability at least $1-\delta$.
The bound on $\remainder_1$  for the case $\regu=0$ is similar and we thus omit the details.

Moreover, for $\remainderterm_2$, we have with probability at least $1-\delta$ that,  for all $\regu\in[0,1]$,
\begin{align*}
  |\remainderterm_2| &
    \leq 
   |\frac{2}{\numfold}\sum_{i=1}^\numfold (\estate_i-\ate)| \cdot \sqrt{\lexp(\atefun(\EstPar(\regu;\dsetobs)); \distexp)}\\
      &\leq
      \polyshort\frac{\sqrt{\log(1/\delta)}}{\sqrt{\nexp}}
      \cdot 
      \sqrt{\lexp(\atefun(\EstPar(\regu;\dsetobs)); \distexp)},
\end{align*}
where the second inequality follows from Eq.~\eqref{eq:concentration_estate_1} in Lemma~\ref{lm:concentration_estate}.
Finally, note that
\begin{align*}
    &\lexp(\atefun(\EstPar(\regu;\dsetobs)); \distexp)\\
    &\qquad\leq 
    2 
    \lexp(\atefun(\EstPar(\regu;\dset)); \distexp)
    +
    2(\atefun(\EstPar(\regu;\dsetobs))-\atefun(\EstPar(\regu;\dset)))^2
    \\
   & \qquad \leq 
   2 
   \lexp(\atefun(\EstPar(\regu;\dset)); \distexp)
   +
    \frac{\polyshort\log(1/\delta)}{\nexp}
\end{align*}   
where the first inequality uses $(a+b)^2\leq 2(a^2+b^2)$, and  the second inequality follows from Lemma~\ref{lm:inverse_first_entry}.
Combining the bounds on $\remainderterm_1$, $\remainderterm_2$ and $\lexp(\atefun(\EstPar(\regu;\dsetobs)); \distexp)$ yields the desired result.


\paragraph{Proof of claim~\eqref{eq:temp_claim_1}.}
Note that $\EstPar(\regu;\dsetobs),\EstPar(\regu;\dset_{-i})$ are empirical risk minimizers. Taking the derivatives with respect to $\theta$, we have
\begin{align*}
    2(1-\regu)(\atefun(\EstPar(\regu;\dset_{-i}))-\estate_{-i})\cdot\nabla_\Par\atefun(\EstPar(\regu;\dset_{-i}))
    + \regu \nabla_\Par\lobs(\EstPar(\regu;\dset_{-i});\dsetobs)
    &=0,\\
      2(1-\regu)(\atefun(\EstPar(\regu;\dsetobs))-\ate)\cdot\nabla_\Par\atefun(\EstPar(\regu;\dsetobs))
    + \regu \nabla_\Par\lobs(\EstPar(\regu;\dsetobs);\dsetobs)
    &=0.
\end{align*}
Introduce the shorthand $\diffpar_i\defn \EstPar(\regu;\dset_{-i})-\EstPar(\regu;\dsetobs)$.
Taking the difference and performing a Taylor expansion yields
\begin{align}
    \hesshort_i(\regu)(\EstPar(\regu;\dset_{-i})
    -
    \EstPar(\regu)
    ) &= 2(1-\regu)(\estate_{-i}-\ate)\cdot \nabla_\Par\atefun(\EstPar(\regu;\dset_{-i}))\nonumber\\
    &=2 (1-\regu)(\estate_{-i}-\ate)\cdot \onehot_1
    ,\label{eq:taylor_diffpar_1}
\end{align} 
where 
\begin{align*}
    \hesshort_i(\regu)
    &\defn
    \regu \int_0^1 \nabla_\Par^2 \lobs(\EstPar(\regu;\dsetobs)+t\diffpar_i;\dsetobs)dt
    \\
    &\qquad +
    (1-\regu) \int_0^1 \nabla_\Par^2 \lexp(\atefun(\EstPar(\regu;\dsetobs))+t\diffpar_i;\distexp)dt\\
    &=
    \regu \int_0^1 \nabla_\Par^2 \lobs(\EstPar(\regu;\dsetobs)+t\diffpar_i;\dsetobs)dt
    + 2(1-\regu)\Eone_{11}
\end{align*} 
with $\onehot_1\defn (1,0,\cdots,0)^\top\in\R^{\Pardim}$ and $\Eone_{11}\in\R^{\Pardim\times\Pardim}$ being the matrix where the $(1,1)$-th entry is one and all other entries are zero. Recall that $\hesshort(\regu)=  \regu  \nabla_\Par^2 \lobs(\EstPar(\regu;\dsetobs);\dsetobs)
+ (1-\regu)\nabla_\Par^2 \lexp(\atefun(\EstPar(\regu;\dsetobs));\distexp)$. By Lemma~\ref{lm:inverse_first_entry}, we have $\vecnorm{\diffpar_i}{2}\leq 
\polyshort |\estate_{-i}-\ate|$. 
Therefore,
\begin{align*}
   \MoveEqLeft{ \vecnorm{\diffpar_i- 2(1-\regu)(\estate_{-i}-\ate)\hesshort(\regu)^{-1} \onehot_1}{2}}\\
   &=
  \vecnorm{2(1-\regu)(\estate_{-i}-\ate)[\hesshort_i(\regu)^{-1}-\hesshort(\regu)^{-1}] \onehot_1}{2}\\
   &=
   \vecnorm{2(1-\regu)(\estate_{-i}-\ate)\hesshort(\regu)^{-1}
   \big(
   \hesshort(\regu)
   -
   \hesshort_i(\regu)
   \big)
   \hesshort_i(\regu)^{-1}
   \onehot_1}{2}\\
   &\leq
  \opnorm{\hesshort(\regu)^{-1}}\opnorm{\hesshort(\regu)-\hesshort_i(\regu)}
  \vecnorm{2(1-\regu)(\estate_{-i}-\ate)\hesshort_i(\regu)^{-1} \onehot_1}{2}\\
  &\leq
  \frac{\boundobsthr}{\lboundobsh}\cdot\vecnorm{\diffpar_i}{2}^2\leq \polyshort |\estate_{-i}-\ate|^2\leq \frac{\polyshortprime\log(1/\delta)}{\nexp},
\end{align*}
for some $\polyshortprime=\polyprime>0$ with probability at least $1-\delta$, 
where the first inequality uses Eq.~\eqref{eq:taylor_diffpar_1} and the last inequality follows from Assumption~\ref{ass:zest_as_ate}. Finally, applying a union bound over all $i\in[\numfold]$ yields the desired result in Eq.~\eqref{eq:temp_claim_1}. The case $\regu=0$ follows immediately from Lemma~\ref{lm:inverse_first_entry}.

\paragraph{Proof of claim~\eqref{eq:temp_claim_15}.}
The proof of claim~\eqref{eq:temp_claim_15} follows from the same arguments as in the proof of claim~\eqref{eq:temp_claim_1} since $\EstPar(\regu;\dset_{-i})$ in the proof of claim~\eqref{eq:temp_claim_1} can be replaced by $\EstPar(\regu;\dset)$ without loss of generality.

\paragraph{Proof of claim~\eqref{eq:temp_claim_2}.}
By the expression of Schur's complement, we have
\begin{align*}
   \MoveEqLeft{[(1-\regu) \onehot_1^\top\hesshort(\regu)^{-1}\onehot_1]^{-1}}\\
    &=
   \frac{1}{1-\regu}\big( \hesshort(\regu)_{11} - \hesshort(\regu)_{1,2:\Pardim}^\top(\hesshort(\regu)_{2:\Pardim,2:\Pardim})^{-1}\hesshort(\regu)_{2:\Pardim,1}
    \big)
    \\
    &\geq 
    \frac{1}{1-\regu}\Big((\regu\lboundobsh + 2(1-\regu))
    -
   \regu\frac{\boundobsh^2}{\lboundobsh}\Big)
   =2+\frac{\regu}{1-\regu}\Big(\lboundobsh
   -\frac{\boundobsh^2}{\lboundobsh}
   \Big).
\end{align*}
Thus, we have $2+\frac{\regu}{1-\regu}\Big(\lboundobsh
-\frac{\boundobsh^2}{\lboundobsh}
\Big)\geq 1$ (and therefore $|(1-\regu) \onehot_1^\top\hesshort(\regu)^{-1}\onehot_1|\leq 1$) when $\regu\leq 1/\polyshort_1$ for some $\polyshort_1=\poly>0$ sufficiently large. On the other hand, when $\regu\geq \polyshort_1$, we have
\begin{align*}
    (1-\regu) \onehot_1^\top\hesshort(\regu)^{-1}\onehot_1
    \leq
    \frac{1-\regu}{\sigma_{\min}(\hesshort(\regu))}
    \leq  \frac{1-\regu}{\regu \lboundobsh}
    \leq \frac{\polyshort_1 }{\lboundobsh}\leq \polyshort.
\end{align*} 
Combining the two cases completes the proof.

\subsubsection{Proof of Lemma~\ref{lm:lexp_cv}}\label{sec:pf_lexp_cv}
By definition of  $\lexp$, we have
\begin{align*}
    \MoveEqLeft{\Big|\frac{1}{\numfold} \sum_{i=1}^\numfold \lexp(\atefun(\EstPar(\regu; \dset_{-i})); \distexp) -
      \lexp(\atefun(\EstPar(\regu; \dset)); \distexp)
      \Big|}
\\
      &=
      \underbrace{\frac{1}{\numfold}\sum_{i=1}^\numfold (\atefun(\EstPar(\regu;\dset_{-i}))
      -
      \atefun(\EstPar(\regu;\dset))
      )^2}_{\revdef\remainderterm_3}\\
      &\qquad
      + \underbrace{\frac{2}{\numfold}\sum_{i=1}^\numfold 
      ( \atefun(\EstPar(\regu;\dset))-\ate)
      (\atefun(\EstPar(\regu;\dset_{-i}))
      -
      \atefun(\EstPar(\regu;\dset))
      )}_{\revdef\remainderterm_4}.
\end{align*}
Similarly to the proof of Lemma~\ref{lm:lexp_finite}, we claim that
\begin{subequations}
\begin{itemize}
    \item[1.] When $\regu\in(0,1]$, we have
\begin{align}
   |\atefun(\EstPar(\regu;\dset_{-i}))-\atefun(\EstPar(\regu;\dset))
   - 
   2(1-\regu)(\estate_{-i}-\estate)\onehot_1^\top\hesshorttil(\regu)^{-1} \onehot_1
   |\nonumber\\
   \qquad \leq \polyshort\frac{\log(\numfold/\delta)}{{\nexp}},
   \label{eq:temp_claim_1_alter}
    \end{align}
    for all $i\in[\numfold]$ 
    for some $\polyshort=\poly>0$ with probability at least $1-\delta$,
    where 
    \begin{align*}
        \hesshorttil(\regu)\defn  \regu  \nabla_\Par^2 \lobs(\EstPar(\regu;\dset);\dsetobs)
+ (1-\regu)\nabla_\Par^2 \lexp(\atefun(\EstPar(\regu;\dset));\expsam_{[\nexp]}).
    \end{align*} 
    Moreover, when $\regu=0$, we have 
    $
        |\atefun(\EstPar(0;\dset_{-i}))-\atefun(\EstPar(0;\dset))|
        =| \estate_{-i} - \estate|.$
        \item[2.] There exists some $\polyshort=\poly>0$ such that
        \begin{align}
           \sup_{\regu\in(0,1]} (1-\regu)\onehot_1^\top\hesshorttil(\regu)^{-1}\onehot_1\leq \polyshort.
            \label{eq:temp_claim_2_alter}
        \end{align}
\end{itemize}
\end{subequations}
The proof of this claim will be given momentarily.

With these two claims at hand and using Eq.~\eqref{eq:concentration_estate_3}~and Assumption~\ref{ass:linear_ate:a}, we have
    \begin{align*}
       \remainderterm_3 
       &
\leq
\frac{4(1-\regu)^2(\onehot_1^\top\hesshorttil(\regu)^{-1} \onehot_1)^2}{\numfold}\sum_{i=1}^\numfold 
(\estate_{-i}-\estate)^2
+\remainder_3 
\leq \frac{\polyshort}{\numfold}\sum_{i=1}^\numfold 
(\estate_{-i}-\estate)^2
+\remainder_3 \\
&
\leq  \frac{\polyshort}{\numfold}\sum_{i=1}^\numfold 
(\estate_{-i}-\trueate)^2 + (\estate-\trueate)^2
+\remainder_3 \leq \frac{\polyshort\log(1/\delta)}{\nexp} + \remainder_3
    \end{align*} for some $\remainder_3$ such that $|\remainder_3|\leq 
    \log^2(\numfold/\delta)/(\nexp)^2$ for all $\regu\in(0,1]$
      with probability at least $1-\delta$. The bound on $\remainderterm_3$  when $\regu=0$ follows similarly.
      Thus we have 
      \begin{align*}|\remainderterm_3|\leq \frac{\polyshort\log(1/\delta)}{\nexp}
        \end{align*}
        for all $\regu\in[0,1]$ with probability at least $1-\delta$ since $\sqrt{\nexp}\geq \polyshort\numfold(\log^{1.5}(\numfold)+\log^{0.5}(1/\delta))$.

        Moreover, for $\remainderterm_4$, we have by the Cauchy-Schwarz
        inequality that
    \begin{align*}
        \remainderterm_4
      &\leq 
      \sqrt{\lexp(\EstPar(\regu;\dset);\distexp)}\cdot 
      \sqrt{\remainderterm_3}
    \leq
\polyshort\sqrt{\frac{\log(1/\delta)}{\nexp}}\cdot
\sqrt{\lexp(\EstPar(\regu;\dset);\distexp)}
    \end{align*}
    for all $\regu\in(0,1]$ with probability at least $1-\delta$. Combining the bounds on $\remainderterm_3$ and $\remainderterm_4$ completes the proof.

\paragraph{Proof of claim~\eqref{eq:temp_claim_1_alter}.}
Since $\EstPar(\regu;\dset),\EstPar(\regu;\dset)$ are both assumed to be the empirical risk minimizer on the respective datasets, we have
\begin{align*}
    2(1-\regu)(\atefun(\EstPar(\regu;\dset_{-i}))-\estate_{-i})\cdot\nabla_\Par\atefun(\EstPar(\regu;\dset_{-i}))
  + \regu \nabla_\Par\lobs(\EstPar(\regu;\dset_{-i});\dsetobs)
  &=0.\\
    2(1-\regu)(\atefun(\EstPar(\regu;\dset))-\estate)\cdot\nabla_\Par\atefun(\EstPar(\regu;\dset))
  + \regu \nabla_\Par\lobs(\EstPar(\regu;\dset);\dsetobs)
  &=0.
\end{align*}
Let $\diffpartwo_i\defn \EstPar(\regu;\dset_{-i})-\EstPar(\regu;\dset)$. Taking the difference and performing a Taylor expansion yields
\begin{align}
  \hesshorttil_i(\regu)(\EstPar(\regu;\dset_{-i})
  -
  \EstPar(\regu;\dset)
  ) 
  &= 2(1-\regu)(\estate_{-i}-\estate)\cdot \onehot_1,\label{eq:taylor_diffpartwo_1}
\end{align}
where
\begin{align*}
    \hesshorttil_i(\regu)
    &\defn
    \regu \int_0^1 \nabla_\Par^2 \lobs(\EstPar(\regu;\dset)+t\diffpartwo_i;\dsetobs)dt\\
    &\qquad
    +
    (1-\regu) \int_0^1 \nabla_\Par^2 \lexp(\atefun(\EstPar(\regu;\dset))+t\diffpartwo_i;\expsam_{[\nexp]})dt
    \\
    &= 
    \regu \int_0^1 \nabla_\Par^2 \lobs(\EstPar(\regu;\dset)+t\diffpartwo_i;\dsetobs)dt
    +
   2 (1-\regu) 
   \Eone_{11}
    \end{align*}
    with $\onehot_1\defn (1,0,\cdots,0)^\top\in\R^{\Pardim}$ and $\Eone_{11}\in\R^{\Pardim\times\Pardim}$ being the matrix where the $(1,1)$-th entry is one and all other entries are zero. Recall that $\hesshorttil(\regu)\defn  \regu  \nabla_\Par^2 \lobs(\EstPar(\regu;\dset);\dsetobs)
    + (1-\regu)\nabla_\Par^2 \lexp(\atefun(\EstPar(\regu;\dset));\expsam_{[\nexp]})$. By Lemma~\ref{lm:inverse_first_entry}, we have $\vecnorm{\diffpar_i}{2}\leq 
    \polyshort |\estate_{-i}-\estate|$. 
    Therefore, similar to the proof of Lemma~\ref{lm:lexp_finite}, 
    \begin{align*}
       \MoveEqLeft{\vecnorm{\diffpartwo_i- 2(1-\regu)(\estate_{-i}-\ate)\hesshorttil(\regu)^{-1} \onehot_1}{2}}
       \\
       &=
      \vecnorm{2(1-\regu)(\estate_{-i}-\estate)[\hesshorttil_i(\regu)^{-1}-\hesshorttil(\regu)^{-1}] \onehot_1}{2}\\
       &=
       \vecnorm{2(1-\regu)(\estate_{-i}-\estate)\hesshorttil(\regu)^{-1}
       \big(
       \hesshorttil(\regu)
       -
       \hesshorttil_i(\regu)
       \big)
       \hesshorttil_i(\regu)^{-1}
       \onehot_1}{2}\\
       &\leq
      \opnorm{\hesshorttil(\regu)^{-1}}\opnorm{\hesshorttil(\regu)-\hesshorttil_i(\regu)}
      \vecnorm{2(1-\regu)(\estate_{-i}-\estate)\hesshorttil_i(\regu)^{-1} \onehot_1}{2}\\
      &\leq
      \frac{\boundobsthr}{\lboundobsh}\cdot\vecnorm{\diffpartwo_i}{2}^2\leq \polyshort |\estate_{-i}-\estate|^2\leq \frac{\polyshortprime\log(1/\delta)}{\nexp},
    \end{align*}
    for some $\polyshortprime=\polyprime>0$ with probability at least $1-\delta$, 
    where the first inequality uses Eq.~\eqref{eq:taylor_diffpar_1} and the last inequality follows from Assumption~\ref{ass:linear_ate:a} and a triangle inequality. Finally, applying a union bound over all $i\in[\numfold]$ gives Eq.~\eqref{eq:temp_claim_1_alter}. The case $\regu=0$ follows immediately from Lemma~\ref{lm:inverse_first_entry}.

    \paragraph{Proof of claim~\eqref{eq:temp_claim_2_alter}.}
    The proof follows from the same argument as the proof of claim~\eqref{eq:temp_claim_2} in the proof of Lemma~\ref{lm:lexp_finite}. We thus omit the details here.

    \subsection{Proof of Corollary~\ref{cor:robustness_atefun_estpar}}\label{sec:pf_cor_robustness_atefun_estpar}

    Note that  the experimental sample size condition~\eqref{eq:sample_size_condition} is satisfied  with $\delta =1/\nexp$ when $\numfold\leq  \polyshort\sqrt{\nexp}/\log^{1.5}\nexp$.
    Therefore, we have by Theorem~\ref{thm:cv_main} that 
    \begin{align*}
        (\atefun(\EstPar(\estregu;\dset))-\trueate)^2= \lexp(\atefun(\EstPar(\estregu;\dset));\distexp)
        \leq \frac{\polyshortprime\log(1/\delta)}{\nexp}
    \end{align*}
    with probability at least $1-\delta$ for any $\delta\geq 1/\nexp$. 
    Let $\event\defn\{(\atefun(\EstPar(\estregu;\dset))-\trueate)^2\geq \polyshortprime\log(\nexp)/\nexp\}$. Then we have $\Prob(\event)\leq 1/\nexp$.
    Thus, 
    \begin{align*}
       &\quad ~\E[ (\atefun(\EstPar(\estregu;\dset))-\trueate)^2]
       \\
       & =\int_0^\infty \Prob((\atefun(\EstPar(\estregu;\dset))-\trueate)^2\geq t)dt\\
        &
        = \int_0^{\polyshortprime\log\nexp/\nexp} \Prob((\atefun(\EstPar(\estregu;\dset))-\trueate)^2\geq t)dt
         \\
    &\qquad +\int_{\polyshortprime\log\nexp/\nexp}^{(\boundate+\boundobspar)^2} \Prob((\atefun(\EstPar(\estregu;\dset))-\trueate)^2\geq t)dt\\
    &\leq
    \int_0^{\polyshortprime\log\nexp/\nexp}
    \exp(-\polyshort \nexp t)dt
    +
    \int_{\polyshortprime\log\nexp/\nexp}^{(\boundate+\boundobspar)^2} \Prob(\event)dt
    \leq \frac{\polyshort}{\nexp}.
    \end{align*}
    This completes the proof.

    \subsection{Proof of Theorem~\ref{thm:mean_estimation_minimax}}\label{sec:pf_thm_mean_estimation_minimax}
    We prove the theorem by contradiction. Let $\smallgap\in[0,\boundmean/2]$ be some value which will be specified later. 
    If there exists some $\estmean\in\minmaxclass_{\meanconsta}$ such that 
    \begin{align}
        \sup_{\truemean\in[-\boundmean,\boundmean]}
        \E_{(\expout_i)_{i=1}^{\nexp},(\obsout_i)_{i=1}^{\nobs}\simiid \cN(\truemean,1)}[(\estmean-\truemean)^2]
        \leq \frac{\meanvala}{\nexp}\label{eq:pf_thm_mean_estimation_minimax:0}
    \end{align}
    for some value $\meanvala>0$, then we have by Chebyshev's inequality that, for $\truemean = 0$,
    \begin{align*}
        \P_{\substack{(\expout_i)_{i=1}^{\nexp}\simiid\cN(\truemean,1),\\(\obsout_i)_{i=1}^{\nobs}\simiid \cN(\truemean,1)}}\big[|\estmean((\expout_i)_{i=1}^{\nexp};(\obsout_i)_{i=1}^{\nobs})-\truemean|\geq \smallgap\big] \leq \frac{\sqrt{\meanvala/\nexp}}{\smallgap }.
    \end{align*}
    Suppose we have chosen $\smallgap$ such that 
    \begin{align}
        \P_{\substack{(\expout_i)_{i=1}^{\nexp}\simiid\cN(\truemean+2\smallgap,1),\\(\obsout_i)_{i=1}^{\nobs}\simiid \cN(\truemean,1)}}\big[|\estmean((\expout_i)_{i=1}^{\nexp};(\obsout_i)_{i=1}^{\nobs})-\truemean|\geq \smallgap\big] \leq \frac{1}{2}.\label{eq:pf_thm_mean_estimation_minimax:1}
    \end{align}
    Then it follows from the triangle inequality that 
    \begin{align*}
    \P_{\substack{(\expout_i)_{i=1}^{\nexp}\simiid\cN(\truemean+2\smallgap,1),\\(\obsout_i)_{i=1}^{\nobs}
    \simiid \cN(\truemean,1)}}\big[|\estmean((\expout_i)_{i=1}^{\nexp};(\obsout_i)_{i=1}^{\nobs})-(\truemean+2\smallgap)|\geq \smallgap\big] \geq \frac{1}{2},
    \end{align*}
    and therefore 
    \begin{align}
    \E_{\substack{(\expout_i)_{i=1}^{\nexp}\simiid\cN(\truemean+2\smallgap,1),\\(\obsout_i)_{i=1}^{\nobs}\simiid \cN(\truemean,1)}}[(\estmean-(\truemean+2\smallgap))^2]\geq\frac{\smallgap^2}{2}.\label{eq:pf_thm_mean_estimation_minimax:2}
    \end{align}
    We will show at the end of the proof that there exist some absolute constants $c_3,c_4>0$ such that, when $\meanvala\leq c_3$, one can choose $\smallgap = \min\{\sqrt{c_4\log(1/\meanvala)}/\sqrt{\nexp},{\boundmean}/{2}\}$ such that Eq.~\eqref{eq:pf_thm_mean_estimation_minimax:1} (and therefore Eq.~\ref{eq:pf_thm_mean_estimation_minimax:2}) holds.

    As a consequence of Eq.~\eqref{eq:pf_thm_mean_estimation_minimax:2}, $\estmean$ does not belong to the class $\minmaxclass_{\meanconsta}$ for  $\meanconsta\leq \min\{c_4\log(1/\meanvala)/2,\\{\nexp}/8\}$ when Eq.~\eqref{eq:pf_thm_mean_estimation_minimax:0} holds. 
     Therefore,
    conversely, for the absolute constant $\widetilde{c}_1\defn c_4\log(1/c_3)/2$ and 
      any $\meanconsta\in[\widetilde{c}_1,\nexp/8]$, Eq.~\eqref{eq:pf_thm_mean_estimation_minimax:0} is \emph{not} satisfied for any $\estmean\in\minmaxclass_{\meanconsta} $ with any $\meanvala< \exp(-2\meanconsta/c_4)\revdef \meanconstb$.  This completes the proof.

    \paragraph{Verification of Eq.~\eqref{eq:pf_thm_mean_estimation_minimax:1}.}
    Let $\truemean=0.$
    Denote the event $\{|\estmean((\expout_i)_{i=1}^{\nexp};(\obsout_i)_{i=1}^{\nobs})-\truemean|\geq \smallgap\}$ by $\event$. Introduce the shorthand notations $\P_{\truemean}$ and $\P_{\truemean+2\smallgap}$ 
    to denote the joint distribution $(\expout_i)_{i=1}^{\nexp}\simiid\cN(\truemean,1),
    (\obsout_i)_{i=1}^{\nobs}\simiid \cN(\truemean,1)$ 
    and  $(\expout_i)_{i=1}^{\nexp}\simiid\cN(\truemean+2\smallgap,1),
    (\obsout_i)_{i=1}^{\nobs}\simiid \cN(\truemean,1)$, respectively. 
    When $\P_{\truemean}(\event) \leq \frac{1}{8}$, 
    we have
    \begin{align*}
    2\nexp\smallgap^2
    &\overset{(i)}{=} 
    \KL(\P_{\truemean+2\smallgap}||\P_{\truemean})
    \\
    &\overset{(ii)}{\geq} 
    \P_{\truemean+2\smallgap}(\event) \log\frac{\P_{\truemean+2\smallgap}(\event)}{\P_{\truemean}(\event)}
    + (1-\P_{\truemean+2\smallgap}(\event)) \log\frac{1-\P_{\truemean+2\smallgap}(\event)}{1-\P_{\truemean}(\event)}
    \\
    &\geq
    \P_{\truemean+2\smallgap}(\event) \log\Big(\frac{1}{\P_{\truemean}(\event)}\Big)
    -
    \log 2
    \geq
    \Big(\P_{\truemean+2\smallgap}(\event)-\frac{1}{3}\Big) \log\Big(\frac{1}{\P_{\truemean}(\event)}\Big)
    \end{align*} 
    where step~(i) follows from the formula of KL divergence between two Gaussian distributions, and step~(ii) follows from data-processing inequality. Therefore, to ensure that $\P_{\truemean+2\smallgap}(\event) \leq \frac{1}{2}$, it suffices to choose $\smallgap$ such that 
    \begin{align}
       \frac{2\nexp\smallgap^2}{\log\frac{1}{\P_{\truemean}(\event)}}
       \leq \frac{2\nexp\smallgap^2}{\log(\smallgap/\sqrt{\meanvala/\nexp})}
       \leq
       \frac{1}{6},\text{ and }  8\sqrt{\frac{\meanvala}{\nexp}}\leq \smallgap\leq \frac{\boundmean}{2}.
       \label{eq:pf_thm_mean_estimation_minimax:3}
    \end{align} It can be verified that there exist  absolute constants $c_3,c_4>0$ sufficiently small such that 
     when $\meanvala\leq c_3$,  $\smallgap = \sqrt{c_4\log(1/\meanvala)}/\sqrt{\nexp}$  satisfies the conditions in Eq.~\eqref{eq:pf_thm_mean_estimation_minimax:3}.

    \subsection{Auxiliary lemmas}\label{sec:auxiliary_lemmas}

    \begin{lemma}\label{lm:inverse_first_entry}
        Let $\EstPar(\regu;\dsetobs)$ be defined as in Eq.~\eqref{eq:estpar_dsetobs_def}. Under the assumptions in Theorem~\ref{thm:cv_main}, when $\regu>0$, we have
        \begin{align*}
         \vecnorm{\EstPar(\regu;\dset_{-i})-\EstPar(\regu;\dsetobs)}{2}
         &\leq
         \Big(1+\frac{\boundobsh}{\lboundobsh}\Big)|\estate_{-i}-\trueate|,
         \\
         \vecnorm{\EstPar(\regu;\dset_{-i})-\EstPar(\regu;\dset)}{2}
        & \leq
         \Big(1+\frac{\boundobsh}{\lboundobsh}\Big)|\estate_{-i}-\estate|,
    \\
    \vecnorm{\EstPar(\regu;\dset)-\EstPar(\regu;\dsetobs)}{2}
    & \leq
    \Big(1+\frac{\boundobsh}{\lboundobsh}\Big)|\estate-\trueate|.
        \end{align*} When $\regu=0$, we have
        \begin{align*}
        |\atefun(\EstPar(\regu;\dset_{-i}))-\atefun(\EstPar(\regu;\dsetobs))|
        &=
        |\estate_{-i}-\trueate|,\\
        |\atefun(\EstPar(\regu;\dset_{-i}))-\atefun(\EstPar(\regu;\dset))|
        &=
        |\estate_{-i}-\estate|,\\
        |\atefun(\EstPar(\regu;\dset))-\atefun(\EstPar(\regu;\dsetobs))|
        &=
        |\estate-\trueate|.
        \end{align*}
    \end{lemma}
    
     See the proof  in Section~\ref{pf:lm:inverse_first_entry}.

     \begin{lemma}[Concentration properties of 
         $\estate$]\label{lm:concentration_estate}
         Under the assumptions in Theorem~\ref{thm:cv_main}, we have with probability at least $1-\delta$ that
         \begin{subequations}
         \begin{align}
          \frac{1}{\numfold}\sum_{i=1}^\numfold (\estate_{i}-\ate)
          &\leq \polyshort\frac{\sqrt{\log(1/\delta)}}{\sqrt{\nexp}},
          \label{eq:concentration_estate_1}
          \\
          \frac{1}{\numfold}\sum_{i=1}^\numfold (\estate_{-i}-\ate)(\estate_{i}-\ate)
          &\leq \polyshort\frac{\log(1/\delta)}{\nexp}
          \label{eq:concentration_estate_2},\\
          \frac{1}{\numfold}\sum_{i=1}^\numfold (\estate_{-i}-\ate)^2
          &\leq \polyshort\frac{\log(1/\delta)}{\nexp}
          \label{eq:concentration_estate_3}
         \end{align} for some constant $\polyshort=\poly>0$.
         \end{subequations}
     \end{lemma}
     See the proof in Section~\ref{sec:pf_concentration_estate}.

     \begin{lemma}[Assumption~\ref{ass:zest_as_ate} implies Assumption~\ref{ass:linear_ate}]\label{lm:linear_expansion_estate}
        Under Assumption~\ref{ass:zest_as_ate}, there exist some constants $\polyshort,\polyshortprime>0$ such that for any $\delta\in(0,1/2)$ and any index set $\samset$ with  $\samsetnum\geq \polyshortprime \log(1/\delta)$, we have with probability at least $1-\delta$ that
         \begin{align*}
             \vecnorm{\estatepar-\trueatepar + [\nabla\PopZfun(\trueatepar)]^{-1} \empmean{\samset}[ \Zfun(\expsam;\trueatepar)]}{2}
             &\leq
              \frac{\polyshort\log(1/\delta)}{\samsetnum},
             \\
               |\estate(\expsam_{\samset})-\trueate
               +\onehot_1^\top
               [\nabla\PopZfun(\trueatepar)]^{-1} \empmean{\samset}[ \Zfun(\expsam;\trueatepar)]  
               | 
               &\leq 
               \frac{\polyshort\log(1/\delta)}{\samsetnum}. 
         \end{align*} 
     \end{lemma}
      Consequently, Assumption~\ref{ass:linear_ate} is satisfied with $\Zfuntilone(\expsam)=\Zfun(\expsam;\trueatepar)$, $\estate(\expsam_{\samset})=\estatepar_1$ and 
    some $(\boundZfuntil,\boundate,\boundatediff,\boundatelin,\boundatenum)$ depending polynomially on the parameters $(\datepar,1/\strongcvx,\\\boundatepar,\bZfunzero,\bZfunone,\bZfuntwo)$ in Assumption~\ref{ass:zest_as_ate}.

     See the proof in Section~\ref{sec:pf_linear_expansion_estate}.

     \begin{lemma}[A sufficient condition for Assumption~\ref{ass:zest_as_ate:b}]\label{lm:convexity_atepar}
        Let Assumptions~\ref{ass:zest_as_ate:a},~\ref{ass:zest_as_ate:c}~and~\ref{ass:convexity_atepar} hold. Then $\sigma_{\min}(\nabla\PopZfun(\trueatepar))\geq \strongcvx$ and there exist some constants $\polyshort,\polyshortprime>0$ such that for any $\delta\in(0,1/2)$ and any index set $\samset$ with  $\samsetnum\geq \polyshortprime\log(1/\delta)$, with probability at least $1-\delta$, 
    \begin{align*}
        \vecnorm{\estatepar-\trueatepar}{2}\leq \frac{\polyshort\sqrt{\log(1/\delta)}}{\sqrt{\samsetnum}}.
    \end{align*}
    \end{lemma}
    See the proof in Section~\ref{sec:pf_convexity_atepar}.

\subsubsection{Proof of Lemma~\ref{lm:inverse_first_entry}}\label{pf:lm:inverse_first_entry}
    We only prove the bounds on $ \EstPar(\regu;\dset_{-i})-\EstPar(\regu;\dsetobs)$ (and $\atefun(\EstPar(\regu;\dset_{-i}))-\atefun(\EstPar(\regu;\dsetobs))$ when $\regu=0$). The bounds on $\EstPar(\regu;\dset_{-i})-\EstPar(\regu;\dset)$ and
    $\EstPar(\regu;\dset)-\EstPar(\regu;\dsetobs)$ follow from similar arguments.
    \\

    \myunder{Case 1: $\regu>0$:}
  from Eq.~\eqref{eq:taylor_diffpar_1}, we have
   \begin{align*}
    \EstPar(\regu;\dset_{-i})-\EstPar(\regu;\dsetobs)
    =
   2(1-\regu)(\estate_{-i}-\ate) \hesshort_i(\regu)^{-1} \onehot_1.
   \end{align*}
   It suffices to show $\vecnorm{(1-\regu) \hesshort_i(\regu)^{-1} \onehot_1}{2}\leq (1+\boundobsh/\lboundobsh)/2$.
Adopt the shorthands $\Htermone_i(\regu)\defn \int_0^1 \nabla_\Par^2 \lobs(\EstPar(\regu;\dsetobs)+t\diffpar_i;\dsetobs)dt$ and $\Htermtwo_{i,\eps}(\regu)\defn \eps\IdMat_{\Pardim}+(2-\eps)\Eone_{11}$ for  $\eps\geq 0$. Then we have
\begin{align*}
 \MoveEqLeft{\quad~  \vecnorm{(1-\regu) (\regu\Htermone_i(\regu)+(1-\regu)\Htermtwo_{i,\eps}(\regu))^{-1}\onehot_1}{2}}\\
    &=
    \vecnorm{\Htermtwo_{i,\eps}(\regu)^{-1}\onehot_1
    -
    (\regu\Htermone_i(\regu)+(1-\regu)\Htermtwo_{i,\eps}(\regu))^{-1}[\regu\Htermone_i(\regu)]\Htermtwo_{i,\eps}(\regu)^{-1}\onehot_1}{2}
    \\
    &\leq
    \vecnorm{\Htermtwo_{i,\eps}(\regu)^{-1}\onehot_1}{2}
    \cdot
    (1+\opnorm{(\Htermone_i(\regu)+(1-\regu)\Htermtwo_{i,\eps}(\regu)/\regu)^{-1}}\opnorm{\Htermone_i(\regu)} ),
\end{align*}
where the first equality follows from Woodbury's matrix identity. Since 
$ \vecnorm{\Htermtwo_{i,\eps}(\regu)^{-1}\onehot_1}{2}=1/2$, $\opnorm{\Htermone(\regu)}\leq \boundobsh$ and 
$\opnorm{(\Htermone_i(\regu)+(1-\regu)\Htermtwo_{i,\eps}(\regu)/\regu)^{-1}}\leq \opnorm{\Htermone_i(\regu)^{-1}}\leq \lboundobsh^{-1}$, it follows that 
\begin{align*}
    \vecnorm{(1-\regu) (\regu\Htermone_i(\regu)+(1-\regu)\Htermtwo_{i,\eps}(\regu))^{-1}\onehot_1}{2}
    \leq \frac{(1+\boundobsh/\lboundobsh)}{2}
\end{align*} for any $\eps\geq 0$. When $\regu>0$, since $\hesshort_i(\regu)=\regu\Htermone_i(\regu)+(1-\regu)\Htermtwo_{i,0}(\regu)$ is non-singular, taking $\eps\to 0$ in the bound above yields the desired result.\\

\myunder{Case 2: $\regu=0$:}
we have $\atefun(\EstPar(\regu;\dset_{-i}))=\estate_{-i}$ and $\atefun(\EstPar(\regu;\dsetobs))=\trueate$. The result follows immediately.

\subsubsection{Proof of Lemma~\ref{lm:concentration_estate}}\label{sec:pf_concentration_estate}
\paragraph{Proof of Eq.~\eqref{eq:concentration_estate_1}.}
Eq.~\eqref{eq:concentration_estate_1} follows by  noting that $\estate_i-\ate,i\in[\numfold]$ are i.i.d. $\polyshort/\sqrt{\nexp/K}$ sub-Gaussian random variables  by Assumption~\ref{ass:linear_ate:a}, and applying Hoeffding's inequality.

\paragraph{Proof of Eq.~\eqref{eq:concentration_estate_2}.}
By Assumption~\ref{ass:linear_ate:b}, it can be verified that 
\begin{align*}
    \frac{1}{\numfold}\sum_{i=1}^\numfold (\estate_{-i}-\ate)(\estate_{i}-\ate)
    &=
    \sum_{i=1}^\numfold 
  \empmean{\fold_i}[\Zfuntilone(\expsam)]
  \empmean{[\nexp]\backslash\fold_i}[\Zfuntilone(\expsam)]
  +\remainder_a
\end{align*}
for some $\remainder_a$ such that $|\remainder_a|\leq \polyshort\log^{1.5}(\numfold/\delta)/(\nexp^{1.5}/\numfold)$ with probability at least $1-\delta$. Moreover, 
\begin{align*}
  \MoveEqLeft{\frac{1}{\numfold}  \sum_{i=1}^\numfold 
  \empmean{\fold_i}[\Zfuntilone(\expsam)]
  \empmean{[\nexp]\backslash\fold_i}[\Zfuntilone(\expsam)]}\\
&\leq
\polyshort (\empmean{[\nexp]}[\Zfuntilone(\expsam)])^2
  +
  \frac{\polyshort}{\numfold^2}\sum_{i=1}^\numfold (\empmean{\fold_i}[\Zfuntilone(\expsam)])^2\\
  &\leq 
  \frac{\polyshort\log(1/\delta)}{\nexp}
  +
  \frac{\polyshort}{\nexp} 
  \Big(\sqrt{\frac{\log(1/\delta)}{\numfold}}+{\frac{\log(1/\delta)}{\numfold}}\Big)\\
  &
  \leq \polyshort\frac{\log(1/\delta)}{\nexp}
\end{align*}
with probability at least $1-\delta$ for some constant $\polyshort=\poly>0$, where the second line follows from 
 Hoeffding's inequality and Bernstein's inequality (noting that $(\empmean{\fold_i}[\Zfuntilone(\expsam)])^2$ is $\polyshort/(\nexp/\numfold)$ sub-Exponential). Putting the pieces together and using the fact that $\sqrt{\nexp}\geq \polyshort\numfold(\log^{1.5}(\numfold)+\log^{0.5}(1/\delta))$ yields Eq.~\eqref{eq:concentration_estate_2}.

 \paragraph{Proof of Eq.~\eqref{eq:concentration_estate_3}.}
 Similarly, we have by Lemma~\ref{lm:linear_expansion_estate} that 
 \begin{align*}
    \frac{1}{\numfold}\sum_{i=1}^\numfold (\estate_{-i}-\ate)^2
    &\leq 
    \frac{2}{\numfold}
     \sum_{i=1}^\numfold 
  (\empmean{[\nexp]\backslash\fold_i}[\Zfuntilone(\expsam)])^2
 +
  \remainder_b
\end{align*}
for some $\remainder_b$ such that $|\remainder_b|\leq \polyshort\log^{2}(\numfold/\delta)/(\nexp)^2$ with probability at least $1-\delta$. Moreover, basic algebra gives
\begin{align*}
  \frac{1}{\numfold}  \sum_{i=1}^\numfold 
  (\empmean{[\nexp]\backslash\fold_i}[\Zfuntilone(\expsam)])^2
&\leq
4\Big[(\empmean{[\nexp]}[\Zfuntilone(\expsam)])^2 + \frac{1}{\numfold^3}\sum_{i=1}^\numfold (\empmean{\fold_i}[\Zfuntilone(\expsam)])^2\Big]\\
&\leq \polyshort\frac{\log(1/\delta)}{\nexp},
\end{align*}
where the second inequality follows from the same argument as in the proof of Eq.~\eqref{eq:concentration_estate_2}. Putting the pieces together and using the fact that $\sqrt{\nexp}\geq \polyshort\numfold(\log^{1.5}(\numfold)+\log^{0.5}(1/\delta))\geq\polyshort (\log \numfold+\log^{0.5}(1/\delta))$ yields Eq.~\eqref{eq:concentration_estate_3}.

\subsubsection{Proof of Lemma~\ref{lm:linear_expansion_estate}}\label{sec:pf_linear_expansion_estate}
    Adopt the shorthand notation $\atepardiff\defn \estatepar-\trueatepar$. 
    By a Taylor expansion on $\sum_{j\in\samset} \Zfun(\expsam_j;\estatepar)- \sum_{j\in\samset} \Zfun(\expsam_j;\trueatepar)$, we have
    \begin{align*} 
      \empmean{\samset}\Big[\int_0^1  \nabla\Zfun(\expsam;\trueatepar+t\atepardiff) dt \Big]\atepardiff 
       &=
        - \empmean{\samset}\Big[ \Zfun(\expsam;\trueatepar)\Big].
    \end{align*}
    Thus, 
    \begin{align*}
      &\quad
      \estatepar-\trueatepar + [\nabla\PopZfun(\trueatepar)]^{-1}\empmean{\samset}[ \Zfun(\expsam;\trueatepar)]\\
        &=
       \Big[[\nabla\PopZfun(\trueatepar)]^{-1} - \empmean{\samset}\Big[\int_0^1  \nabla\Zfun(\expsam;\trueatepar+t\atepardiff) dt\Big]^{-1}\Big] \empmean{\samset}\Big[ \Zfun(\expsam;\trueatepar)\Big]\\
       &=
       [\nabla\PopZfun(\trueatepar)]^{-1} \Big(\empmean{\samset}\Big[\int_0^1  \nabla\Zfun(\expsam;\trueatepar+t\atepardiff) dt \Big]
       -\nabla\PopZfun(\trueatepar)\Big)\\
       &~~~~\cdot
       \Big[\empmean{\samset}\Big[\int_0^1  \nabla\Zfun(\expsam;\trueatepar+t\atepardiff) dt \Big]\Big]^{-1} \empmean{\samset}\Big[ \Zfun(\expsam;\trueatepar)\Big].
    \end{align*}
 Recall that $\vecnorm{\atepardiff}{2}\leq \frac{\polyshort\sqrt{\log(1/\delta)}}{\sqrt{\samsetnum}}$ with probability at least $1-\delta$ by Assumption~\ref{ass:zest_as_ate}.
    We claim that  there exist some constants $\polyshort,\polyshortprime>0$ such that when $\samsetnum\geq \polyshortprime \log(1/\delta)$, with probability at least $1-\delta$, 
    \begin{subequations}
    \begin{align}
        \opnorm{\empmean{\samset}[  \nabla\Zfun(\expsam;\trueatepar) ]-\nabla\PopZfun(\trueatepar)} 
        &
        \leq \polyshort\sqrt{\frac{\log(1/\delta)}{\samsetnum}}\leq \frac{\strongcvx}{4},\label{eq:first_order_apprx_claim_0}
         \\
\vecnorm{\empmean{\samset}\Big[\Zfun(\expsam;\trueatepar)\Big]}{2}
& \leq 
\frac{\polyshort\sqrt{\log(1/\delta)}}{\sqrt{\samsetnum}},\label{eq:first_order_apprx_claim_1}
\\
\opnorm{\empmean{\samset}\Big[\int_0^1  \nabla\Zfun(\expsam;\trueatepar+t\atepardiff) dt\Big]-\empmean{\samset}\Big[\nabla\Zfun(\expsam;\trueatepar)\Big]}
&\leq \polyshort\vecnorm{\atepardiff}{2}\leq \frac{\strongcvx}{4}.\label{eq:first_order_apprx_claim_2}
    \end{align}
    \end{subequations}
   Putting the claims together,  noting that  $\sigma_{\min}(\nabla\PopZfun(\trueatepar))\geq\strongcvx$ and applying the triangle inequality and a union bound, we have
\begin{align*}
\MoveEqLeft{\quad\vecnorm{\estatepar-\trueatepar + [\nabla\PopZfun(\trueatepar)]^{-1} \empmean{\samset}[ \Zfun(\expsam;\trueatepar)]}{2}}\\
      &\leq
      \frac{\polyshort}{\strongcvx}
      \cdot 
      \Big(\vecnorm{\atepardiff}{2}+\frac{\sqrt{\log(1/\delta)}}{\sqrt{\samsetnum}}\Big)
      \cdot \frac{1}{\strongcvx} \cdot \sqrt{\frac{\log(1/\delta)}{\samsetnum}}\\
      &\leq \polyshort\cdot\frac{{\log(1/\delta)}}{{\samsetnum}},
\end{align*}
with probability at least $1-\delta$ when $\samsetnum\geq \polyshortprime\log(1/\delta)$ for some constant $\polyshortprime = \polyprimez>0$ sufficiently large. The bound on $\estate(\expsam_{\samset})-\trueate$ follows immediately from taking the first coordinate of $\estatepar-\trueatepar$.

    \paragraph{Proof of the claims.}
    Claim~\eqref{eq:first_order_apprx_claim_0} follows from
    applying Hoeffding's inequality to each element of the matrix and a union bound; claim~\eqref{eq:first_order_apprx_claim_1} again follows from Hoeffding's inequality for each element of the vector and a union bound; claim~\eqref{eq:first_order_apprx_claim_2} uses the assumption that $\opnorm{\nabla^2\Zfun(\expsam;\atepar)}\leq \bZfuntwo$.

\subsubsection{Proof of Lemma~\ref{lm:convexity_atepar}}\label{sec:pf_convexity_atepar}
    First, $\sigma_{\min}(\nabla\PopZfun(\trueatepar))\geq \strongcvx$ since condition~\ref{ass:convexity_atepar} assumes $\nabla\PopZfun(\trueatepar)\succeq \gamma \IdMat$.
The proof of the second part of this lemma follows from standard nonasymptotic analysis of the maximum likelihood estimator (MLE) (see e.g.,~Lemma~9 in~\citet{lin2024worthwhile}). Namely, we will show the following claims:
\begin{subequations}
    \begin{itemize}
        \item[1.] There exists some constant $\polyshort>0$ such that  for any $\delta\in(0,1/2)$, 
\begin{align}
\sup_{\atepar\in\ateparspace}\vecnorm{\PopZfun(\atepar)-\empmean{\samset}[ \Zfun(\expsam;\atepar)]}{2}\leq \polyshort\cdot\frac{\sqrt{\log(1/\delta)}}{\sqrt{\samsetnum}}\label{eq:convexity_atepar_claim_1}
\end{align}
with probability at least $1-\delta$.
        \item[2.]
\begin{align}
   \inprod{\PopZfun(\atepar)}{\atepar-\trueatepar}\geq 
   \begin{cases}
    \frac{\strongcvx}{2}\cdot\vecnorm{\atepar-\trueatepar}{2}^2, & \text{if } \vecnorm{\atepar-\trueatepar}{2}\leq \frac{\strongcvx}{2\bZfuntwo},\\
    \frac{\strongcvx^2}{4\bZfuntwo}\cdot\vecnorm{\atepar-\trueatepar}{2}, & \text{if } \vecnorm{\atepar-\trueatepar}{2}> \frac{\strongcvx}{2\bZfuntwo}.
   \end{cases}
   \label{eq:convexity_atepar_claim_2}
\end{align}
\end{itemize}
\end{subequations}
From claim~\eqref{eq:convexity_atepar_claim_1}, we have
\begin{align*}
    \MoveEqLeft{\inprod{\PopZfun(\estatepar)}{\estatepar-\trueatepar} = 
\inprod{\PopZfun(\estatepar)-\empmean{\samset}[ \Zfun(\expsam;\estatepar)]}{\estatepar-\trueatepar}}
\\
&\leq
\sup_{\atepar\in\ateparspace}| \inprod{\PopZfun(\atepar)-\empmean{\samset}[ \Zfun(\expsam;\atepar)]}{\estatepar-\trueatepar}
\\
& 
   \leq
   \sup_{\atepar\in\ateparspace}\vecnorm{\PopZfun(\atepar)-\empmean{\samset}[ \Zfun(\expsam;\atepar)]}{2}\cdot\vecnorm{\estatepar-\trueatepar}{2}
   \leq
   \polyshort\cdot\frac{\sqrt{\log(1/\delta)}}{\sqrt{\samsetnum}}\cdot\vecnorm{\estatepar-\trueatepar}{2}.
\end{align*}Combining this with claim~\eqref{eq:convexity_atepar_claim_2} and noting  $\samsetnum\geq \polyshort\log(1/\delta)$ for some $\polyshort>0$ sufficiently large yields Lemma~\ref{lm:convexity_atepar}.

\paragraph{Proof of claim~\eqref{eq:convexity_atepar_claim_1}.}
Let $\PopZfun_i(\atepar)$ denote the $i$-th element of the vector $\PopZfun(\atepar)$. For each $i\in[\datepar]$, let
$\PopZfprocess^i_{\atepar}\defn \PopZfun_i(\atepar)-\empmean{\samset}[ \Zfun_i(\expsam;\atepar)],~\atepar\in\ateparspace$ is a sub-Gaussian process with respect to the metric 
$\metric(\atepar_a,\atepar_b)\defn \frac{\bZfunzero\cdot\vecnorm{\atepar_a-\atepar_b}{2}}{\sqrt{\samsetnum}}$. Thus, by Dudley's entropy integral bound (see e.g.,~Theorem~5.22 in~\citet{wainwright2019high}), we have
\begin{align*}
\MoveEqLeft{~\quad\E[\sup_{\atepar\in\ateparspace}|\PopZfun_i(\atepar)-\empmean{\samset}[ \Zfun_i(\expsam;\atepar)]|]
    \leq
  c\int_0^{\boundatepar\bZfunzero/\sqrt{\samsetnum}}\log \covernum(\epsilon;\metric,\ateparspace) d\epsilon}\\
  & = 
  \frac{c\bZfunzero}{\sqrt{\samsetnum}}\int_0^{\boundatepar} \log \covernum\Big(\frac{\bZfunzero}{\sqrt{\samsetnum}}\cdot t;\metric,\ateparspace\Big) d t
  = 
  \frac{c\bZfunzero}{\sqrt{\samsetnum}}\int_0^{\boundatepar} \log \covernum( t;\vecnorm{\cdot}{2},\ateparspace) d t\\
  &
 \overset{(i)}{\leq}
  \frac{c\bZfunzero}{\sqrt{\samsetnum}}\cdot\int_0^{\boundatepar} \datepar \log\Big(1+2\frac{\boundatepar}{t}\Big) d t
  \leq 
  \frac{c\bZfunzero\cdot\datepar}{\sqrt{\samsetnum}}\cdot\boundatepar\leq \frac{\polyshort}{\sqrt{\samsetnum}},
\end{align*}
where step~(i) follows from the fact that $\covernum( t;\vecnorm{\cdot}{2},\ateparspace)\leq (1+2\boundatepar/t)^{\datepar}$ (see e.g., example~5.8 in~\citet{wainwright2019high}). Combining this with a concentration inequality for functions with bounded differences (see e.g.,~Corollary~2.21 in~\citet{wainwright2019high}), we arrive at 
\begin{align*}
\sup_{\atepar\in\ateparspace}|\PopZfun_i(\atepar)-\empmean{\samset}[ \Zfun_i(\expsam;\atepar)]|
    \leq \frac{\polyshort}{\sqrt{\samsetnum}}\cdot\Big(1+{\sqrt{\log(1/\delta)}}\Big)\leq \polyshort\cdot\frac{\sqrt{\log(1/\delta)}}{\sqrt{\samsetnum}}
\end{align*} 
with probability at least $1-\delta$ for some constant $\polyshort=\polyz>0$.

\paragraph{Proof of claim~\eqref{eq:convexity_atepar_claim_2}.}
When  $\vecnorm{\atepar-\trueatepar}{2}\leq \frac{\strongcvx}{2\bZfuntwo}$, we have
\begin{align*}
   \MoveEqLeft{\quad ~\inprod{\PopZfun(\atepar)}{\atepar-\trueatepar}}
   \\
    &=
    \inprod{\PopZfun(\atepar)-\PopZfun(\trueatepar)}{\atepar-\trueatepar}\\
    &=
    (\atepar-\trueatepar)^\top{\nabla\PopZfun(\trueatepar)}(\atepar-\trueatepar)\\
    &\qquad
    +
    (\atepar-\trueatepar)^\top\Big[\int_0^1 \nabla\PopZfun(\trueatepar+t(\atepar-\trueatepar)) dt-\nabla\PopZfun(\trueatepar)\Big](\atepar-\trueatepar)\\
    &\geq
    \strongcvx\cdot\vecnorm{\atepar-\trueatepar}{2}^2 -\bZfuntwo \vecnorm{\atepar-\trueatepar}{2}^3
    \geq \frac{\strongcvx}{2}\cdot\vecnorm{\atepar-\trueatepar}{2}^2.
\end{align*}
This proves the first case.
Introduce the unit-norm vector $\normaldiff\defn \frac{\atepar-\trueatepar}{\vecnorm{\atepar-\trueatepar}{2}}$.
Similarly, when $\vecnorm{\atepar-\trueatepar}{2}> \frac{\strongcvx}{2\bZfuntwo}$, we have
\begin{align*}
    \inprod{\PopZfun(\atepar)}{\normaldiff}
    &=
    \inprod{\PopZfun(\atepar)-\PopZfun(\trueatepar)}{\normaldiff}
   =
    \inprod{\int_0^{\vecnorm{\atepar-\trueatepar}{2}} ~\nabla\PopZfun(\trueatepar+t\normaldiff) dt
    ~ \normaldiff}{\normaldiff}\\
    &\overset{(i)}{\geq}
    \inprod{\int_0^{\gamma/(2\bZfuntwo)} ~\nabla\PopZfun(\trueatepar+t\normaldiff) dt
    ~ \normaldiff}{\normaldiff}
    \overset{(ii)}{\geq} \frac{\strongcvx^2}{4\bZfuntwo}\cdot\vecnorm{\normaldiff}{2}^2= \frac{\strongcvx^2}{4\bZfuntwo},
\end{align*}
where step~(i) uses $\nabla\PopZfun(\atepar)\succeq \bzero$  for all $\atepar\in\ateparspace$ and step~(ii) follows from  
\begin{align*}
    \sigma_{\min}(\nabla\PopZfun(\trueatepar+t\normaldiff))
    &\geq 
    \sigma_{\min}(\nabla\PopZfun(\trueatepar)) - \opnorm{\nabla\PopZfun(\trueatepar+t\normaldiff)-\nabla\PopZfun(\trueatepar)}\\
    &
    \geq 
    \strongcvx - t\vecnorm{\normaldiff}{2}\bZfuntwo
    =
    \strongcvx - t\bZfuntwo\geq \frac{\strongcvx}{2}
\end{align*} when $t\leq \gamma/(2\bZfuntwo)$.

\end{document}